\documentclass[10pt,journal,compsoc,review]{IEEEtran}
\ifCLASSOPTIONcompsoc
  \usepackage[nocompress]{cite}
\else
   \usepackage{cite}
\fi
\ifCLASSINFOpdf
  \usepackage[pdftex]{graphicx}
\else
\fi
\usepackage{url}

\usepackage{amsmath}
\usepackage{multirow}
\usepackage{dblfloatfix}
\usepackage{cite}
\usepackage{eurosym}
\usepackage{graphicx}
\usepackage{amssymb}
\usepackage{algorithm}
\usepackage{algpseudocode}
\usepackage{mathtools}
\usepackage{enumitem}
\usepackage{setspace}
\usepackage{hyphenat}
\usepackage[mathscr]{euscript}

\newcommand{\RR}{I\!\!R}
\def\bsq#1{
\lq{#1}\rq}
\def\eg{\textit{e.g.}}
\def\ie{\textit{i.e.}}

\begin{document}
%
\title{Expert and Crowd-Guided Affect Annotation and Prediction}
%
%
%
%

\author{Ramanathan~Subramanian,~\IEEEmembership{Senior Member,~IEEE,}
	      Yan~Yan,~\IEEEmembership{Member,~IEEE,}
        Nicu~Sebe,~\IEEEmembership{Senior Member,~IEEE}

\IEEEcompsocitemizethanks{\IEEEcompsocthanksitem Ramanathan Subramanian is with the University of Canberra, Bruce, ACT 2617. (Email: ramanathan.subramanian@ieee.org) 

\IEEEcompsocthanksitem Yan Yan is with the Department of Computer Science, Illinois Institute of Technology, 60616 Illinois, USA. (Email: yyan34@iit.edu)

\IEEEcompsocthanksitem Nicu Sebe is with the  Dept. of Information Engineering and Computer Science, University of Trento, 38123 Trento, Italy. (Email: sebe@disi.unitn.it)

} 
\thanks{}}

%
%

\markboth{IEEE Transactions on Affective Computing, July~2015}%
{IEEE Transactions on Affective Computing, July~2015}
\IEEEtitleabstractindextext{%
\begin{abstract}
We employ crowdsourcing to acquire time-continuous affective annotations for movie clips, and refine noisy models trained from these crowd annotations incorporating expert information within a Multi-task Learning (MTL) framework. We propose a novel \textbf{e}xpert \textbf{g}uided MTL (EG-MTL) algorithm, which minimizes the loss with respect to both crowd and expert labels to learn a set of weights corresponding to each movie clip for which crowd annotations are acquired. We employ EG-MTL to solve two problems, namely, \textbf{\texttt{P1}}: where dynamic annotations acquired from both experts and crowdworkers for the \textbf{Validation} set are used to train a regression model with audio-visual clip descriptors as features, and predict dynamic arousal and valence levels on 5--15 second snippets derived from the clips; and \textbf{\texttt{P2}}: where a classification model trained on the \textbf{Validation} set using dynamic crowd and expert annotations (as features) and static affective clip labels is used for binary emotion recognition on the \textbf{Evaluation} set for which only dynamic crowd annotations are available. Observed experimental results confirm the effectiveness of the EG-MTL algorithm, which is reflected via improved arousal and valence estimation for \textbf{\texttt{P1}}, and higher recognition accuracy for \textbf{\texttt{P2}}.     
\end{abstract}

\begin{IEEEkeywords}
Crowdsourcing, Affect annotation and Prediction, Multi-task learning, Expert-guided.
\end{IEEEkeywords}}

\maketitle

\IEEEdisplaynontitleabstractindextext

%
\IEEEpeerreviewmaketitle

\IEEEraisesectionheading{\section{Introduction}\label{sec:introduction}}

%
%
%
%
\IEEEPARstart{A}ffective video tagging has been acknowledged as an important multimedia problem for long, given its utility for applications such as personalized media recommendation and video summarization. For over a decade, many affective tagging methodologies have been proposed in literature-- these approaches can be broadly classified as \textit{content-centric}~\cite{HanjalicITM2005,wang2006affective}, which analyze audio-visual content in the stimulus to predict the conveyed emotion, or \textit{user-centric}~\cite{MAHNOB-HCI,DEAP,abadi2015decaf,PatrasIVC,DBLP:conf/acii/AbadiKRAS13,10.1145/1873951.1874047,10.1145/3242969.3242988} that infer the evoked emotion by examining the viewer's facial expressions and physiological responses (\eg, brain signals, eye and muscle movements). While human behavioral signals such as cognitive and emotional states are inherently dynamic and constantly evolve over time, most affective tagging approaches only associate a single, static emotional label to the stimulus.

The above limitation in affective computing can be attributed to a number of factors. Firstly, \textit{interpreting} and \textit{measuring} emotion in terms of \textit{arousal} and \textit{valence}\footnote{We will only discuss the dimensional or circumplex model of affect~\cite{russell1980circumplex} in this work.} is an inherently difficult problem-- emotion is a highly subjective feeling, and the discrepancy between the emotion \textit{envisioned} by the content creator versus the actual emotion \textit{evoked} in viewers has been highlighted by many works. This subjectivity results in noisy or \bsq{ill-generalizable} machine learning models~\cite{MihalisTPAMI}, and state-of-the-art static emotion tagging methods can only achieve less than 70\% accuracy employing multimodal user signals recorded in controlled lab conditions. Secondly, acquiring a sufficient number of time-continuous emotional annotations to model the dynamic \bsq{ground-truth} is an extremely tedious and difficult task as humans are better at rating attributes (especially behavioral) in relative rather than absolute terms~\cite{soleymaniCMM13,AngelikiFG13}. 

One attractive solution to acquire large amounts of dynamic annotations is \emph{crowdsourcing} (CS). Recently, CS has become popular for performing tedious and large-scale annotation tasks via human collaboration via the Internet. When it is difficult to employ a sufficient number of experts for analyzing extensive data, CS is an efficient alternative as many individuals work on small data chunks to provide useful information in the form of tags. CS has been successfully employed to develop data-driven solutions for computationally difficult problems in multiple domains like natural language processing~\cite{williams2011crowd} and computer vision~\cite{YuenRLT09}. Two reasons mainly contribute to the success of CS-- (1) crowd workers are paid a fraction of the wages that experts are entitled to, thereby achieving cost efficiency, and (2) the experimenter's task becomes scalable when the original task is split into smaller and manageable micro-tasks and distributed among crowdworkers. Nevertheless, cost-effectiveness is achieved at the expense of expertise; crowdworkers may lack the motivation and/or technical and cognitive skills to effectively perform a given task~\cite{ross2010crowdworkers}. Therefore, efficient machine learning techniques robust to noisy data are crucial to the success of CS approaches. 

This paper is an extension of the work presented in~\cite{Abadi2014}, where Multi-task learning (MTL) was employed to learn a crowd-based model for predicting dynamic arousal (A) and valence (V) levels in movie snippets. Given a set of \textit{related} tasks, MTL \textit{simultaneously} learns all tasks by modeling the similarities as well as differences among them to build task-specific classification or regression models. This joint learning accounting for task relationships is more efficient than learning each task independently. Given a set of movie clips whose emotional attributes are known, say via an oracle, one can expect clips of similar nature (\eg, high valence) to have some similarities in terms of content, and conversely, in terms of the responses they evoke from viewers. Therefore, MTL can be seen as a naturally useful tool for large-scale affective media tagging. We extend the MTL idea by incorporating \textit{expert} knowledge in the learning process to refine crowd models.   

In this work, we seek to improve the efficacy of crowd-based models by employing a small amount of \textit{expert} data to guide the learning process-- we employed 16 experts familiar with emotional attributes to provide dynamic A,V ratings for the clips annotated by crowdworkers\footnote{It is reasonable to expect experts to provide more consistent annotations as compared to crowdworkers, and this is reflected in the form of considerably higher inter-expert agreement scores-- see Section~\ref{ExpRes}.}, and enhance crowd models employing expert knowledge via a novel \textbf{e}xpert-\textbf{g}uided MTL (EG-MTL) algorithm. The EG-MTL algorithm seeks to simultaneously minimize the loss with respect to both crowd and expert labels in the optimization framework and learns a set of weights corresponding to each of the movie clips for which crowd annotations were sought. The learning framework is flexible, \ie, it is guided by both \textit{a-priori} knowledge embedded in the form of a graph, as well as data descriptors.  

More specifically, we employ EG-MTL to solve problems \textbf{\texttt{P1}} and \textbf{\texttt{P2}} described below:
\begin{itemize}[noitemsep]
\item[\textbf{\texttt{P1}}:] For a dataset comprising both crowd and expert dynamic emotion annotations (termed \textbf{Validation} set), refine a regression model trained using crowd labels and audio-visual clip descriptors via expert knowledge\footnote{We also show regression performance employing annotations from only 7 experts, which is insufficient for training a model as such.} for dynamic A,V prediction on 5--15 second snippets (derived from the same clips).  
\item[\textbf{\texttt{P2}}:] Train a binary classification (emotion recognition) model on the \textbf{Validation} set using static \bsq{ground-truth} A,V labels and the dynamic A,V crowd ratings as features. Incorporate expert knowledge to refine this model and enhance recognition on a second \textbf{Evaluation} set, for which only dynamic crowd ratings are available.   
\end{itemize}

We show that the performance of crowd-based models trained for \textbf{\texttt{P1}} and \textbf{\texttt{P2}} can be significantly improved upon incorporating expert information via EG-MTL. Overall, this work makes the following research contributions:
\begin{itemize}[noitemsep]
\item[1.] While other works such as~\cite{MihalisTPAMI,Raykar2010} have focused on fusing information from multiple annotators to obtain a representative \bsq{gold standard} annotation, this is the first affective computing work that focuses on enhancing noisy crowd-based models via expert knowledge.  
\item[2.] This is also the first work to employ Multi-task Learning for continuous/static emotion prediction. MTL can effectively reveal latent relationships between related tasks, and we employ the same to investigate a) similarities among time-continuous annotations of multiple raters, and b) similarities among audio-visual features corresponding to high/low arousal and valence content. The proposed EG-MTL algorithm is flexible as the learning is guided by both \textit{a-priori} knowledge defining task relationships in the form of a graph, as well as data descriptors employed for model training.
\item[3.] Different from prior works such as~\cite{AngelikiFG13,soleymaniCMM13} which have cursorily examined the relationship between dynamic and static A,V labels, this work more thoroughly examines the influence of dynamic emotion changes on the overall emotion evoked by a movie scene. As typified by problem \textbf{\texttt{P2}}, we explicitly attempt static A,V prediction from dynamic A,V profiles.
\item[4.] Given that movie clips are unique stimuli in the sense that they convey emotions more dynamically and effectively than other media (\eg, music videos as shown in~\cite{abadi2015decaf}), we believe that the compiled dynamic expert and crowd annotations will be of great value to the affective computing community. These annotations will be made publicly available for future research\footnote{Upon acceptance of this manuscript.}.  
\end{itemize}

\begin{sloppypar}
The paper is organized as follows: Section~\ref{RelWorks} overviews the literature. Description of the stimuli and protocol employed for acquiring affective annotations is presented in Section~\ref{Prot}. A brief overview of various MTL baselines along with a description of the proposed EG-MTL algorithm is provided in Section~\ref{MTL}. Annotation data analysis and emotion prediction experiments are detailed in Section~\ref{ExpRes}. Conclusions are stated in Section~\ref{Con}.
\end{sloppypar}

\section{Related work}\label{RelWorks}

This section examines related work on (1) Crowdsourcing for media processing, (2) Affective analysis and tagging, (3) CS for affective Computing and (4) Multi-task learning.

\subsection{Crowdsourcing for media processing}
Steiner \emph{et al.}~\cite{Steiner2011} defined three types of video events, namely, visual events, occurrence events and Internet-based events, and showed that these events can be detected from video sequences via crowdsourcing upon combining textual, visual and behavioral cues. Vondrick \emph{et al.}~\cite{VondrickRP10} argued that frame-by-frame video annotation is essential for a variety of tasks, as in the case of time-continuous emotion measurement, even if it is tedious to accomplish for human annotators. An online framework for collecting valid facial responses to media content was proposed by McDuff \emph{et al.}~\cite{mcduff2012crowdsourcing}, who found significant differences between subgroups who liked/disliked or were familiar/unfamilar with a particular commercial.

\subsection{Affective analysis and tagging}
A primary issue in affective multimedia analysis is the inherent difficulty in finding a sufficient number of knowledgable annotators so as to generate reliable ground-truth labels for model training. Consequently, only a few annotators are used in a majority of affective studies~\cite{soleymani2008affective,wang2006affective}. Also, emotion perception varies with individual traits such as personality~\cite{Kehoe2012a}, and considerable differences may be observed in affective ratings compiled from different persons over a small population. To address this problem, a number of works have turned to crowdsourcing or large-scale user studies. In a  seminal study affective movie study, Gross \emph{et al.}~\cite{gross1995emotion} compiled a benchmark collection of movie clips for evoking eight emotional states such as anger, disgust, fear and neutral, based on emotion ratings compiled for 250 movie clips from 954 subjects.

DEAP~\cite{DEAP}, MAHNOB~\cite{MAHNOB-HCI} and DECAF~\cite{abadi2015decaf} are three recent works that have attempted user-centric static affect recognition via physiological responses compiled from 27--40 users to music and movie stimuli. Given the aforementioned differences in emotion perception and the rather small number of participants involved, it is unsurprising that these state-of-the-art approaches do not achieve high recognition accuracies, even with user data compiled under controlled lab conditions. Challenges in acquiring time-continuous emotion annotations on a large database are discussed in~\cite{AngelikiFG13}. This work highlights some inherent problems concerning dynamic annotation of emotional attributes such as (i) definition of fuzzy attributes such as A,V being less-intuitive to some annotators, (ii) humans being more adept at rating relative rather than absolute attributes, and (iii) the existence of subject-specific annotation time-delays between occurrence of an emotional event and its annotation for continuous ratings. Among the handful of works that have attempted continuous emotion prediction, Nicolau \emph{et al.}~\cite{Nicolaou2012} proposed a output-associative relevance vector machine based regression framework for A,V estimation from multiple non-verbal cues such as facial expressions, shoulder movements and audio cues.  

\subsection{Crowdsourcing for affective computing}
Soleymani \emph{et al.}~\cite{soleymani2009collaborative} performed CS on a limited scale to collect 1300 affective annotations from 40 volunteers for 155 Hollywood movie clips. In another CS-based affective video annotation study, Soleymani \emph{et al.}~\cite{pun011} compiled annotations for the MediaEval 2010 Affect Task Corpus on AMT, and asked workers to self-report their boredom levels. In a recent CS-based media tagging work, Soleymani \emph{et al.}~\cite{soleymaniCMM13} presented a dataset of 1000 songs for music emotion analysis, each annotated continuously over time by at least 10 users. Nevertheless, movies best approximate the real world and are more effective in eliciting emotions from viewers as compared to musical content as shown in in~\cite{abadi2015decaf}, which is why we believe continuous emotion annotation and prediction in movie stimuli is valuable in the context of affective media representation and modeling. Also, fusing noisy annotations from multiple crowdworkers to obtain a representative ground-truth annotation becomes crucial in the context of CS-- Raykar 
\emph{et al.}~\cite{Raykar2010} proposed an annotation fusion mechanism where the multiple annotations are assumed to be Gaussian distributed, whose mean represents the true label and whose variance is denoted by the annotation noise. Nicolaou \emph{et al.}~\cite{MihalisTPAMI} proposed an improved annotation fusion methodology for continuous annotations employing probabilistic canonical correlation analysis (PCCA). Their approach also included a latent time warping process to account for annotator-specific temporal lags. 

\subsection{Multi-task learning}
\begin{sloppypar}
Recently, multi-task learning (MTL) has been employed in several computer vision applications such as image classification~\cite{yuan}, head pose estimation~\cite{yan2013no} and multi-view action recognition~\cite{yan2014multi}. Given a set of related tasks, MTL~\cite{caruana1998multitask} seeks to simultaneously learn a set of task-specific classification or regression models. The intuition behind MTL is simple: a joint learning procedure which accounts for task relationships is expected to lead to more accurate models as compared to learning each task separately. While MTL has been used previously for learning from noisy crowd annotations~\cite{AAAIHiro}, MTL has not been hitherto used for  affective media tagging. 
\end{sloppypar}

\subsection{Analysis of related work}

A careful examination of related literature suggests that (i) Due to the subjectivity in emotion perception, a vast majority of affective computing studies have involved small or mid-sized user populations, and only recently, has attention been devoted to compiling a large repository of affective annotations via crowdsourcing. (ii) Most affective studies have been restricted to predicting the overall emotion evoked by stimuli rather than the continuous emotion profile-- this is mainly due to the fact that acquiring reliable dynamic affective annotations to serve as ground-truth is both difficult and tedious. Of late, there has been increasing interest in continuous emotion prediction. (iii) Fusion of multiple annotations to synthesize a representative annotation that can be used for training generalizable models is itself a non-trivial task. 

This work represents one of the first attempts to take affective computing \bsq{beyond the lab}, and focuses on refining models trained from dynamic crowd annotations using a small amount of expert knowledge. Note that a similar framework can be employed to harness non-verbal behavioral cues as well. The following section details the stimuli used and protocol adopted for compiling affective crowd annotations.      

\section{Experimental setup}\label{Prot}
For this study, 
%
we compiled the \textbf{Validation} (\textbf{Val}) and \textbf{Evaluation} (\textbf{Eval}) datasets containing arousal (A) and valence (V) annotations from \textit{crowdworkers} and/or \textit{experts}. Characteristics of the \textbf{Val} and \textbf{Eval} sets are summarized in Table~\ref{Overview}. For both the \textbf{Val} and \textbf{Eval} sets, time-continuous A,V annotations were acquired for 12 affective movie clips used in DECAF~\cite{abadi2015decaf}. The main differences between crowdworkers and experts are as follows: (1) Experts were individuals familiar with affective research and the definition of dimensional emotional attributes such as valence and arousal, but crowdworkers were not; (2) Experts were explicitly asked to view the movie clips as many times as required in order to familiarize themselves with the emotional dynamics prior to clip rating, but crowdworkers were given no such instruction, and (3) Experts were asked to provide ratings that \bsq{reflected the level of A/V meant to be elicited in the viewer} during each time instant so as to mimic the movie director's intentions, while crowdworkers were asked to report \bsq{how they felt} through the course of the scene.  

For the \textbf{Val} set, dynamic emotional annotations were acquired from both crowdworkers (via the CrowdFlower platform as detailed below) and 16 experts (via the GTrace interface~\cite{Gtrace}) which records A,V ratings in the range $[-1,1]$. For the purpose of sanity checking and to examine the relationship between dynamic and static emotional ratings\footnote{This is left to future work and not investigated in this study.}, crowdworkers were also asked to report their overall affective impression of the clip. For the \textbf{Eval} set, only continuous A,V annotations were obtained using GTrace from 35 subjects, mainly undergraduate and graduate students naive to the purpose of this study-- they are also referred to as a \bsq{crowd} in this paper since their characteristics and instructions provided were identical to those of the crowdworkers mentioned above. We now describe some salient aspects of the data acquisition process, focusing mainly on the framework employed for acquiring crowd annotations for the \textbf{Val} set.  

\subsection{Movie clips}

All movie clips used in this work were adopted from DECAF~\cite{abadi2015decaf}, and for each of the \textbf{Val} and \textbf{Eval} sets, 12 clips equally distributed among the four quadrants, namely, high A high V (HAHV). low A high V (LAHV), low A low V (LALV) and high A high V (HAHV) in the AV space were used as shown in Table~\ref{table1}. The movie clips are about 1--1.5 minutes long, and the DECAF study also provides a characteristic emotion tag and a mean static A,V rating for each clip in the range $[-2,2]$, based on self-reports from 42 viewers-- these ratings were used to determine static, binary \bsq{ground-truth} labels for the \textbf{Val} and \textbf{Eval} movie clips.

\begin{table}[t]
\fontsize{7}{7}\selectfont
\renewcommand{\arraystretch}{1.2}
\caption{\label{Overview} Overview of the \textbf{Validation} (Val) and \textbf{Evaluation} (Eval) datasets. }
\vspace{-.2cm}
\centering
\begin{tabular}{|p{3.6cm}|c|c|}
\hline
\centering{\textbf{Attribute}} & \textbf{Val} & \textbf{Eval} \\ \hline
\centering{\textbf{Crowd annotations}} & yes & yes \\
\centering{\textbf{Expert annotations}} & yes & no \\
\centering{\textbf{No. of clips}}& 12 & 12 \\
\centering{\textbf{No. of crowd annotations/clip}} & 15+ & 35 \\
\centering{\textbf{No. of expert annotations/clip}} & 16 &  - \\
\centering{\textbf{Type of crowd annotations}} & Dynamic, static A,V & Dynamic A,V \\
\centering{\textbf{Type of expert annotations}} & Dynamic A,V & - \\
\hline
\end{tabular}
\end{table}

\begin{table*}[t]
\small 
\renewcommand{\arraystretch}{1.2}
\caption{\label{table1} Statistics concerning movie clips used in the study. Valence, arousal ratings are on a scale of $[-2 , 2]$, while clip length is denoted in seconds. Emotion tag refers to the most strongly evoked emotion among observers for a movie clip as per~\cite{abadi2015decaf}.}
\vspace{-.2cm}
\centering
\begin{tabular}{|p{3cm}|c|c|c|c|c|c|c|c|}
\hline
& \multicolumn{4}{c|}{\textbf{Val}} & \multicolumn{4}{c|}{\textbf{Eval}} \\ \hline
&  HAHV  & LAHV &  LALV & HALV & HAHV  & LAHV &  LALV & HALV \\
\hline  \hline
\textbf{No. of video clips}  & 3& 3 & 3& 3 & 3& 3 & 3& 3\\
\textbf{Emotion tags} & \textit{amusing}&\textit{happy}& \textit{angry} & \textit{disgust}, \textit{fear}&\textit{funny}&\textit{happy}&\textit{sad}&\textit{disgust}, \textit{shock}\\
\textbf{Length \ ($\mu \pm \sigma$)}  & 95.8$\pm$9.3 & 75.2$\pm$16.2 & 92.1$\pm$14.7&85.5$\pm$6.5&64.1$\pm$7.9
&76.8$\pm$15.3&71.4$\pm$13.3&91.8$\pm$34.2\\
\textbf{Valence ($\mu \pm \sigma$)}&  1.1$\pm$0.1 & 1.1$\pm$0.3  & -0.8$\pm$0.3 & -1.3$\pm$0.4&1.2$\pm$0.3
&0.9$\pm$0.1&-0.9$\pm$0.1&-0.7$\pm$1.6\\
\textbf{Arousal ($\mu \pm \sigma$)}&  1.1$\pm$0.1 & -0.9$\pm$0.8 & -0.2$\pm$1 & 1.1$\pm$0.6&0.9$\pm$0.3
&-1.2$\pm$0.8&-0.9$\pm$0.4&0.7$\pm$0.5\\
\hline
\end{tabular}
\end{table*}

\subsection{Crowd annotation protocol (Val set)}

\begin{figure*}[!ht]
\centerline{\includegraphics[width=0.55\linewidth]{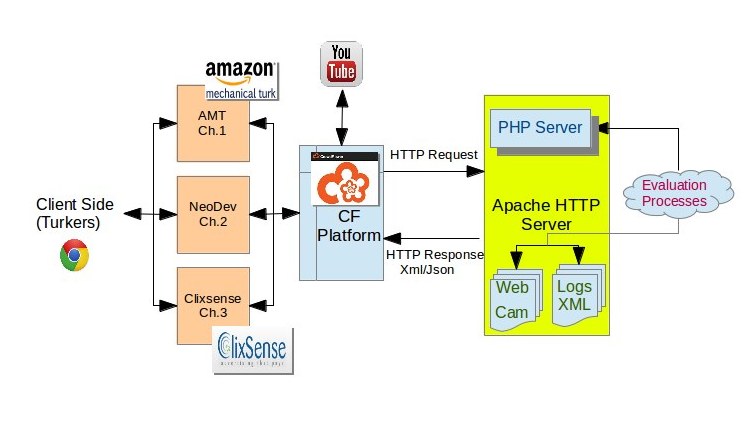}\hspace{.05\linewidth}\includegraphics[width=0.4\linewidth]{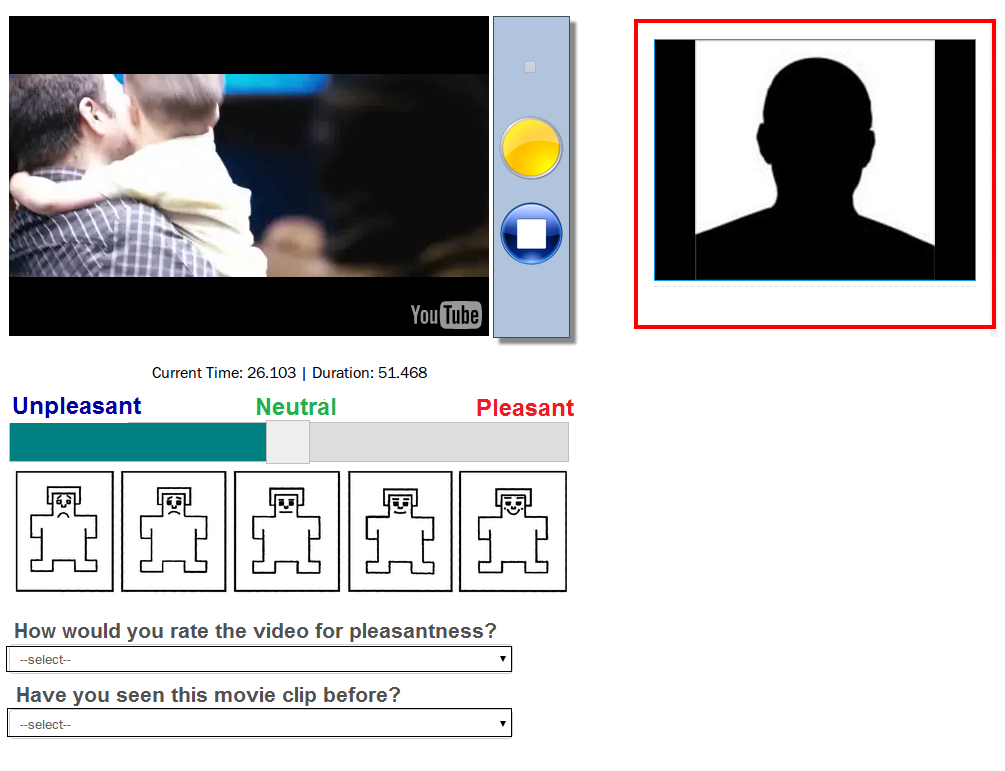}}
\centerline{(a)\hspace{0.5\linewidth}(b)}\vspace{0.1cm}
 \includegraphics[width=.33\linewidth,height=3cm]{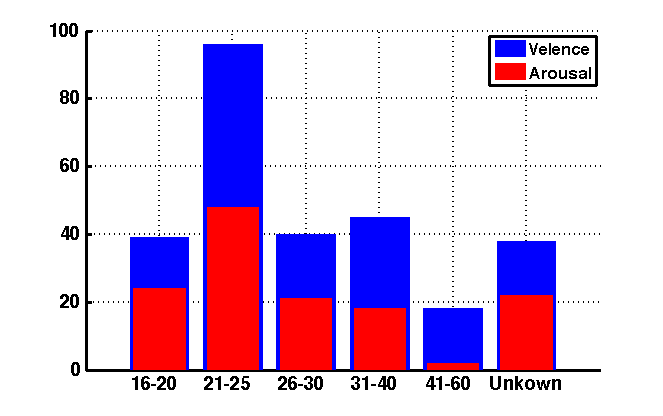}\hspace{0.01\linewidth}
   \includegraphics[width=.33\linewidth,height=3cm]{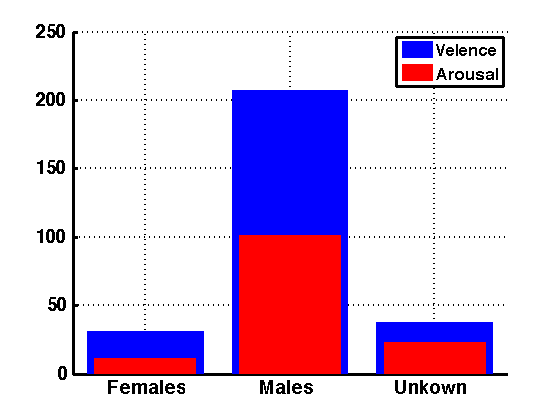}\hspace{0.01\linewidth}\includegraphics[width=.33\linewidth,height=3cm]{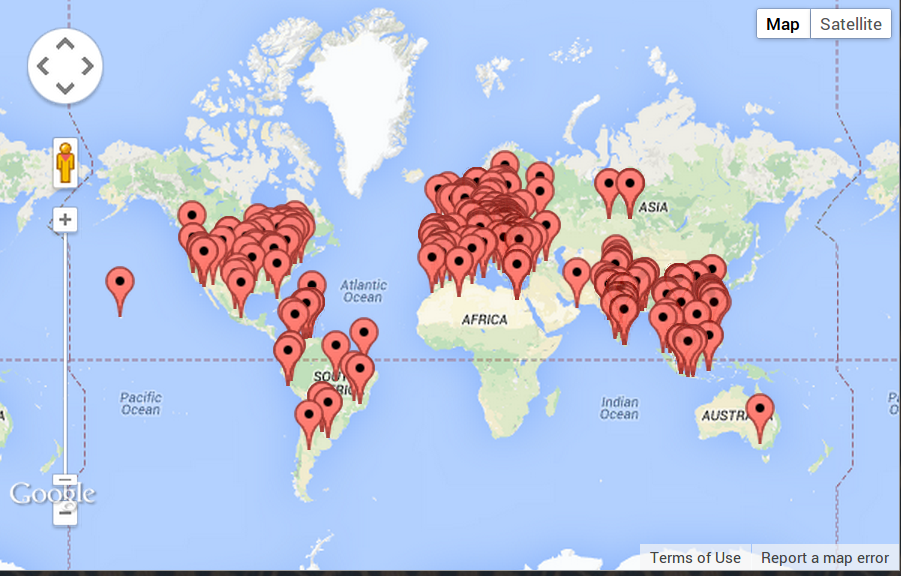}
	\centerline{(c)\hspace{0.33\linewidth}(d)\hspace{0.33\linewidth}(e)}
\vspace{-.3cm}
\caption{{Acquiring crowd annotations for the \textbf{Val} dataset: (a) Architecture of the designed Crowdflower platform (\textit{Turkers} refer to the AMT crowdworkers). (b) User-interface for recording workers' emotional ratings and facial expressions. (c) Age, (d) gender and (e) locality distributions of crowdworkers.}}
\label{f1}
\end{figure*}
\vspace{-.1cm}
We posted the annotation task for \textbf{Val} clips on Amazon Mechanical Turk (AMT) and other CS channels via the CrowdFlower (CF) platform. CF is an intermediate platform for posting the AMT task on our behalf. Moreover, CF provides a simple qualification mechanism to discard outliers. If workers passed the qualification test, they were considered qualified enough to perform a given task. However, pre-designed tests are very generic and limited to simple tasks, which do not allow for trivially discarding bad-quality annotations. So, we performed PHP server-side scripting and redirection, collection and evaluation of all annotations real-time on our server via HTTP requests, before letting workers submit the task. The architecture of the designed CS platform is shown in Fig.~\ref{f1}(a).

To ensure high-quality annotations, each crowdworker could only annotate 5 movie clips, and at least 15 dynamic A,V annotations were collected for each movie clip. We also recorded facial expressions of crowdworkers (not used in this work) as they performed the annotations. Informed consent was obtained from workers and, workers had to provide their demographics (\textit{age}, \textit{gender} and \textit{location}) prior to the task. Time-continuous A/V ratings were compiled from workers over separate sessions (a worker need not annotate for both valence and arousal for the same clip under this setting), and workers were also required to rate each clip for static/overall arousal or valence. Each worker was paid 10 cents/video upon successful task completion.



Workers did not get paid if their annotations and webcam facial videos were not recorded on our server. To evaluate the annotation quality, each video annotation was logged in XML format and analyzed. A continuous slider was used to record emotional rating, and if the slider had not moved for more than 80\% of the clip duration, or if more than 20\% of the data was lost, the annotation was automatically discarded. Also, files smaller than a threshold size were discarded. If the annotation task was left incomplete, a warning message notified the worker about the same. Workers could then re-annotate the video and get paid. Furthermore, some constraints were implemented to maintain annotation quality such as: (1)
Workers could not play (or rate) multiple video clips simultaneously. (2) Workers could annotate a video as many times as they wanted to (the most recent annotation overrode previous ones in such cases). (3) Workers were allowed to use only the Chrome browser for annotation due to unavailability of HTML5 technology support in other browsers. (4) Media player controllers were removed from the interface so that workers could not fast forward/rewind the movie clips, and finally, (5) If the annotation was stopped midway, it had to be redone from scratch.
\subsubsection{Annotation Mechanism}
\begin{sloppypar}
A screen shot of the user interface for recording annotations on AMT is presented in Fig.~\ref{f1}(b). The following components were part of the continuous annotation and facial expression recording process. \\ \\
\textbf{Video Player}: To provide an uninterrupted video stream for workers with low bandwidth, we uploaded the \textbf{Val} movie clips onto \textit{\textbf{YouTube}}. On the client-side, YouTube JavaScript player API was integrated and used in our web-based user interface. \\ \\
\textbf{Slider}: A slider was used to record time-continuous A,V ratings of workers watching movie clips. The slider values ranged from -2 to 2 for both factors (\textsl{very unpleasant} to \textit{very pleasant} for valence, and \textit{calm} to \textit{highly excited} for arousal). In order to facilitate workers' decision making, a standard visual scale {Self Assessment Manikin} (SAM) image was displayed. \\ \\
\textbf{Webcam Panel}: To upload workers' facial expressions in real-time, we used HTML5 technology to buffer the worker's webcam recording on the client-side once the \textit{play} button was pressed. The buffered video was compressed and uploaded on our server upon clip completion. \\ \\
\textbf{Questionnaires}: To ensure that workers were fully engaged while performing the annotation task, they needed to report (1) their overall A/V rating for each viewed clip on a scale of -2 to 2, and (2) their familiarity with the clip to examine if it biased their ratings.
\end{sloppypar}

\subsubsection{Annotation statistics and pre-processing}
Overall, 1012 and 527 workers provided continuous valence and arousal ratings respectively for the \textbf{Val} clips. Their age, gender and locality distributions are shown in Fig.~\ref{f1}(c),(d) and (e). As a preliminary step to eliminate bad-quality annotations, we discarded those annotations with (1) more than threshold missing values, (2) less than standard deviation threshold, and (3) missing or inconsistent static ratings, where sign of the overall rating was opposite to that of the maximum continuous annotation value.  

\section{Expert-guided Multi-task learning (EG-MTL)}\label{MTL}

In this section, we first provide a brief description of multi-task learning (MTL) and the MTL baselines used for performance evaluation, before moving on to describe the proposed EG-MTL algorithm.

Multi-task learning (MTL) exploits the relationships among a set of associated tasks to learn both inter-task similarities as well as task-specific differences, which is more beneficial as compared to learning task-specific models. Given a set of tasks $t = 1...T$, with $X_t$ denoting training data for the task $t$ and $Y_t$ their corresponding labels, MTL seeks to jointly learn a set of weights $W = [W_1..W_T]$, where $W_t$ models task $t$. 
In Section~\ref{ExpRes}, we will examine the relationship between (i) continuous and static affective movie clip labels, and (ii) continuous A,V labels and audio-visual features via MTL variants available as part of the MALSAR library~\cite{zhou2012mutal}, which are described below: 
\\

\noindent \textbf{Multi-task Lasso (MT-Lasso):} extends the Lasso algorithm~\cite{tibshirani1996regression} to MTL, and assumes that sparsity is shared among all tasks. It attempts to minimize the objective function $\sum_{t=1}^{T} \Vert W_t^T X_t -Y_t\Vert_F^2+\alpha \Vert W \Vert_{1}+ \beta \Vert W \Vert^2_F$, where $\Vert.\Vert_F$ and $\Vert.\Vert_1$ denote the Frobenius (L2) and L1 norm respectively. Regularization parameter $\alpha$ controls sparsity, while $\beta$ controls the $\ell_2$-norm penalty. \\

\noindent {$\mathbf{\ell_{21}}$} \textbf{norm-regularized MTL} ($\mathbf{\ell_{21}}$\textbf{-MTL~\cite{Argyrious07}}): minimizes the objective function $\sum_{t=1}^{T} \Vert W_t^T X_t -Y_t\Vert_F^2+\alpha \Vert W \Vert_{2,1}+ \beta \Vert W \Vert^2_F$, where $\Vert.\Vert_{2,1}$ denotes the matrix $\ell_{21}$ norm. The underlying assumption in this model is that \textit{all} tasks are related, which is not always true, and can negatively impact model performance. $\alpha,\beta$ denote regularization parameters controlling group and L2-norm sparsity respectively.  \\


\noindent \textbf{Dirty MTL~\cite{Jalali}:} The key idea of dirty MTL is to decompose $W$ into $P$ and $Q$ such that $W \ = P+Q$, where $P$ and $Q$ denote the group and task-wise sparse components. The algorithm attempts to minimize the objective function $\sum_{t=1}^{T} \Vert W_t^T X_t -Y_t\Vert_F^2+\rho_1 \Vert P \Vert_{1,\infty}+ \rho_2 \Vert Q \Vert_{1} $, where $\rho_1$ controls the group sparsity on $P$, while $\rho_2$ controls sparsity on $Q$. \\

\noindent \textbf{Robust MTL~\cite{Gong12}:} Robust MTL assumes that the model $W$ can be decomposed into two components: a shared feature structure $P$ that captures task-relatedness, and a group-sparse structure $Q$ that detects outliers. If the task is not an outlier, then it falls into the joint feature structure $P$ with its corresponding column in $Q$ being a zero vector; if not, then the $Q$ matrix has non-zero entries at the corresponding column. The algorithm minimizes the objective function $\sum_{t=1}^{T} \Vert W_t^T X_t -Y_t\Vert_F^2+\rho_1 \Vert P \Vert_{2,1}+ \rho_2 \Vert Q \Vert_{2,1}$, such that $W = P + Q$, where regularization parameter $\rho_1$ controls joint feature learning, while parameter $\rho_2$ controls the column-wise group sparsity on $Q$ that detects outliers. \\

\noindent \textbf{Sparse graph regularization (SR-MTL):} where \textit{a-priori} knowledge concerning task-relatedness is modeled in terms of a graph $R$ in the objective function. This way, similarity is only enforced between $W_t$'s corresponding to related tasks. The minimized objective function in this case is $\sum_{t=1}^{T} \Vert W_t^T X_t -Y_t\Vert_F^2+\alpha \Vert WR \Vert^2_{F}+ \beta \Vert W \Vert_1+\gamma \Vert W \Vert^2_F$, where $R$ is the graph encoding task relationships, and $\alpha, \beta, \gamma$ denote regularization parameters as above. 

\subsection{EG-MTL description}\label{our_mtl}
Crowdsourcing has recently become a popular and effective methodology for procuring large amounts of training data in supervised learning approaches. Nevertheless, crowdworkers come from varied backgrounds and may lack the motivation and/or technical and cognitive skills required to effectively accomplish a given task, resulting in noisy outputs. Especially since emotion is a highly subjective phenomenon and the time lag between emotion perception and annotation can vary among crowdworkers, one can expect very diverse dynamic A,V crowd annotations for the presented movie clips. 

In order to improve the efficacy and generalizability of models trained using crowd A,V annotations, we leverage on a small number of annotations provided by more reliable experts\footnote{Considerably higher agreement for A,V is noted among experts as compared to crowdworkers for \textbf{Val} clips-- see Section~\ref{ExpRes}.}. Intuitively, combining expert labels with crowd labels can facilitate better model training than solely using crowd labels; however, to our knowledge, none of the CS-based affect prediction works have attempted to clean crowd models via this perspective.

\subsection{Problem Formulation}

\textbf{Notation:} We denote with $\|\cdot\|_F$ and $\|\cdot\|_1$ the Frobenius and the $\ell_1$ norms respectively. $(\cdot)'$ indicates the transpose operator, while $|.|$ denotes a set cardinality. For problems \textbf{\texttt{P1}} and \textbf{\texttt{P2}} described in Section~\ref{sec:introduction}, we model each movie clip as a task, and for each task (clip) $t$ we denote the set of dynamic A,V crowdworker annotations using the matrix $\mathbf{X}_t \in \RR ^ {N_t \times D}$, 
$\mathbf{X}_t=[\mathbf{x}^t_{_1} ,...,\mathbf{x}^t_{N_t}]'$, where $N_t$ is the crowdworkers that have annotated clip $t$ for A/V and $D$ is the length of dynamic A,V annotations (we only analyze the final 50 seconds of clip time in this work). We also define the matrix $\mathbf{X} \in \RR ^{N \times D}$, $\mathbf{X} = [\mathbf{X}_1',\dots,\mathbf{X}_{R}']'$, 
where $N = \sum_{t = 1}^R {N_t } $ denotes the total number of training examples from the $R$ tasks. For each training sample, we construct a binary label indicator vector $\mathbf{y}^t_i \in \RR ^ {RC} $ as $\mathbf{y}^t_i=[\underbrace {0,0,...,0}_{Task
\ 1},\underbrace {0,1,...,0}_{Task \ 2},...,\underbrace {0,0,...,0}_{Task \ R}]$, where $C$ is the number of class labels ($C=2$ for \textbf{\texttt{P2}} and the number of discrete A/V levels in $[-1,1]$ for \textbf{\texttt{P1}}). The position of the non-zero element indicates the task and class membership of the corresponding training sample. A label matrix $\mathbf{Y} \in \RR^{N \times RC}$ is then obtained concatenating the $\mathbf{y}^t_i$'s for all training samples. We also use some additional expert-labeled
instances to improve model quality. If there are $N_e$ expert labels, then the expert training set can be defined as $\mathbf{P} \in \RR ^{N_e \times D}$ and a label matrix $\mathbf{V} \in \RR^{N_e \times RC}$, in a similar way as defined for the crowd. \\ \\

\noindent \textbf{Methodology:} We propose to solve the following optimization problem:
\vspace{-.15cm}
\begin{eqnarray}
\label{eqn:1}
\mathop {\min }\limits_\mathbf{W} \frac{1}{2} \left\| \mathbf{U}^{1 \mathord{\left/
 {\vphantom {1 2}} \right.
 \kern-\nulldelimiterspace} 2} \mathbf{(Y - XW)} \right\|_F^2  + \lambda _1 \left\| { \mathbf{V - PW}} \right\|_F^2 \nonumber \\+ \lambda _2 \left\| {\mathbf{E W'} } \right\|_F^2  + \lambda _3 \left\| \mathbf{W} \right\|_1 
\end{eqnarray}

\noindent where $\mathbf{W} \in \Re  ^ {D \times RC} $ is the learnt weight matrix, $\mathbf{E} \in \Re ^ {|\mathcal{E}| \times RC}$ is an edge-vertex incident matrix where $\mathbf{E}_{q=(i,j),h}=\gamma_{ij}$ if $i=h$, $\mathbf{E}_{q=(i,j),h}=-\gamma_{ij}$ if $j=h$ and $\mathbf{E}_{q=(i,j),h}=0$ otherwise. $\mathbf{U} \in \Re ^ {N \times N} $ is a diagonal matrix to control the reliability of crowd workers, while $\lambda_1, \lambda_2, \lambda_3$ are appropriate regularization parameters. 

The proposed objective function has three effects. Task relatedness is defined via the graph regularization term (we assume all clips to be related in this work), and knowledge from one task can be utilized by the other related tasks. Prior knowledge regarding the required level of feature sharing is embedded in the learning framework through $\gamma_{ij}$'s (set to 1 in this work). Sparsity is enforced in the learning process, which emphasizes the contribution of discriminative features and de-emphasizes the contribution of less discriminative features. Finally, by incorporating loss functions pertaining to both the crowd and expert labels in the optimization, we \bsq{refine} the learned weights $\mathbf{W}$ so as to agree with the expert knowledge, which is closer to the ground-truth as compared to crowd labels as per our assumption. This model refinement reflects in the form of enhanced classification/regression performance as discussed in Section~\ref{ExpRes}.

\subsection{Optimization}

To solve Eqn.(\ref{eqn:1}), we propose to adopt the Fast Iterative Shrinkage-Thresholding Algorithm (FISTA) \cite{Beck09}. The objective function is convex and is the sum of a smooth term $f(\cdot)$ and a non-smooth term $g(\cdot)$ if we define:
\vspace{-.2cm}
\begin{eqnarray}
f(\mathbf{W}) = \frac{1}{2}\left\|  \mathbf{U}^{{1 \mathord{\left/
 {\vphantom {1 2}} \right.
 \kern-\nulldelimiterspace} 2}} { \mathbf{(Y - XW)} } \right\|_F^2  + \lambda _1 \left\| \mathbf{ V - PW} \right\|_F^2  \nonumber
\end{eqnarray}

\begin{equation}
g(\mathbf{W}) = \lambda _2 \left\| \mathbf{EW' } \right\|_F^2 + \lambda _3 \left\| \mathbf{W} \right\|_1 
\nonumber
\end{equation}

\noindent If we calculate the derivative of $f(\mathbf{W})$ with respect to $\mathbf{W}$, we have gradient descent:

\begin{eqnarray}
\frac{{\partial f(\mathbf{W})}}{{\partial \mathbf{W}}} = \mathbf{X' U (Y - XW)}  + 2\lambda _1 \mathbf{P' (V - PW)} \nonumber
\end{eqnarray}

\noindent which is used in the update equation for $\mathbf{W}$ in Algorithm 1 outlined below. For the non-smooth term $g(\mathbf{W})$, we can adopt Soft Thresholding \cite{Boyd} to solve the non-smooth $\ell_1$-norm term. The optimization procedure is outlined in Algorithm 1.



\begin{algorithm}[!htbp]
  \caption{\small{Accelerated Gradient Descent for solving (1)}}
\begin{small}
 \begin{algorithmic}
\State
\textbf{INPUT}: Crowd and Expert feature matrix $\mathbf{X}$ and $\mathbf{P}$, Crowd and Expert label matrix $\mathbf{Y}$ and $\mathbf{V}$, $\lambda_1$, $\lambda_2$, $\lambda_3$,  $\mathbf{E}$.
\State
  Initialize $\mathbf{W}_0$, $\alpha_0=1$, line search parameter $L_0=1$.
\State  \textbf{LOOP:} \State
\indent $\alpha_k=\frac{1}{2}(1+\sqrt{1+4\alpha^2_{k-1}})$ \State
\indent $L_k=2L_{(k-1)}$ \State
\indent $\mathbf{\hat W} = \mathbf{W}_k  - \frac{2}{L_k}  \mathbf{X' U ( \mathbf{Y} - \mathbf{X}\mathbf{W}_k)  } - \frac{2 \lambda _1}{L_k}  \mathbf{P' (V - PW_k)} $ \State
\indent Solving $\mathbf{W}_{k+\frac{1}{2}} \leftarrow \mathop {\min }\limits_\mathbf{W} \left\| {\mathbf{W} -  \mathbf{\hat W}} \right\|_F^2  + \hat \lambda _1 \left\| {\mathbf{EW'}} \right\|_F^2  + \indent \indent \indent \hat \lambda _2 \left\| \mathbf{W} \right\|_1 $ based on Soft Thresholding \cite{Boyd}.
\State 
\indent $\mathbf{W}_{k+1}= (1+\frac{\alpha_{k-1}-1}{\alpha_k}) \mathbf{W}_{k+\frac{1}{2}} - \frac{\alpha_{k-1}-1}{\alpha_k} \mathbf{W}_k$ \State
\textbf{Until Convergence}
\State 
 \textbf{Output:} $\mathbf{W}$ 
\end{algorithmic}
\end{small}
 \label{fista}
 \end{algorithm}

\section{Data Analysis and Experiments}\label{ExpRes}

In this section, we first examine the dynamic annotations acquired from crowdworkers and experts, before focusing our attention on the experiments concerning problems \textbf{\texttt{P1}} and \textbf{\texttt{P2}}. We computed the Kendall's coefficient of concordance (Kendall's W) to determine inter-annotator agreement for the crowd and expert populations on the \textbf{Val} set, as well as for the crowd on the \textbf{Eval} set-- the mean and standard deviation values for arousal and valence are presented in Tables~\ref{table_KA} and~\ref{table_KV}. Apart from computing W for A,V over the entire (or {Full}) clip, we also determined W for the first (1st) and second (2nd) halves of each clip. This is because all clips began with a relatively neutral segment as reported in~\cite{abadi2015decaf}. Furthermore, based on the static emotion labels available for the clips in~\cite{abadi2015decaf}, we repeated the above analyses for the HA, LA, HV and LV clips.

Tables~\ref{table_KA} and~\ref{table_KV} clearly reveal that the A,V agreement is higher among experts as compared to crowdworkers for the \textbf{Val} set-- the fact that experts are likely to provide more consistent annotations forms the basis of our EG-MTL algorithm, and this result is also on expected lines since experts were conversant with emotional attributes, and familiarized themselves with the emotional scene dynamics by viewing the clips multiple times prior to annotating them (whereas crowdworkers could view the clips only while performing the annotation task). Nevertheless, only a moderate level of agreement is noted among experts for both A and V considering Ws computed over the entire clip duration. Deeper analysis reveals higher agreement on the A,V ratings among experts for the emotionally salient second half of the clip as compared to the first half. Also, both experts and crowdworkers agreed considerably more on the A ratings for HA clips as compared to LA clips (considering full clips); On the other hand, while very similar Ws are noted for expert V ratings in HV and LV clips, crowdworkers agreed considerably more on V ratings for HV clips.

We performed a 3-way ANOVA test on the concordance scores with the population type (expert/crowd), clip-half (1st/2nd) and attribute intensity (HA/LA or HV/LV) as factors. ANOVA on W scores for arousal revealed the main effect of population type (F$_{(1,47)}=11.1,p<0.05$) and attribute intensity (F$_{(1,47)}=21.25,p<0.000001$), while no other main or interaction effects were significant. ANOVA on valence W scores again revealed the main effect of population type (F$_{(1,47)}=6.51,p<0.05$) and attribute intensity (F$_{(1,47)}=7.37,p<0.01$), while also showing up the interaction effect of these two factors (F$_{(1,47)}=6.58,p<0.01$). We also computed Pearson correlations between expert and crowd annotators for the \textbf{Val} set to investigate if there was higher agreement on the trend of the dynamic A,V profiles as compared to the actual A,V ratings. However, there was little difference between the Kendall Ws and Pearson coefficients for both experts and crowdworkers. Finally, much higher agreement was noted among crowd annotations for the \textbf{Eval} set as compared to the \textbf{Val} set. Whether this increased agreement can be attributed to the nature of the \textbf{Eval} clips is unclear due to the unavailability of expert annotations for \textbf{Eval} clips. However, the fact that a majority of the 35 \textbf{Eval} annotators performed annotations in a rather controlled academic environment could have contributed to greater rating consistency.    

\begin{figure}[!htbp]
\centerline{\includegraphics[width=0.23\linewidth]{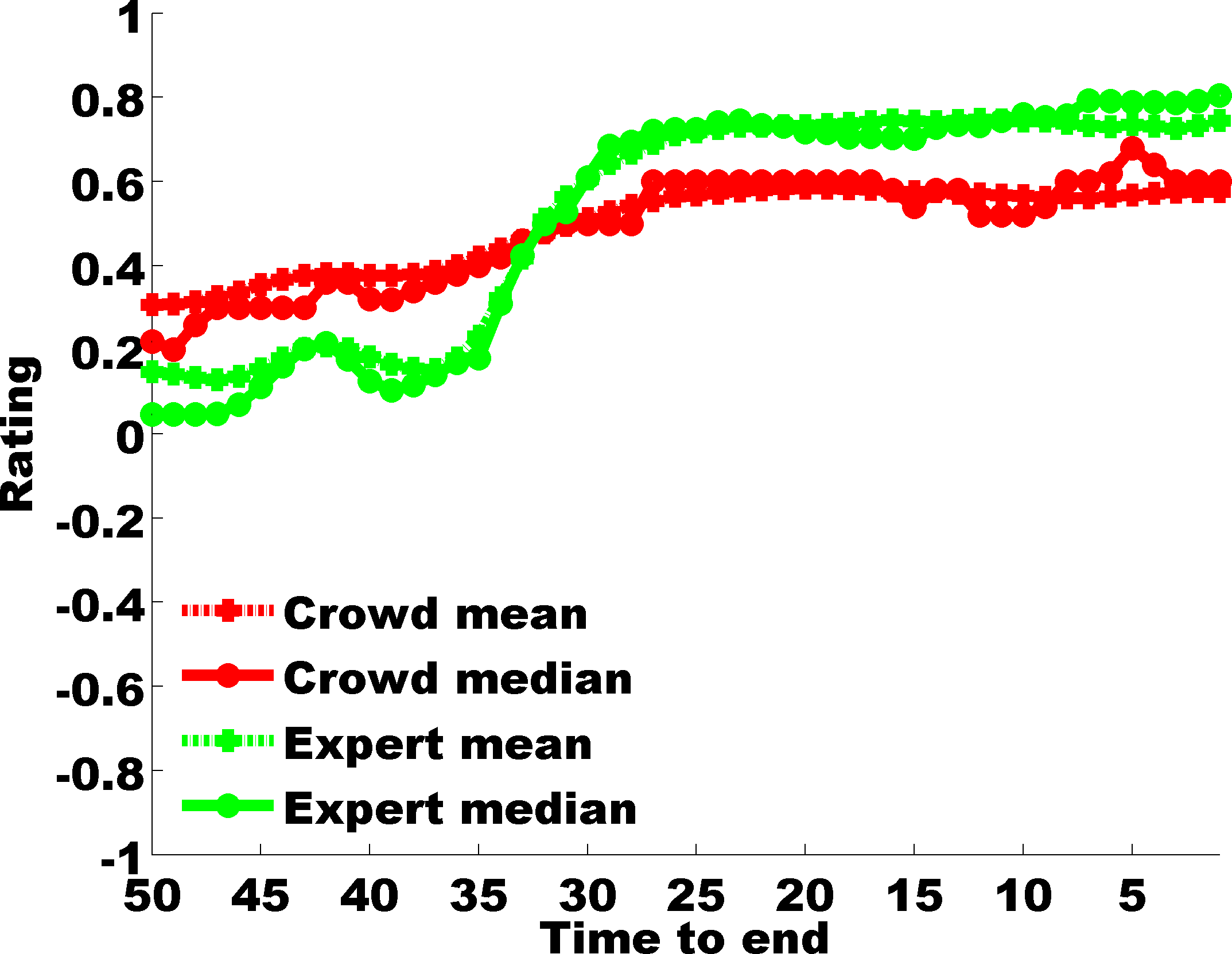}\hspace{.1cm}\includegraphics[width=0.23\linewidth]{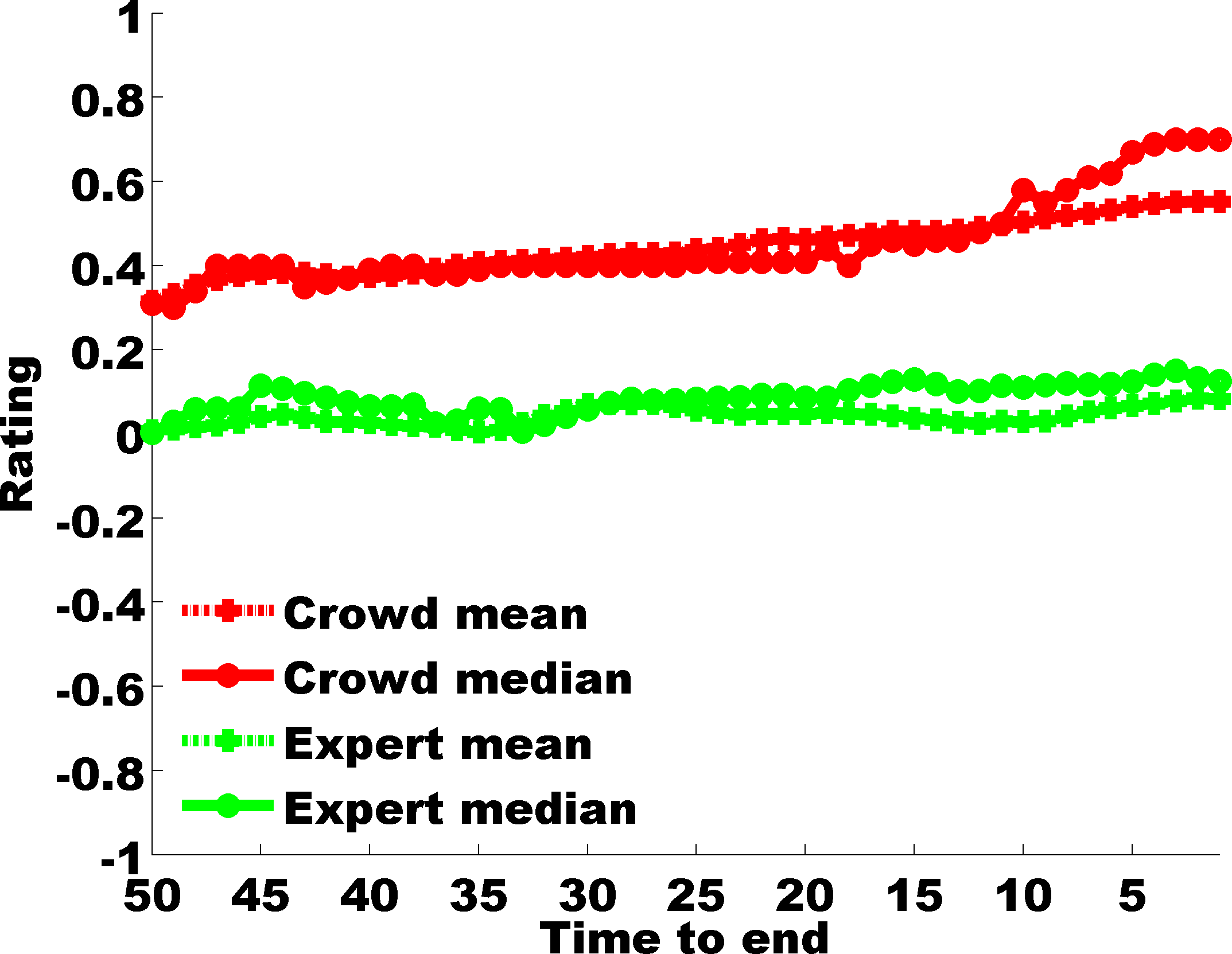}\hspace{.1cm}\includegraphics[width=0.23\linewidth]{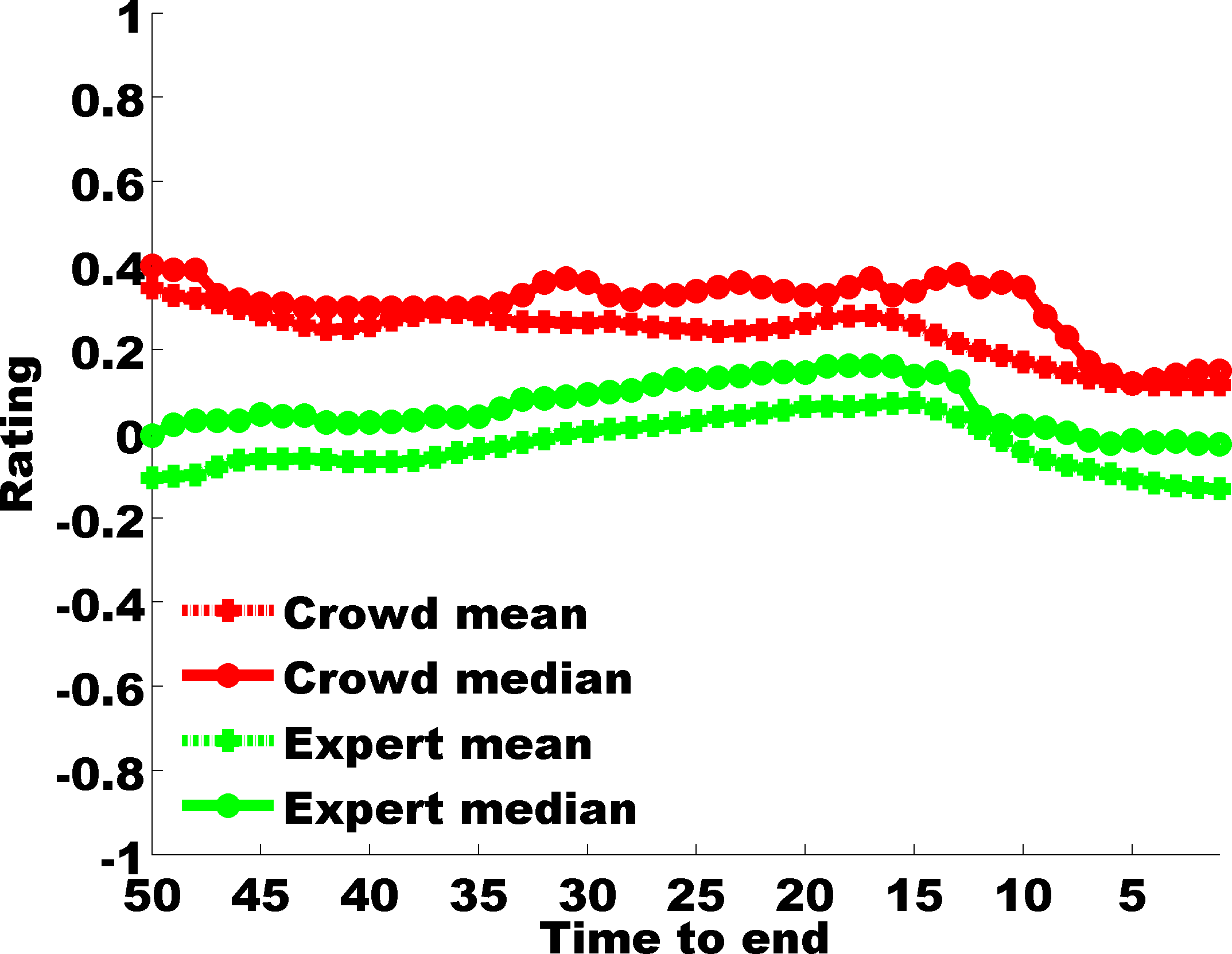}\hspace{.1cm}\includegraphics[width=0.23\linewidth]{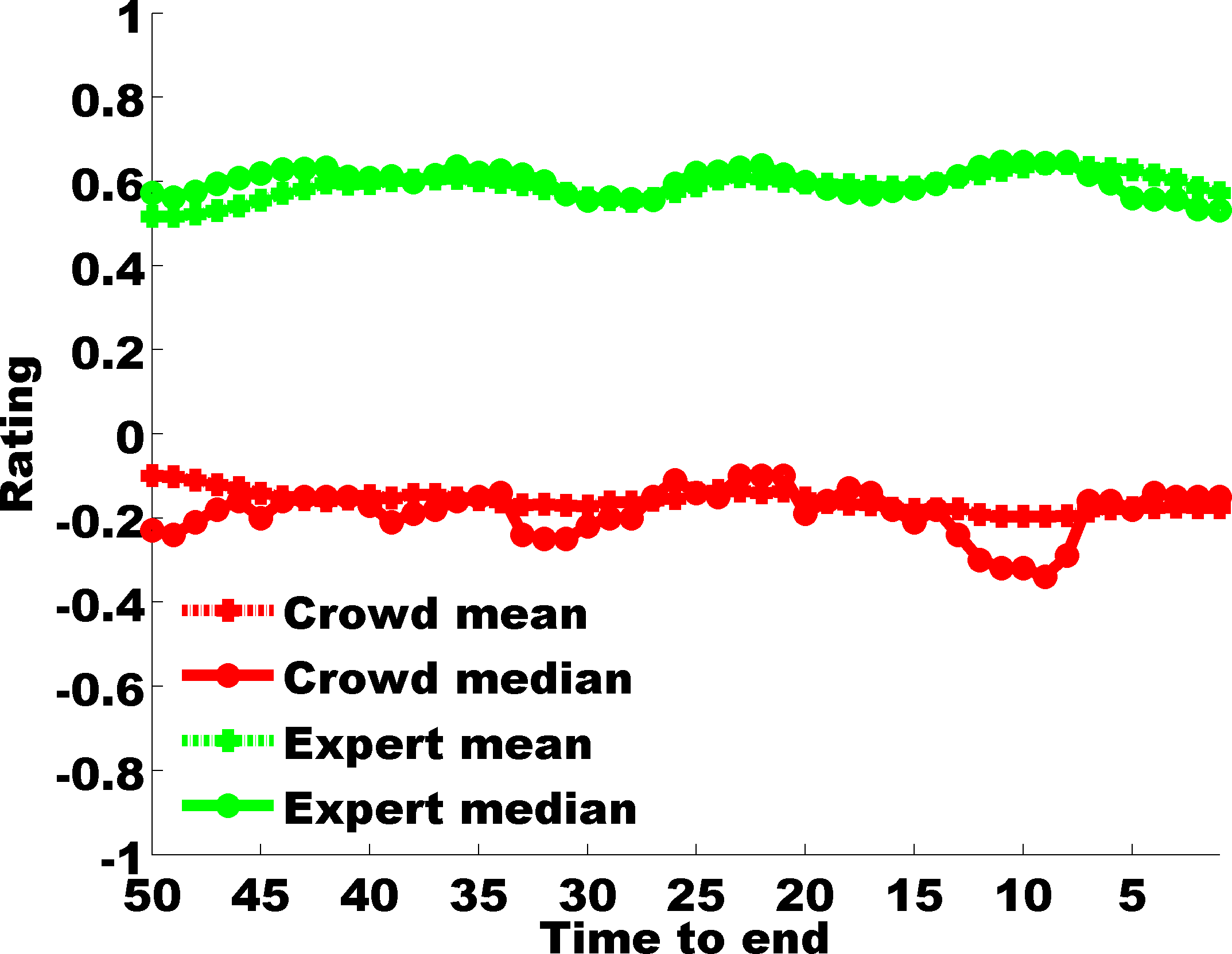}}\vspace{0.05cm}
\centerline{\includegraphics[width=0.23\linewidth]{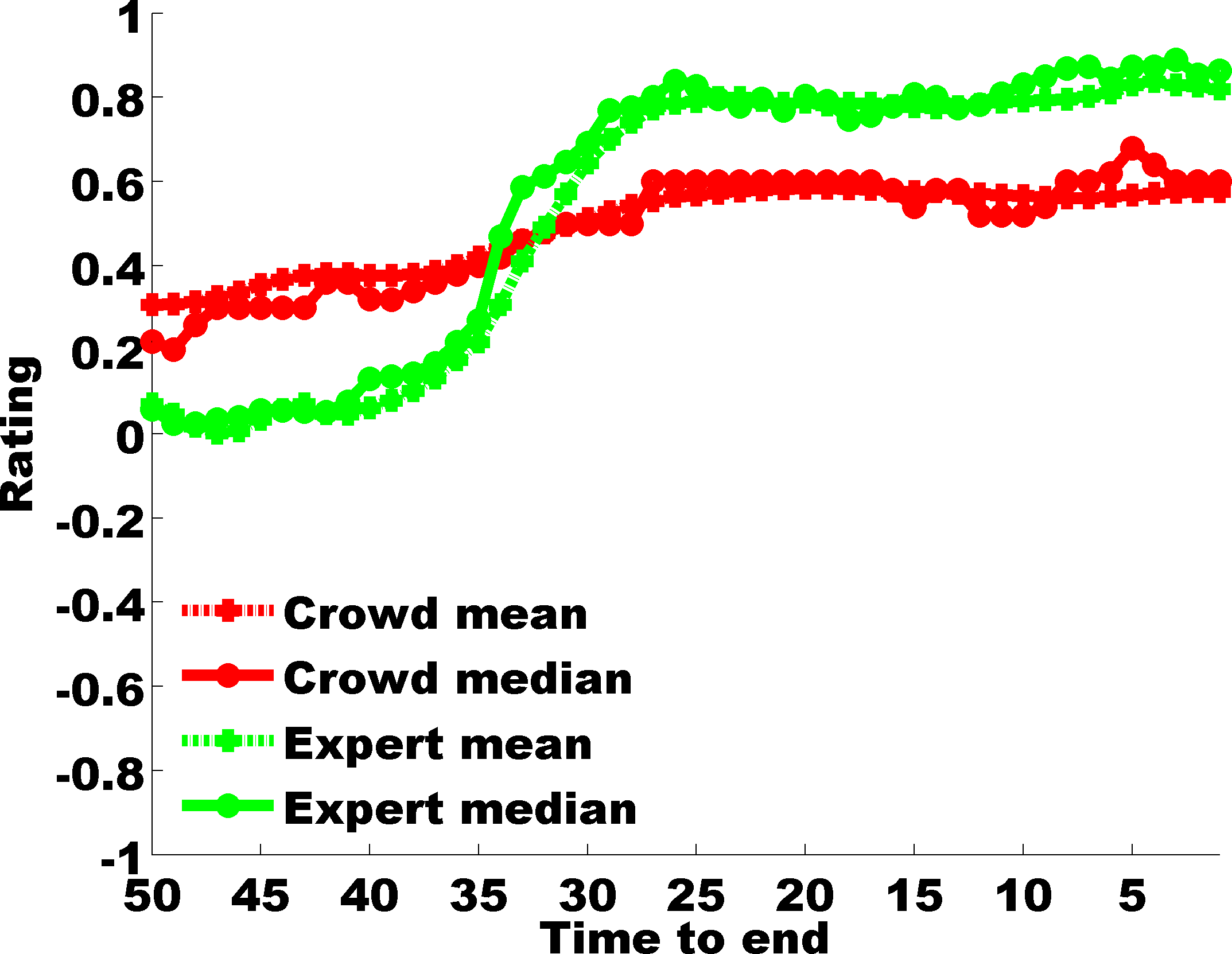}\hspace{.1cm}\includegraphics[width=0.23\linewidth]{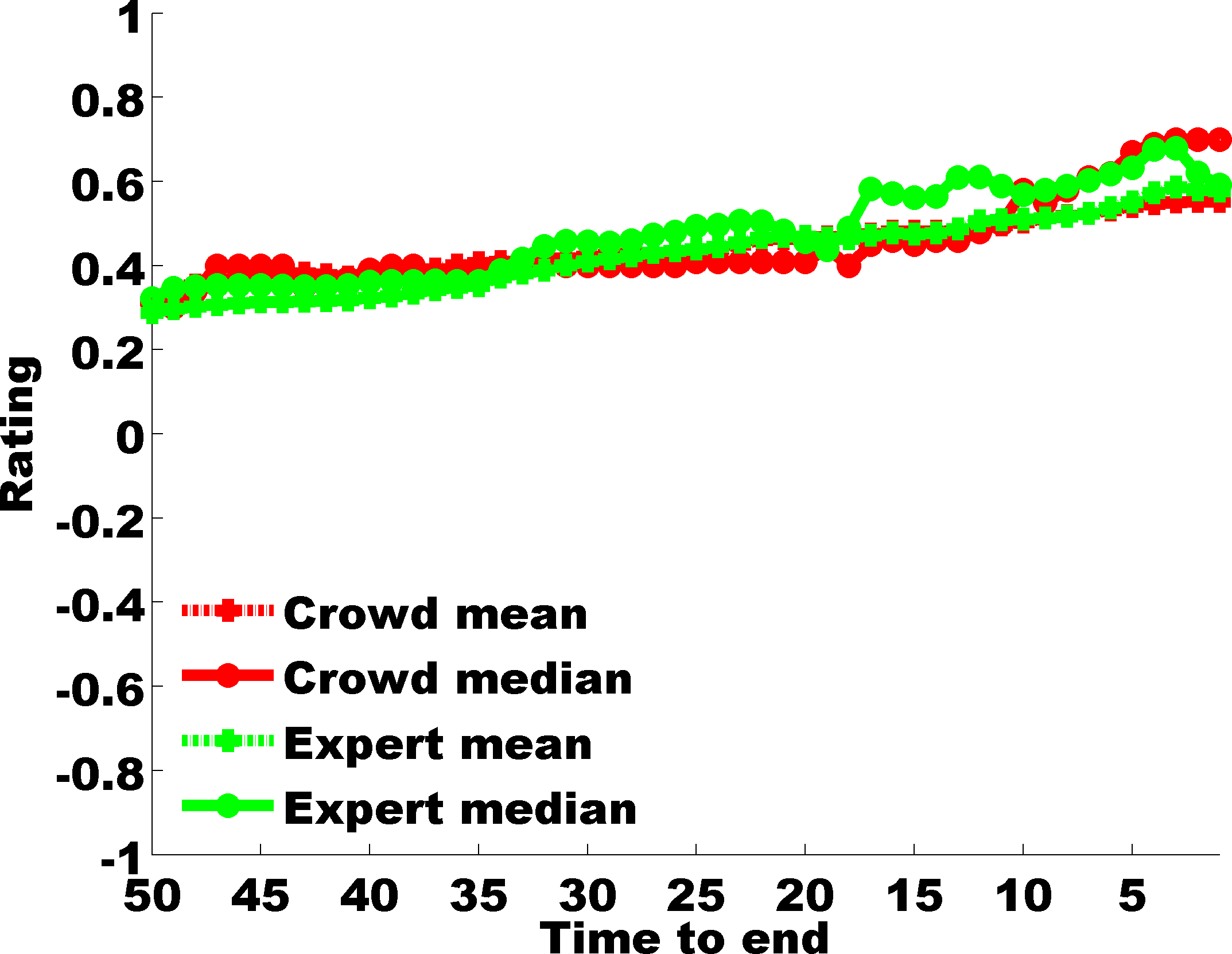}\hspace{.1cm}\includegraphics[width=0.23\linewidth]{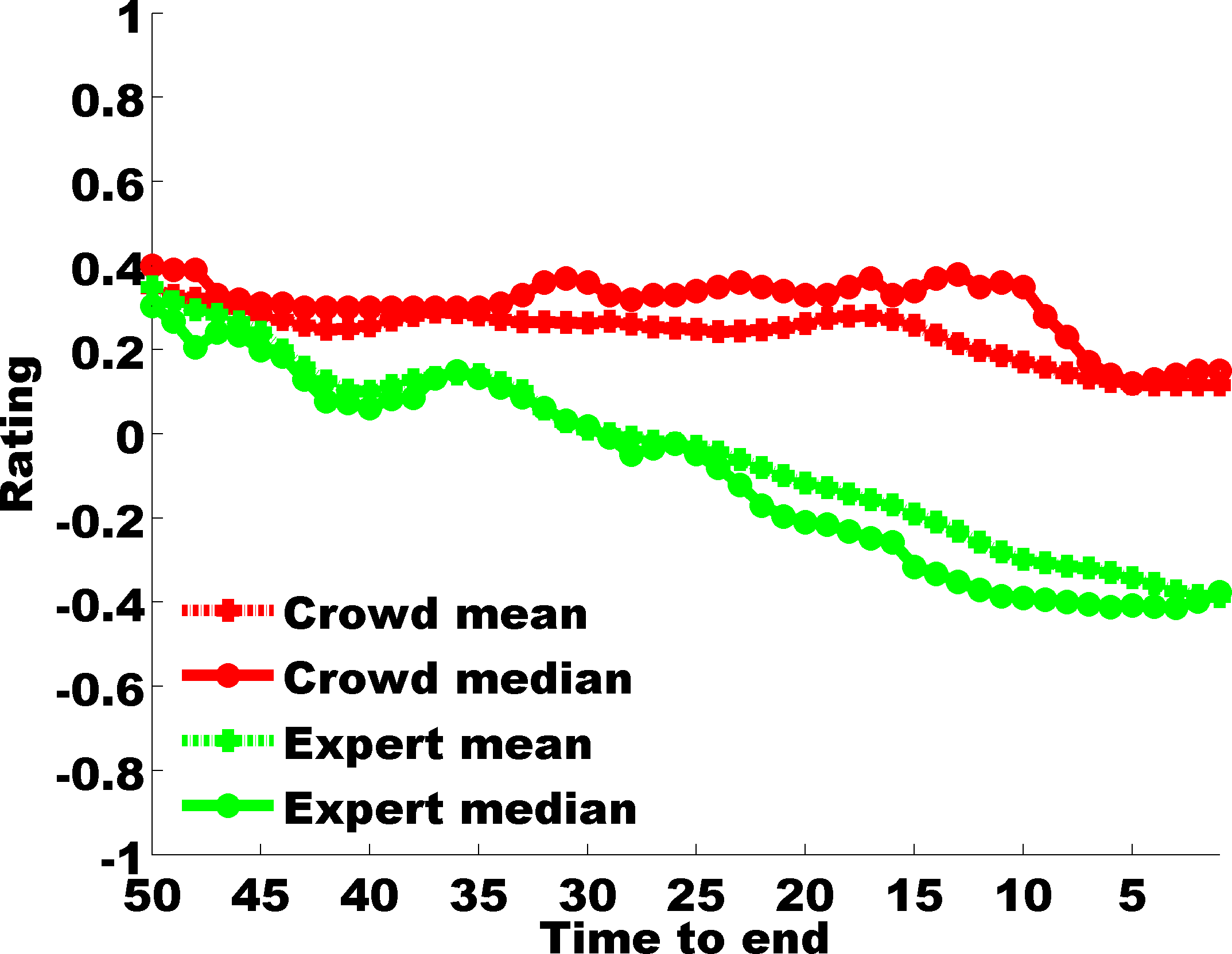}\hspace{.1cm}\includegraphics[width=0.23\linewidth]{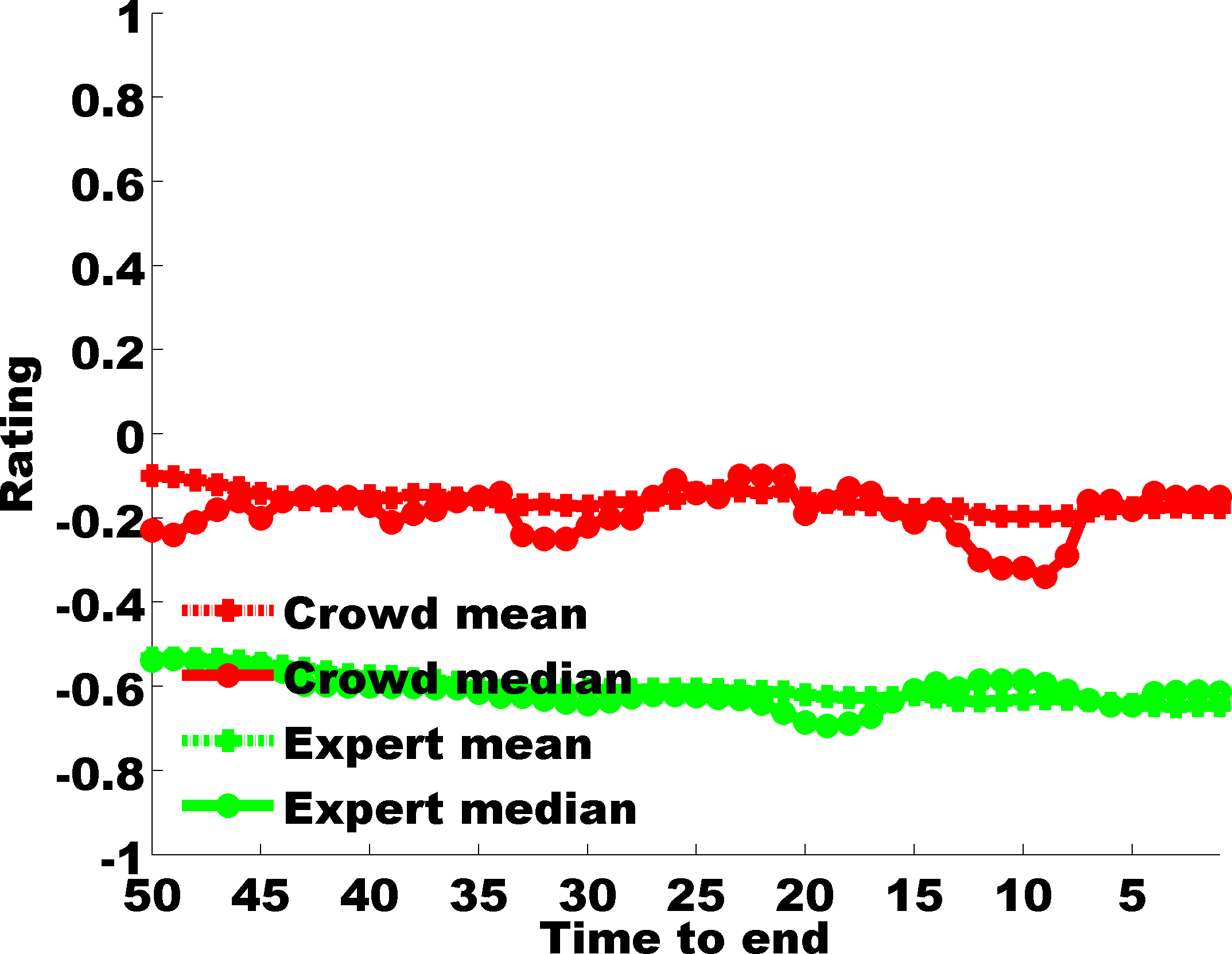}}
\centerline{\small{HAHV} \hspace{0.13\linewidth} \small{LAHV} \hspace{0.14\linewidth} \small{LALV} \hspace{0.14\linewidth} \small{HALV}} 
\vspace{-.05cm}
\caption{Dynamic arousal (top) and valence (bottom) ratings provided by crowd and expert users over the last 50 seconds for four exemplar clips. Best viewed under zoom.}
\label{Crowd_Exp}
\end{figure}

Given that the Multi-task learning framework requires dimensional consistency of the input data, and the time-continuous annotations were of varying length (due to varying clip lengths), we used the A,V ratings provided over the last 50 seconds by experts/crowdworkers in all our experiments. This choice was made owing to two reasons (1) As reported in~\cite{abadi2015decaf}, the latter part of the clips was emotionally more salient, which reflects in higher agreement among even the expert raters for the second half of the movie clips and (2) As shown in Figure~\ref{Crowd_Exp} where the median and mean of the crowd (in red) and expert (in green) A,V profiles are plotted for an exemplar \textbf{Val} clip corresponding to each of the four AV quadrants, the dynamic A/V profile \bsq{tends to} the static emotional label typical within the final 50 second time frame. Before moving on to show how MTL captures latent relationships between tasks in the context of affect prediction, we first describe the audio-visual features extracted from the clips for analysis.

\begin{table*}[!htbp]
\fontsize{7}{7}\selectfont
\renewcommand{\arraystretch}{1.2}
\caption{\label{table_KA} Kendall's W ($\mu \pm \sigma$) for dynamic arousal ratings acquired from expert and crowd populations.}
\vspace{-.2cm}
\centering
\begin{tabular}{|c|c|c|c|c|c|c|c|c|c|c|}
\hline 
& & \multicolumn{3}{c|}{\textbf{All}}&\multicolumn{3}{c|}{\textbf{HA}}&\multicolumn{3}{c|}{\textbf{LA}}\\ \hline
& & Full & 1st & 2nd & Full & 1st & 2nd & Full & 1st & 2nd  \\
\textbf{Val}& Expert & 0.39$\pm$0.27&0.31$\pm$0.24& 0.33$\pm$0.23 & 0.55$\pm$0.21 & 0.44$\pm$0.26 & 0.46$\pm$0.23 & 0.23$\pm$0.23 & 0.18$\pm$0.15 & 0.19$\pm$0.13 \\
\textbf{Val}& Crowd & 0.21$\pm$0.17 &0.18$\pm$0.08 & 0.17$\pm$0.11 & 0.34$\pm$0.14 & 0.24$\pm$0.08 & 0.25$\pm$0.10 & 0.09$\pm$0.04 & 0.12$\pm$0.03 &0.10$\pm$0.05 \\
\textbf{Eval}& Crowd & 0.38$\pm$0.18&0.28$\pm$0.22&0.27$\pm$0.16&0.49$\pm$0.14&0.43$\pm$0.23&0.30$\pm$0.16&0.27$\pm$0.16&0.14$\pm$0.10&0.23$\pm$0.18 \\ \hline
\end{tabular}
\vspace{0.2cm}
\fontsize{7}{7}\selectfont
\renewcommand{\arraystretch}{1.2}
\caption{\label{table_KV} Kendall's W ($\mu \pm \sigma$) for dynamic valence ratings acquired from expert and crowd populations.}
\vspace{-.2cm}
\centering
\begin{tabular}{|c|c|c|c|c|c|c|c|c|c|c|}
\hline 
& & \multicolumn{3}{c|}{\textbf{All}}&\multicolumn{3}{c|}{\textbf{HV}}&\multicolumn{3}{c|}{\textbf{LV}}\\ \hline
& & Full & 1st & 2nd & Full & 1st & 2nd & Full & 1st & 2nd  \\
\textbf{Val}& Expert &  0.47$\pm$0.21 & 0.34$\pm$0.19 & 0.40$\pm$0.24 &0.46$\pm$0.21 & 0.32$\pm$0.15 & 0.42$\pm$0.29 & 0.49$\pm$0.22 & 0.35$\pm$0.24 & 0.38$\pm$0.19 \\
\textbf{Val}& Crowd & 0.22$\pm$0.18&0.27$\pm$ 0.22&0.19$\pm$0.18&0.37$\pm$ 0.10& 0.32$\pm$0.17 & 0.42$\pm$0.16 & 0.06$\pm$0.05& 0.06$\pm$0.05 & 0.11$\pm$0.16\\
\textbf{Eval}& Crowd & 0.40$\pm$0.15 & 0.31$\pm$0.16 & 0.30$\pm$0.16 & 0.30$\pm$0.07 & 0.36$\pm$0.13 & 0.24$\pm$0.15 & 0.27$\pm$0.19 & 0.27$\pm$0.19 & 0.36$\pm$0.17 \\ \hline
\end{tabular}
\end{table*}

\subsection {Audio-visual feature extraction}
For the purpose of problem \textbf{\texttt{P1}} where an MTL model needs to be trained for continuous emotion prediction from the dynamic A/V ratings and audio-visual clip features, we extracted audio and visual features that have been found to correlate well with the A,V dimensions by prior affective studies~\cite{wang2006affective,picard2000affective,HanjalicITM2005,DEAP}. These features were extracted on a per-second basis for our regression experiments.

\subsubsection{Audio features}
Sound information in the form of loudness of speech (energy of sound) is related to arousal, while rhythm and average pitch in speech relates to valence~\cite{picard2000affective}. Mel-frequency cepstrum components (MFCCs)~\cite{li2001classification} are representative of the short-term sound power spectrum and have been commonly employed for emotion recognition. Commonly used features in audio and speech processing~\cite{li2001classification} were extracted from the audio channels. To extract  MFCCs, we divided the audio segment into 20 divisions and then extracted the first 13 MFCC components from each division. Using the sequence of MFCC components over a segment, we computed 13 derivatives of MFCC, DMFCC, and mean auto correlation, AMFCC proposed in~\cite{li2001classification}. Upon calculating MFCC, DMFCC and AFCC (13 values each), we used their means as features. To chracterize emotional speech, we extract formants up to 4400Hz over the audio segment, and formant means were used as features~\cite{mustafa2006robust}. Also, we used the ACA toolbox~\cite{lerch2012introduction} to calculate mean and standard deviation (std) of (i) spectral flux, (ii) spectral centroid and (iii) time-domain zero crossing rate~\cite{li2001classification} over 20 audio segment divisions. We computed the power spectral density of the audio signal and the bandwidth, band energy ratio (BER) and density spectrum magnitude (DSM) according to~\cite{li2001classification}. Finally, we computed the mean proportion of silence as defined in~\cite{chen2006mixed}. Overall, 56 audio features listed in Table~\ref{Features} were extracted.

\subsubsection{Visual features}
Lighting key and color variance~\cite{wang2006affective} are well-known video features known to evoke emotions. Therefore, we extracted lighting key from each frame in the HSV space by multiplying the mean by the standard deviation of V values. Color variance~\cite{DEAP} is defined as the determinant of the covariance matrix of L, U, and V in the CIE LUV color space. Also, the amount of motion in a movie scene is indicative of its excitement level~\cite{DEAP}. Therefore, we computed the optical flow~\cite{lucas1981iterative} in consecutive frames of a video segment to motion magnitude for each frame. As the proportions of colors are important elements for evoking emotions~\cite{valdez1994effects}, A 20-bin color histogram of hue and lightness values in the HSV space was computed for each frame of a segment and averaged over all frames. The mean of the bins reflect the variation in the video content. For each frame in a segment, the median of the L and S values in HSL space were computed; their average for all the frames in a segment is an indication of the segment lightness and saturation. We also used the definitions in~\cite{wang2006affective} to calculate shadow proportion, visual excitement, grayness and visual detail. Extracted video features are listed in Table~\ref{Features}.

\begin{table}[!ht]
\centering
\fontsize{8}{8}\selectfont
\renewcommand{\arraystretch}{1.3}
\caption{\label{Features}Extracted audio-visual features from each movie clip (feature dimension listed in parenthesis). }\vspace{-0.25cm}
\begin{tabular}{p{.4\linewidth}|p{0.55\linewidth}}
\hline
\hline
\small
\textbf{Audio features}					&  \textbf{Description}\\
  \hline
\textbf{MFCC features (39)}					& MFCC coefficients~\cite{li2001classification}, Derivative of MFCC, MFCC Autocorrelation (AMFCC)  \\
\textbf{Energy (1) and Pitch (1)}					& Average energy of audio signal~\cite{li2001classification} and first pitch frequency\\
\textbf{Formants (4)}					& Formants up to 4400Hz \\
 \textbf{Time frequency (8)}					& mean and std of: MSpectrum flux, Spectral centroid, Delta spectrum magnitude, Band energy ratio \cite{li2001classification}\\
\textbf{Zero crossing rate (1)}					& Average zero crossing rate of audio signal~\cite{li2001classification}\\
\textbf{Silence ratio (2)}					& Mean and std of proportion of silence in a time window~\cite{li2001classification,chen2006mixed}\\
\hline
\hline
\textbf{Video features}					&  \textbf{Description}\\
\hline
\textbf{Brightness (6)}					&  Mean of: Lighting key, shadow proportion, visual details, grayness, median of Lightness for frames, mean of median saturation for frames\\
\textbf{Color Features (41)}					&   Color variance, 20-bin histograms for hue and lightness in HSV space \\
\textbf{VisualExcitement (1)}					& Features as defined in~\cite{wang2006affective}\\
\textbf{Motion (1)}					& Mean inter-frame motion~\cite{lucas1981iterative}\\
\hline
\end{tabular}
\end{table}

\subsection{Capturing latent task relationships via MTL}

We now illustrate the utility of MTL, and show how it can capture relationships between related tasks in the form of features shared by the various tasks. To this end, we use the dynamic expert ratings on the \textbf{Val} set since they are noted to be the most consistent from the concordance analyses presented earlier. As a starting point, we perform an emotion recognition experiment modeling each expert as a task, \ie, the dynamic ratings provided by each expert (over the last 50 sec) for all of the movie clips are fed as a feature matrix with the binary ground-truth A,V labels for each movie clip (as derived from~\cite{abadi2015decaf}) denoting the data labels. This learning framework can answer the following question: \textit{Are there any similarities among the dynamic rating patterns of the experts, and which similarities (in terms of time points) most influence the static emotional labels of the movie clips?}

\begin{figure}[!ht]
\centerline{\includegraphics[width=0.49\linewidth]{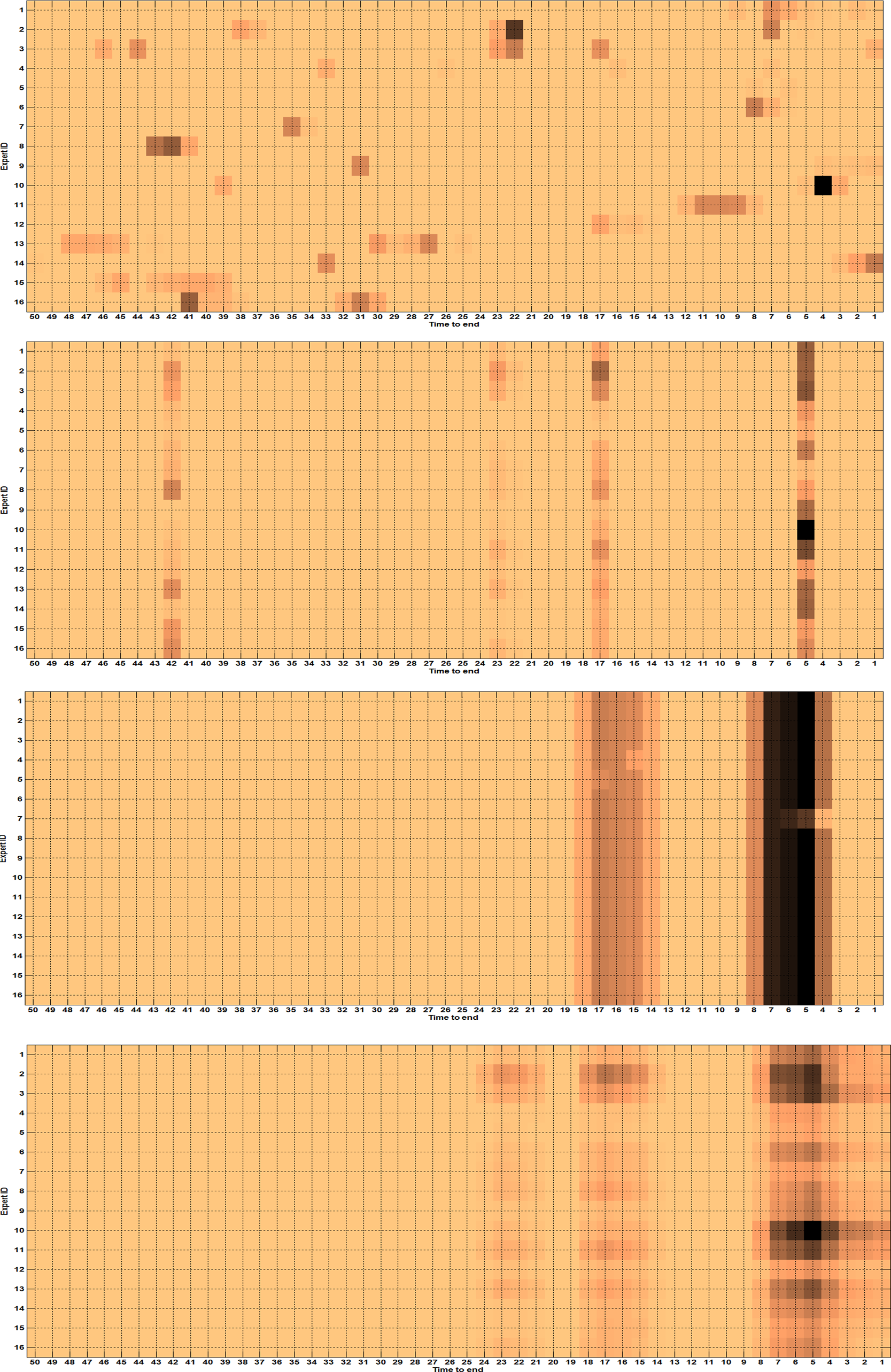}\hspace{.02cm}\includegraphics[width=0.49\linewidth]{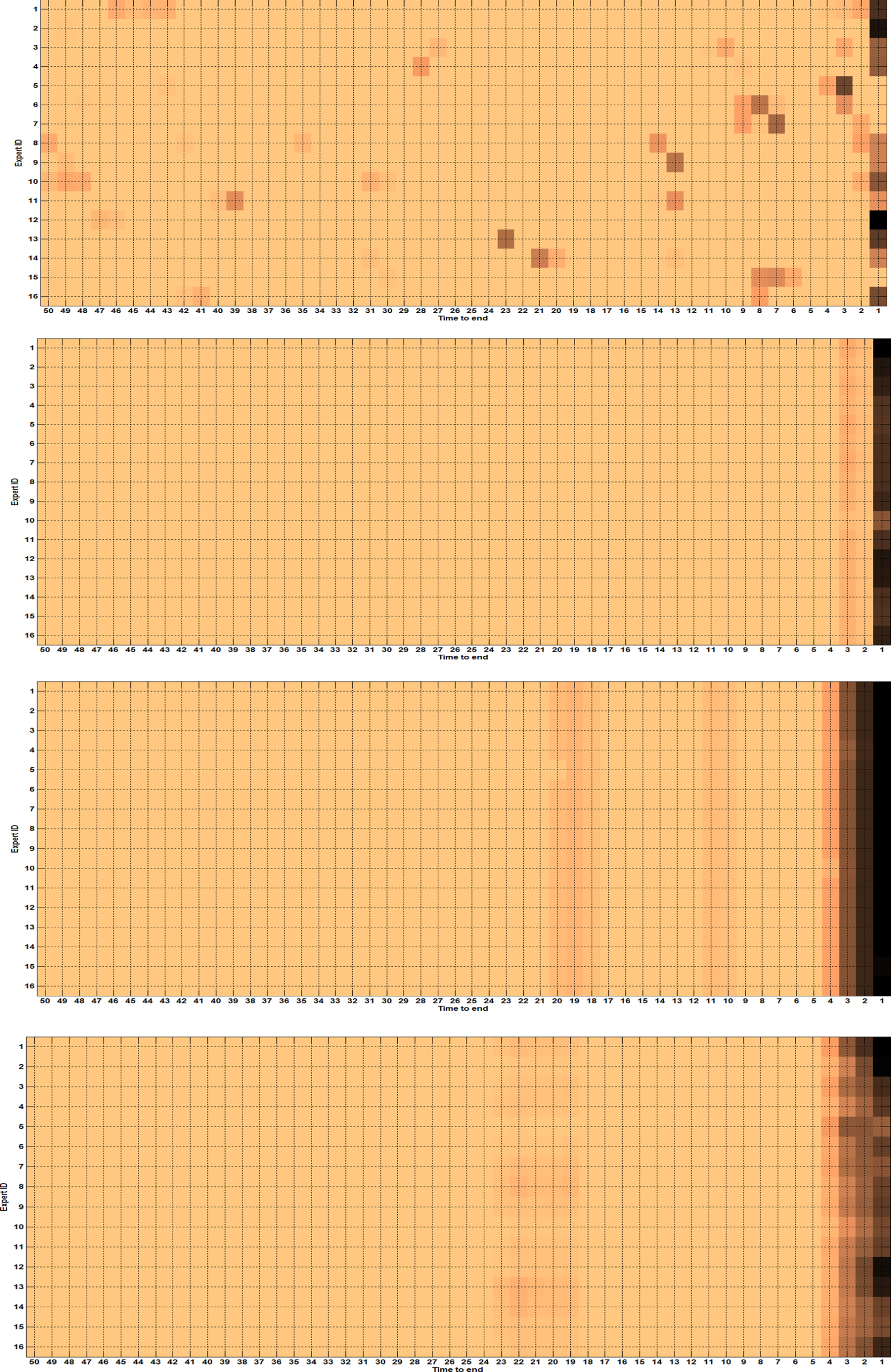}}
\centerline{{Arousal} \hspace{0.45\linewidth} {Valence}} 
\vspace{.1cm}
\caption{MTL with \textbf{\textit{experts as tasks}} and the arousal (left) and valence (right) ratings provided by them as features. (From top to bottom) $W$ matrices obtained with \textit{Lasso}, $\ell_{2,1}$, \textit{Dirty} and \textit{Robust} MTL. Larger weights are denoted using darker shades. Best viewed under zoom. }
\label{Experts_tasks}
\vspace{0.2cm}
\centerline{\includegraphics[width=0.49\linewidth,height=2cm]{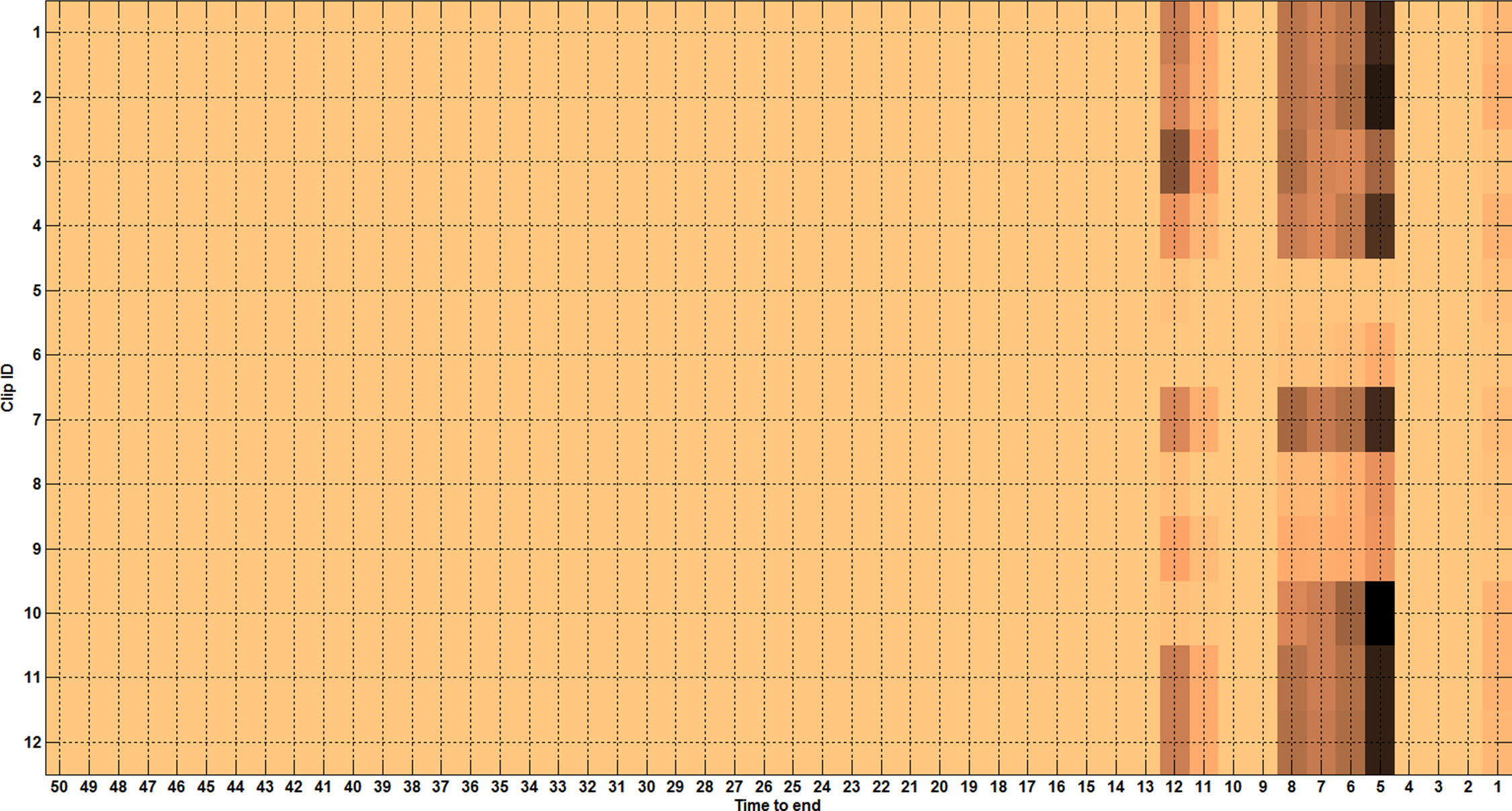}\hspace{.02cm}\includegraphics[width=0.49\linewidth,height=2cm]{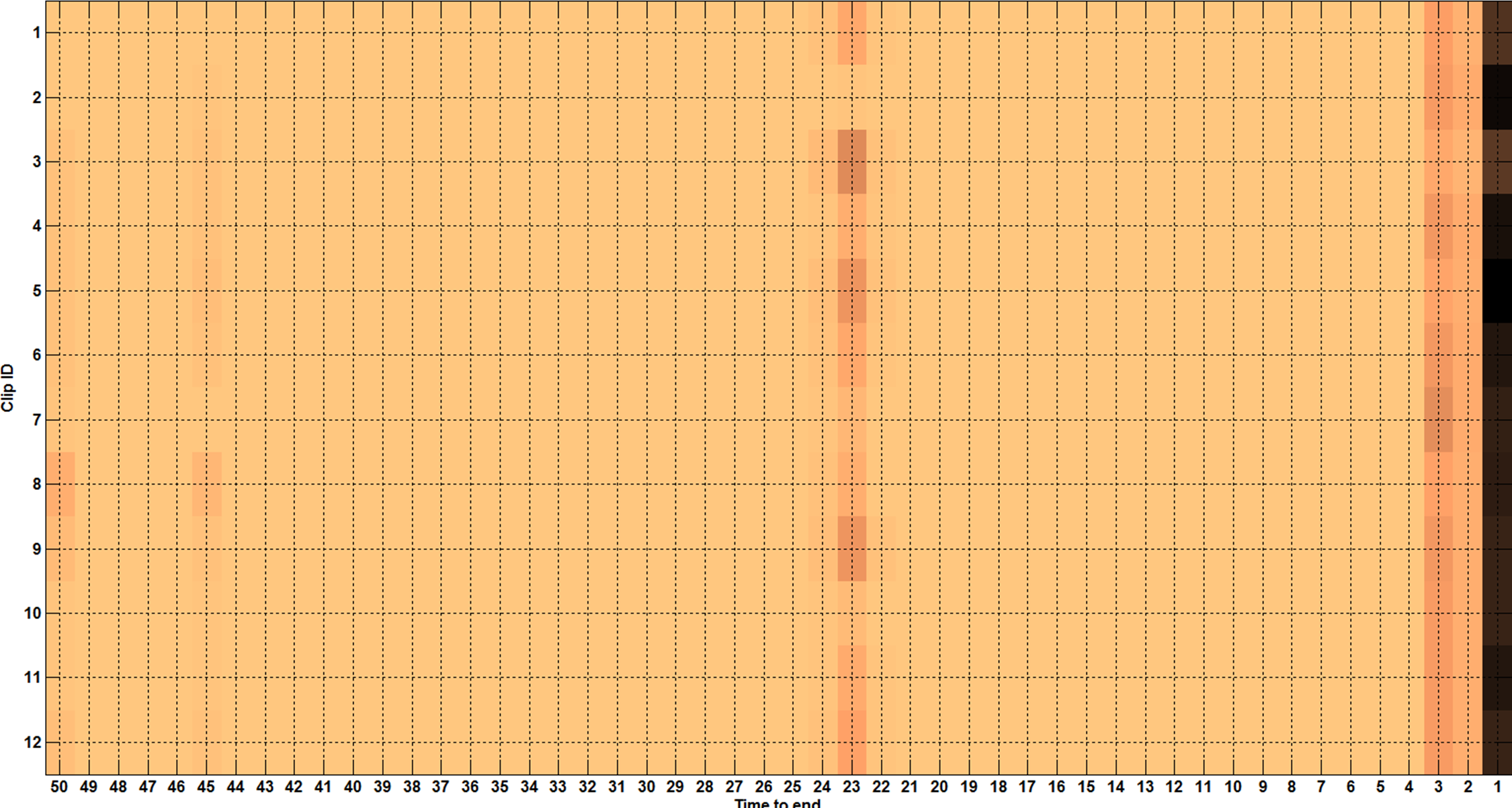}}\vspace{0.1cm}
\centerline{\includegraphics[width=0.49\linewidth,height=2cm]{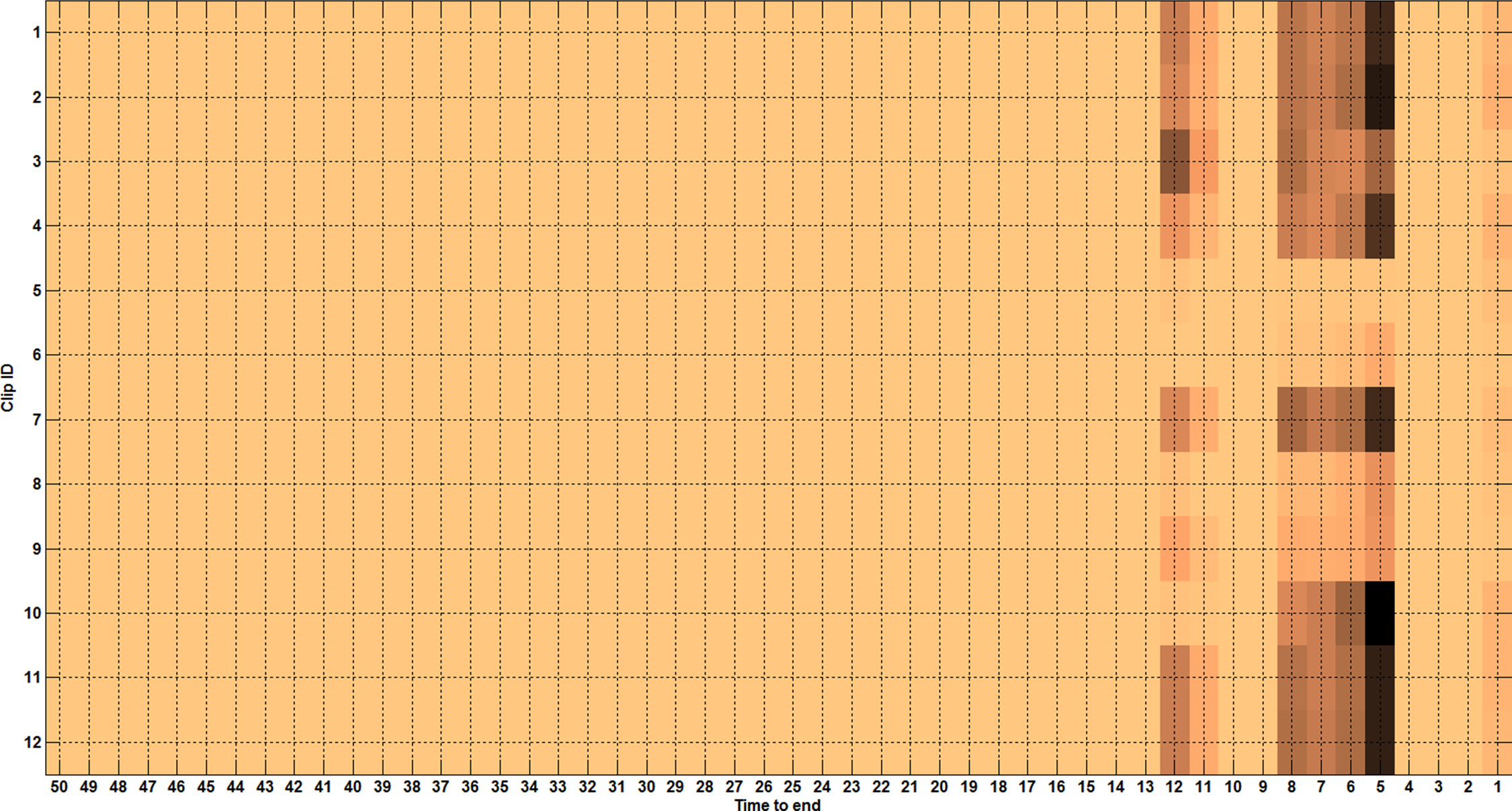}\hspace{.02cm}\includegraphics[width=0.49\linewidth,height=2cm]{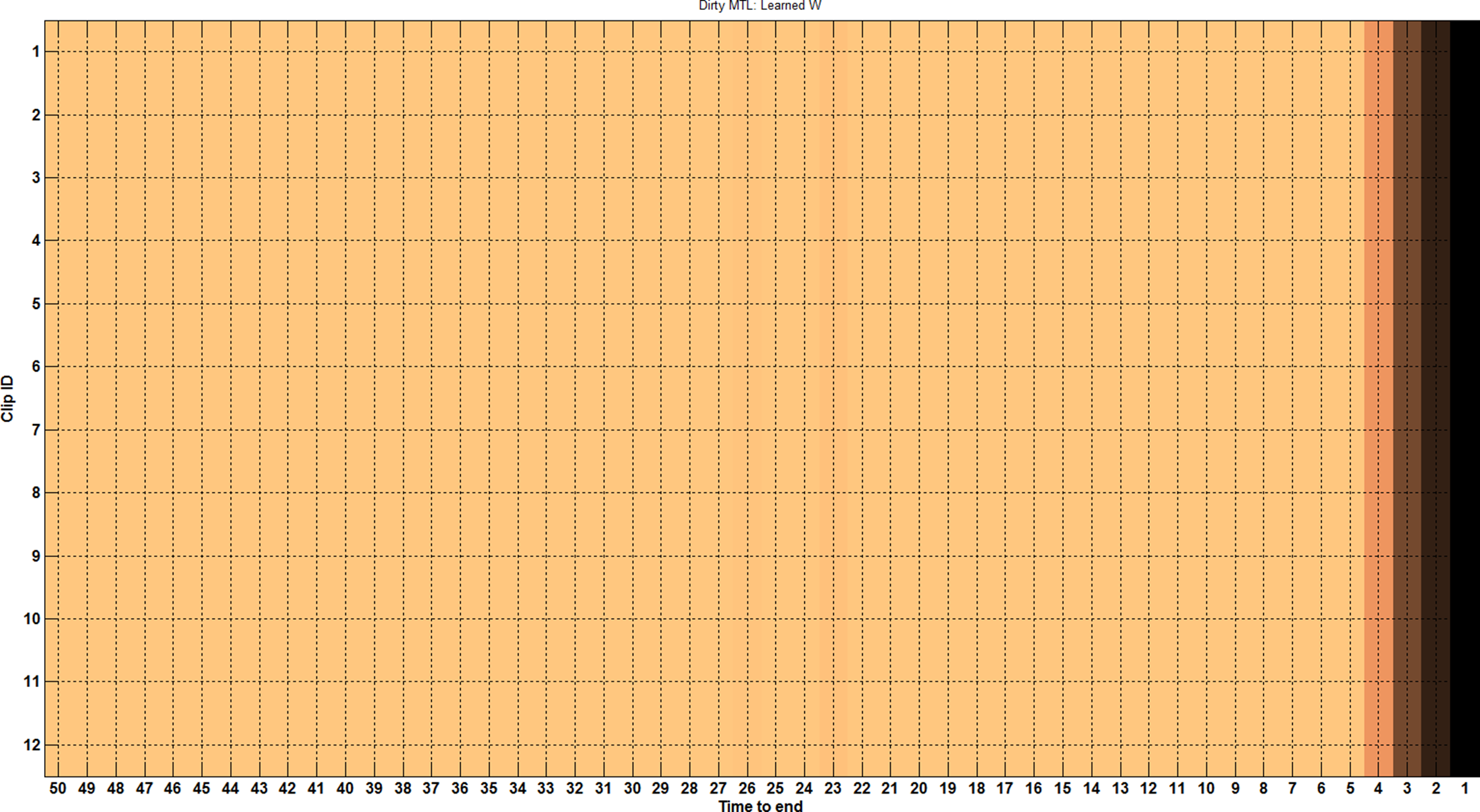}}\vspace{0.1cm}
\centerline{\includegraphics[width=0.49\linewidth,height=2cm]{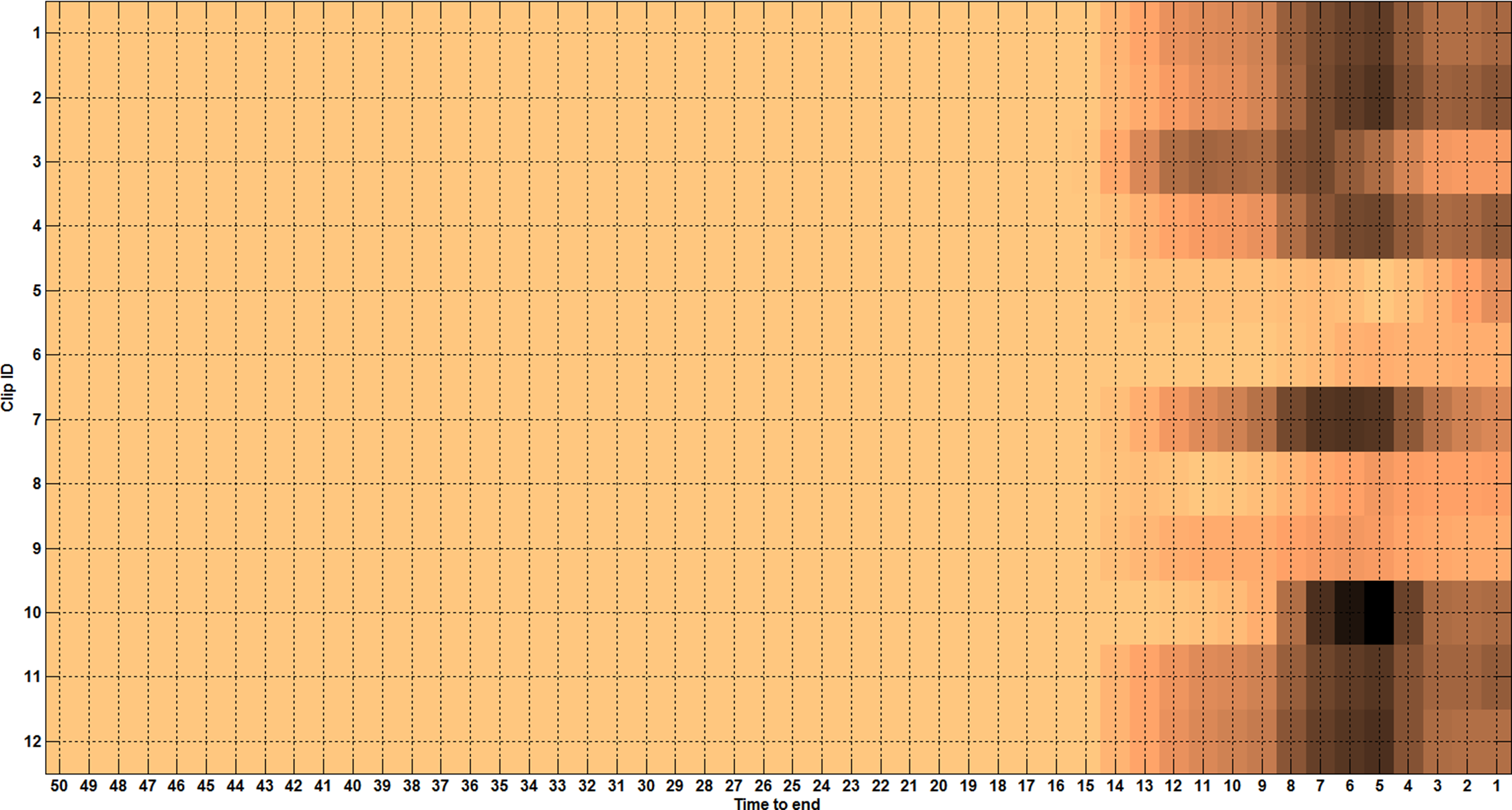}\hspace{.02cm}\includegraphics[width=0.49\linewidth,height=2cm]{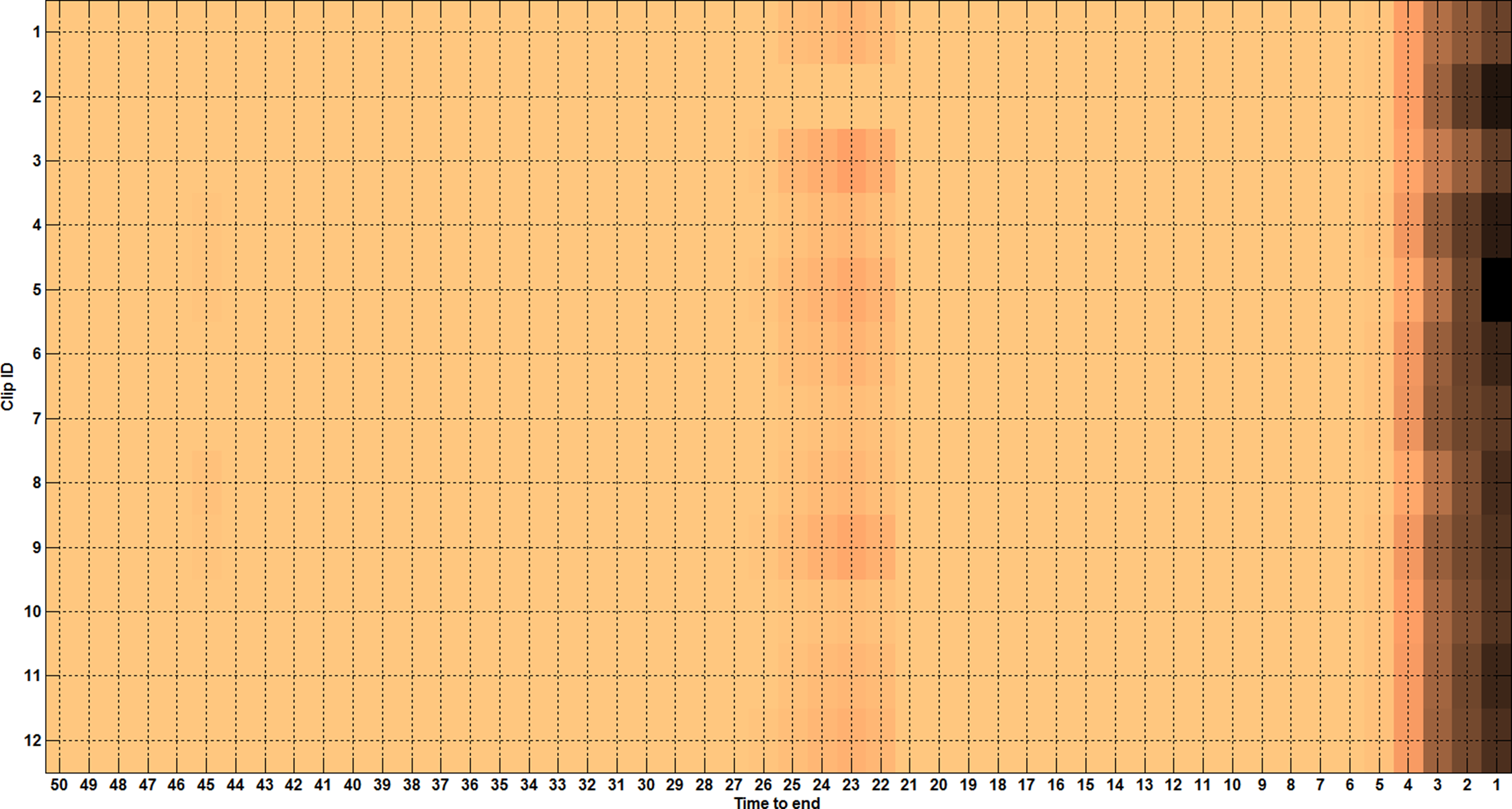}}\vspace{0.1cm}
\centerline{\includegraphics[width=0.49\linewidth,height=2cm]{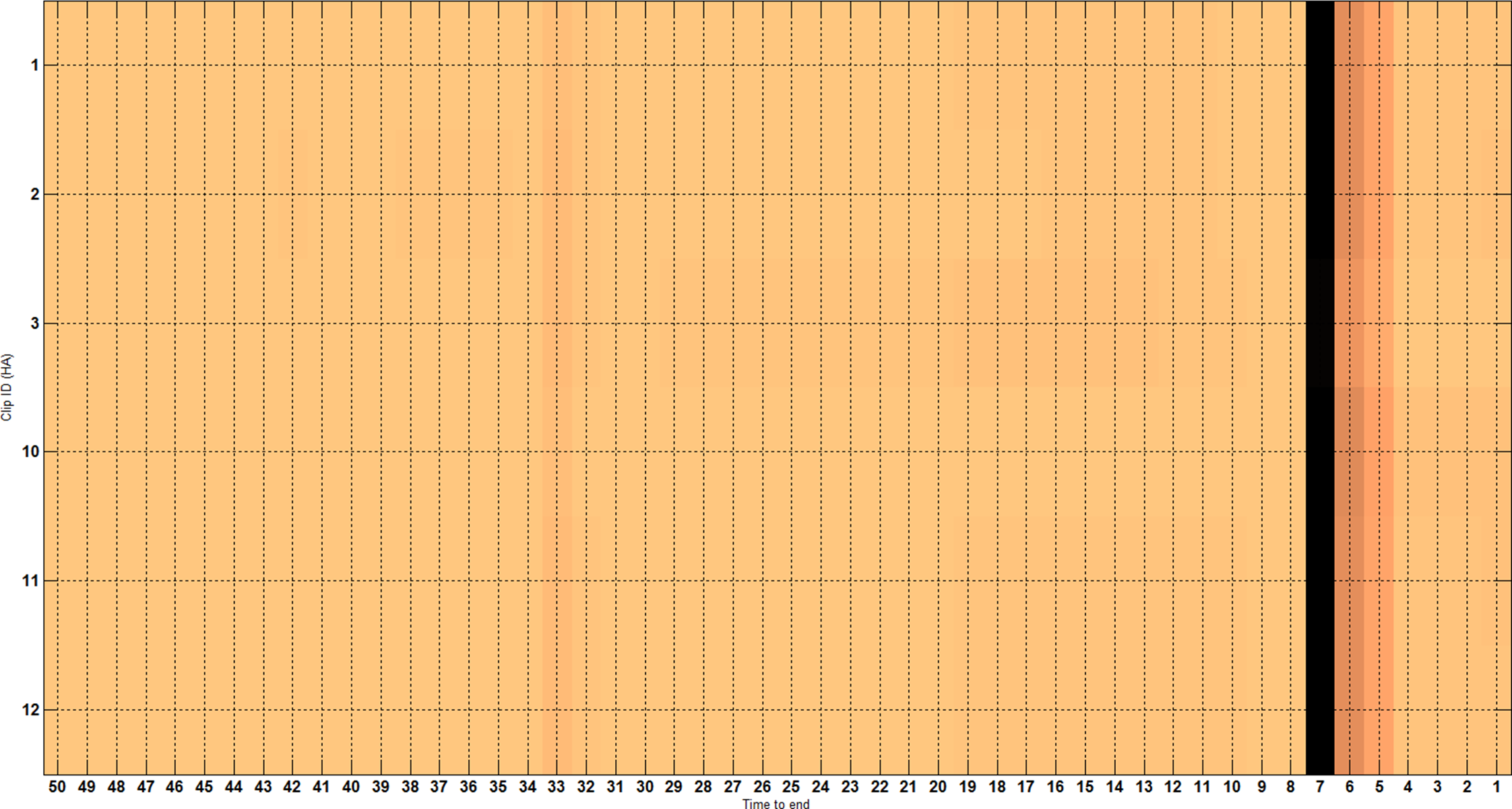}\hspace{.02cm}\includegraphics[width=0.49\linewidth,height=2cm]{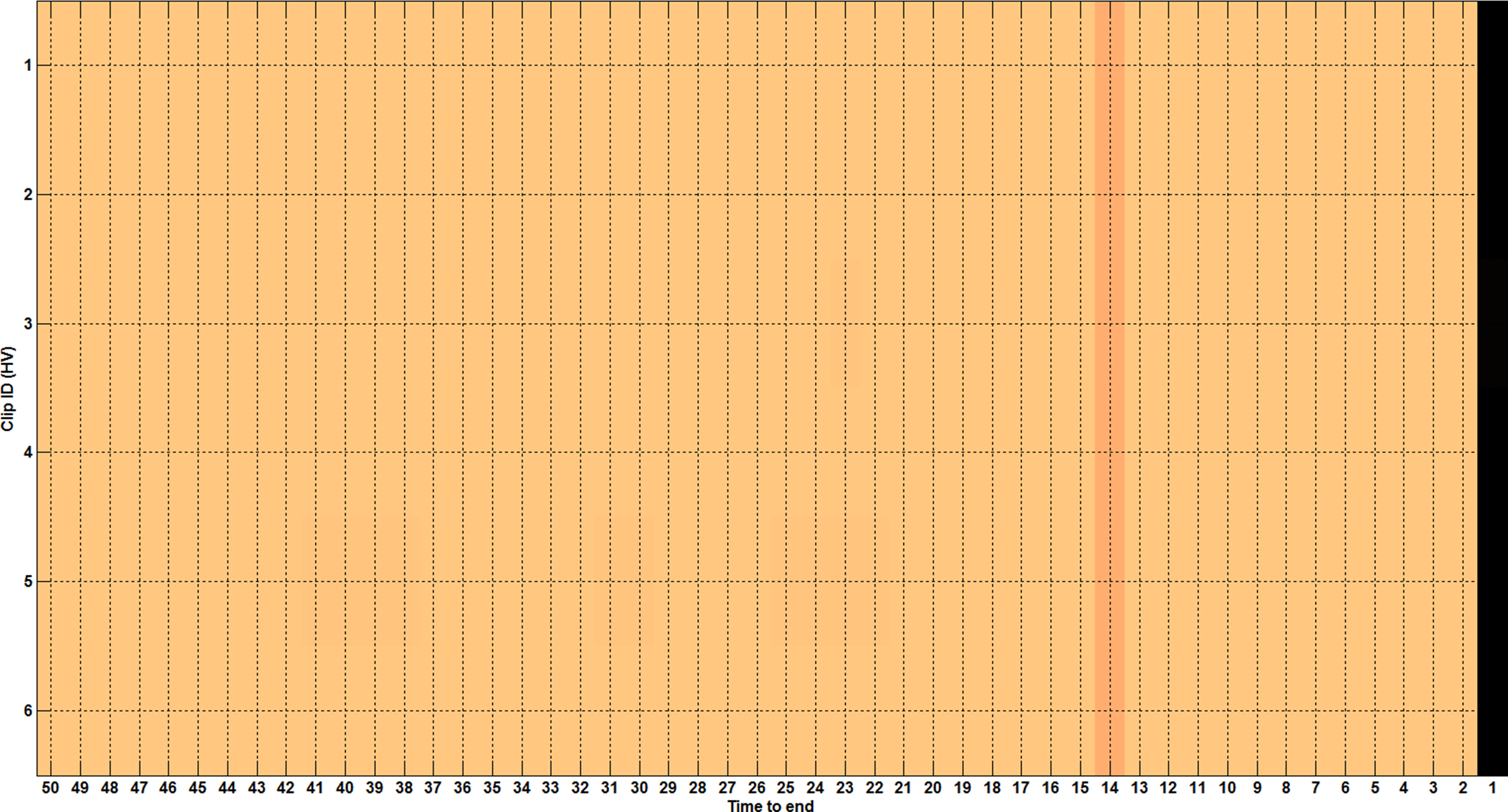}}\vspace{0.1cm}
\centerline{\includegraphics[width=0.49\linewidth,height=2cm]{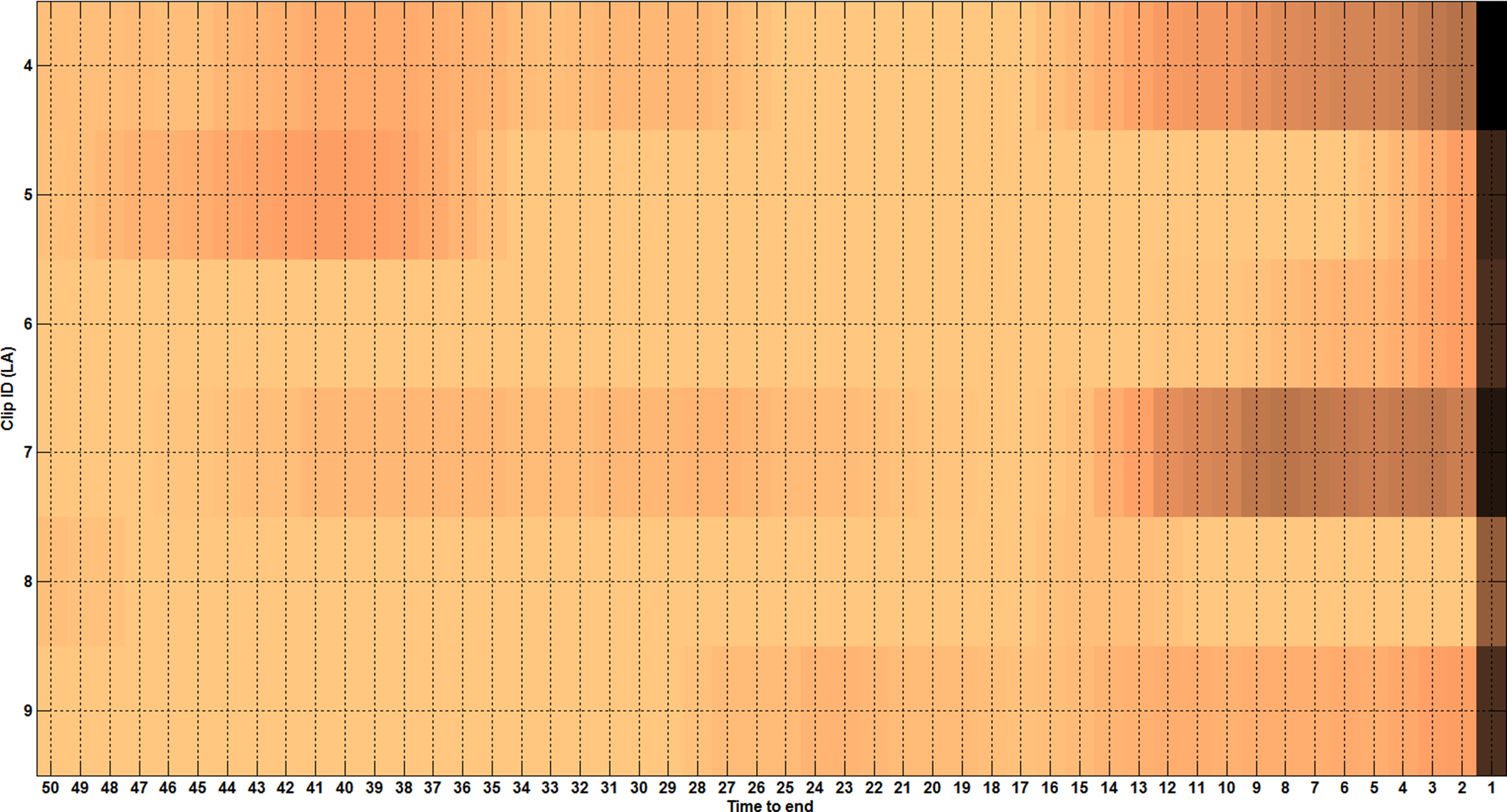}\hspace{.02cm}\includegraphics[width=0.49\linewidth,height=2cm]{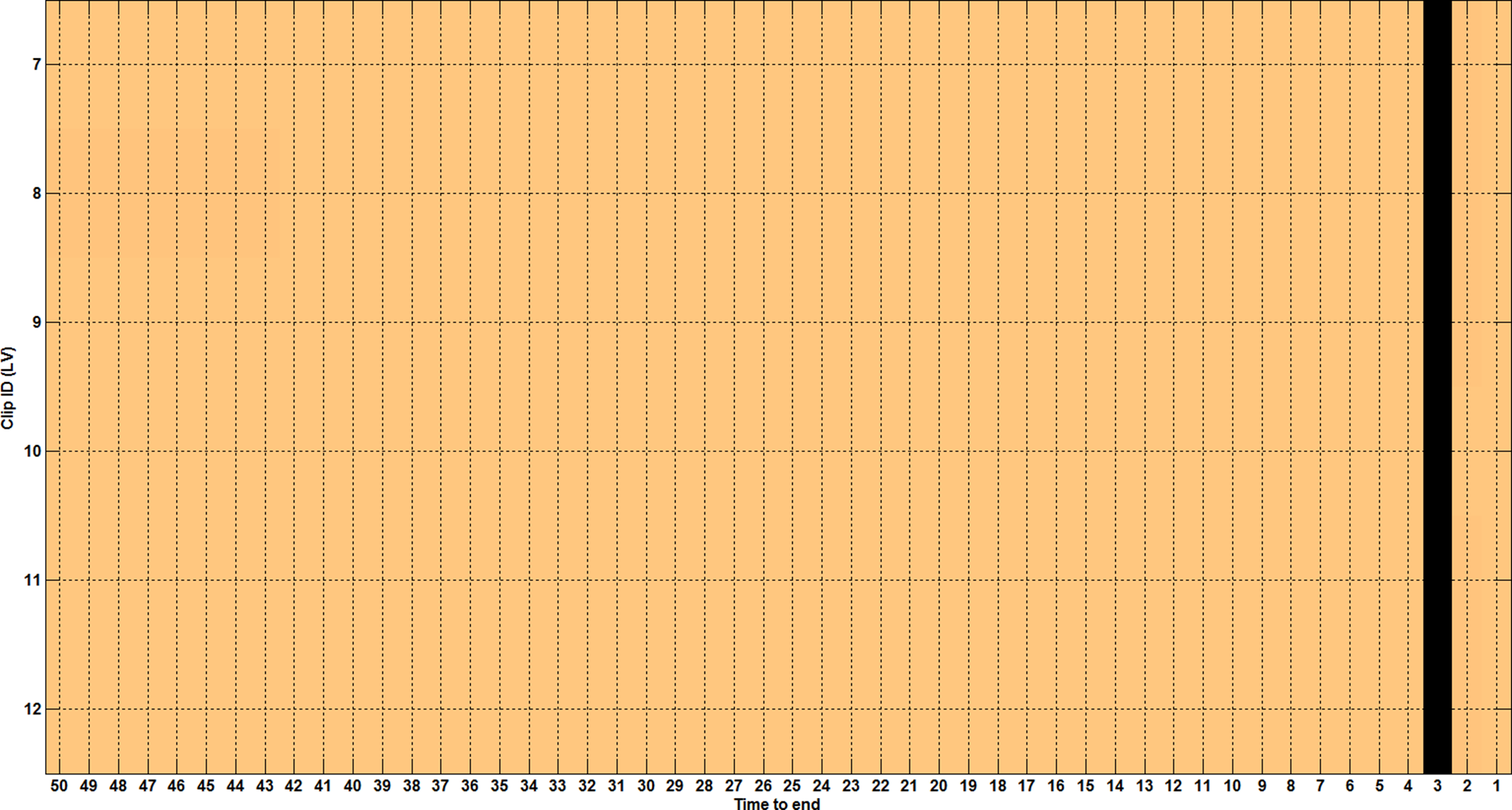}}\vspace{0.05cm}
\centerline{Arousal \hspace{0.44\linewidth} Valence} 
\vspace{-0.1cm}
\caption{MTL with \textbf{Val} \textbf{\textit{movie clips as tasks}} and experts' arousal (left), valence (right) ratings as features. (From top to bottom) $W$ matrices obtained with  $\ell_{2,1}$, \textit{Dirty}, \textit{Robust} and \textit{Graph regularized} MTL (last two rows). Larger weights are denoted using darker shades. Best viewed under zoom.}
\label{Clips_tasks}
\end{figure}

Figure~\ref{Experts_tasks} presents the learned weight ($W$) matrices for the above problem using the Lasso, $\ell_{2,1}$, Dirty and Robust MTL baselines detailed in Section~\ref{MTL}. Apart from Lasso-MTL, all other methods attempt to learn a \textit{group} component, and it is evident from figure rows 2--4 that there are greater similarities in the expert rating patterns for the latter half of the clips as also typified by the higher Kendall W values for the second halves. Also, maximum consistency among expert V ratings is noted at the very end of the movie clips (this also maximally influences the overall clip V label), while the influential time points corresponding to similar A ratings are more distributed.   

For problems \textbf{\texttt{P1}} and \textbf{\texttt{P2}} which require prediction of dynamic or static V,A clip labels, it is more appropriate to model each clip as a task. As a preliminary step, we attempted static emotion recognition modeling each clip as a task, \ie, the MTL baselines are trained with a feature matrix containing dynamic A/V expert ratings and static \textbf{Val} clip A/V labels as data labels. The above learning framework reveals \textit{those time points in the continuous emotion profile that contribute the most to the static emotional clip labels}. Figure~\ref{Clips_tasks} presents the learned $W$s using the $\ell_{2,1}$, Dirty, Robust and SR-MTL methods. Evidently, dynamic A,V ratings towards the end of the movie clip have the greatest impact on its static emotional label-- this result reinforces the observation made from Figure~\ref{Crowd_Exp}. The last two rows show the learned weights using graph-regularized MTL for HA, HV, LA and LV clips (clips with identical static A/V labels are defined as related tasks, and these task relationships are embedded in the form of a graph). Note that the most influential time-points corresponding to HA clips occur a few seconds before the end, while a larger sequence of time points are required to accurately predict the arousal label for LA clips (possibly due to the larger variance in A ratings for LA clips). Conversely, very few time points determine the static clip V for both HV and LV clips.

As a supplement to Figure~\ref{Clips_tasks}, Table~\ref{table_ca} presents static A,V recognition accuracies\footnote{computed on the training data} when the different MTL algorithms are trained with the dynamic \textbf{Val} affective ratings provided by experts and crowdworkers. Sparsity levels of the learned $W$ matrices are also specified in braces. Reflecting the higher consistency in dynamic V annotations and the greater agreement among expert ratings, superior recognition accuracy in general is achieved for V and with expert annotations with the learned $W$ matrices of comparable sparsity.

\begin{figure*}[!htbp]
\centerline{\includegraphics[width=0.23\linewidth]{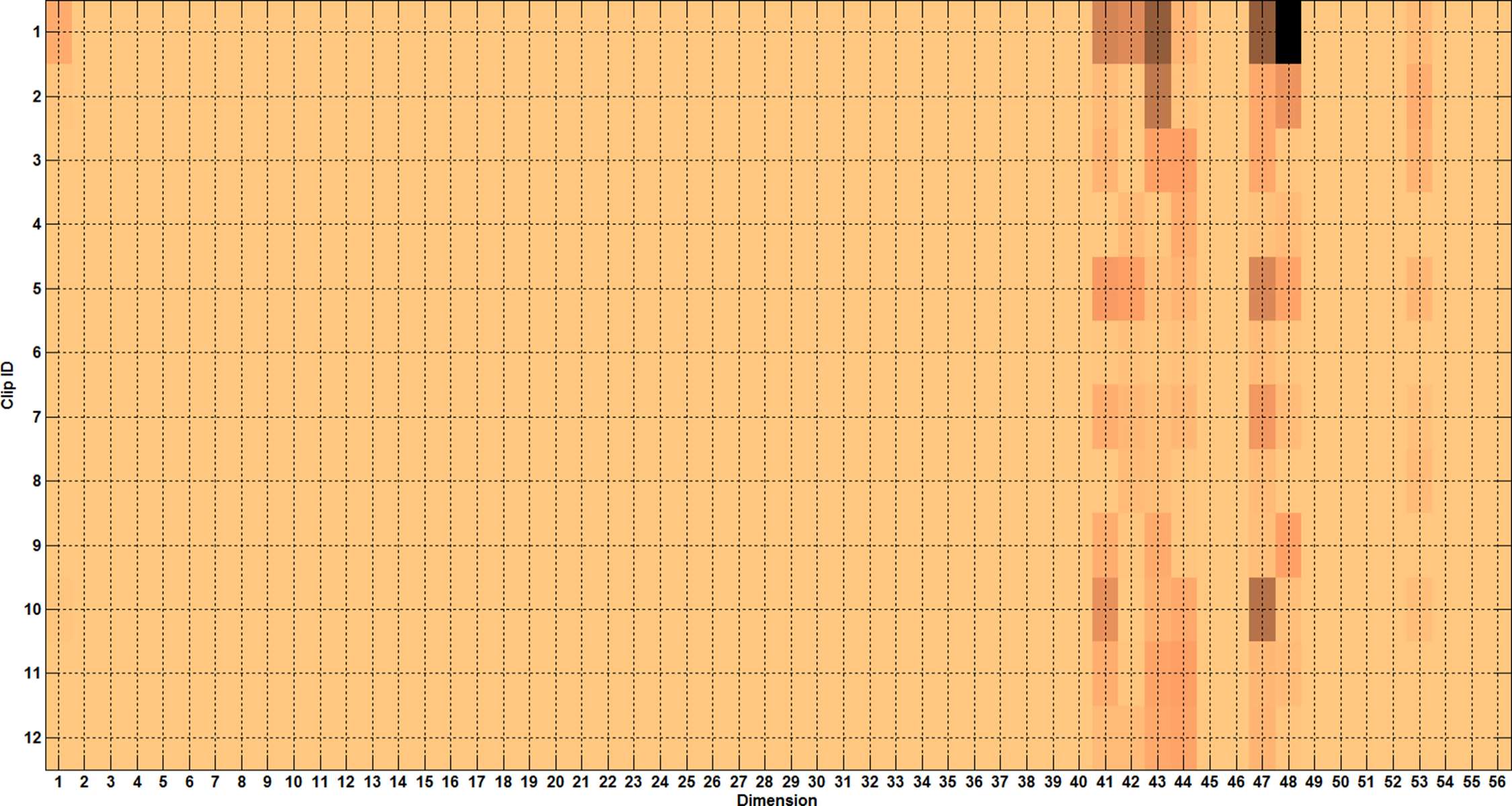}\hspace{.02cm}\includegraphics[width=0.23\linewidth]{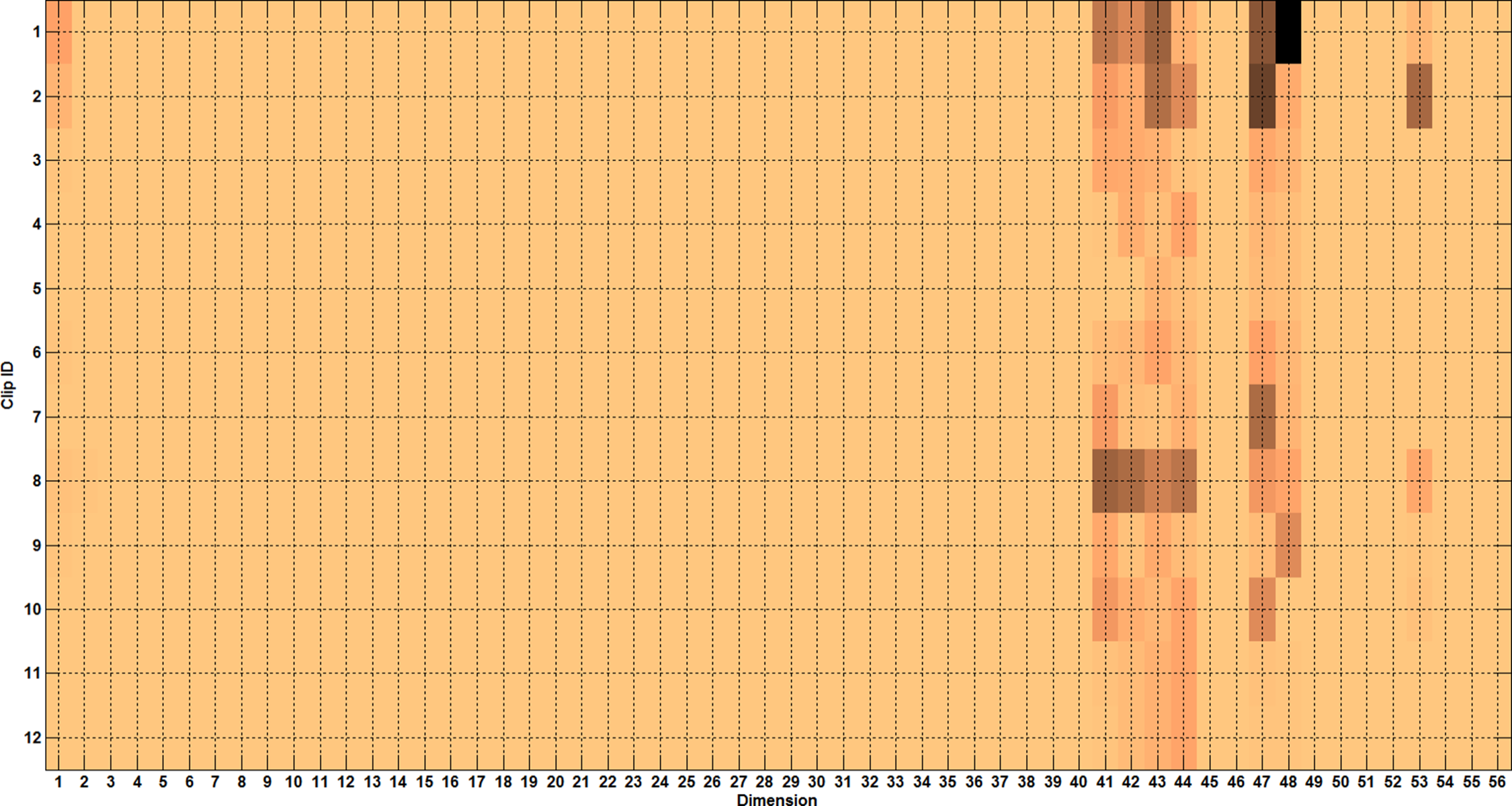}\hspace{.02cm}\includegraphics[width=0.23\linewidth]{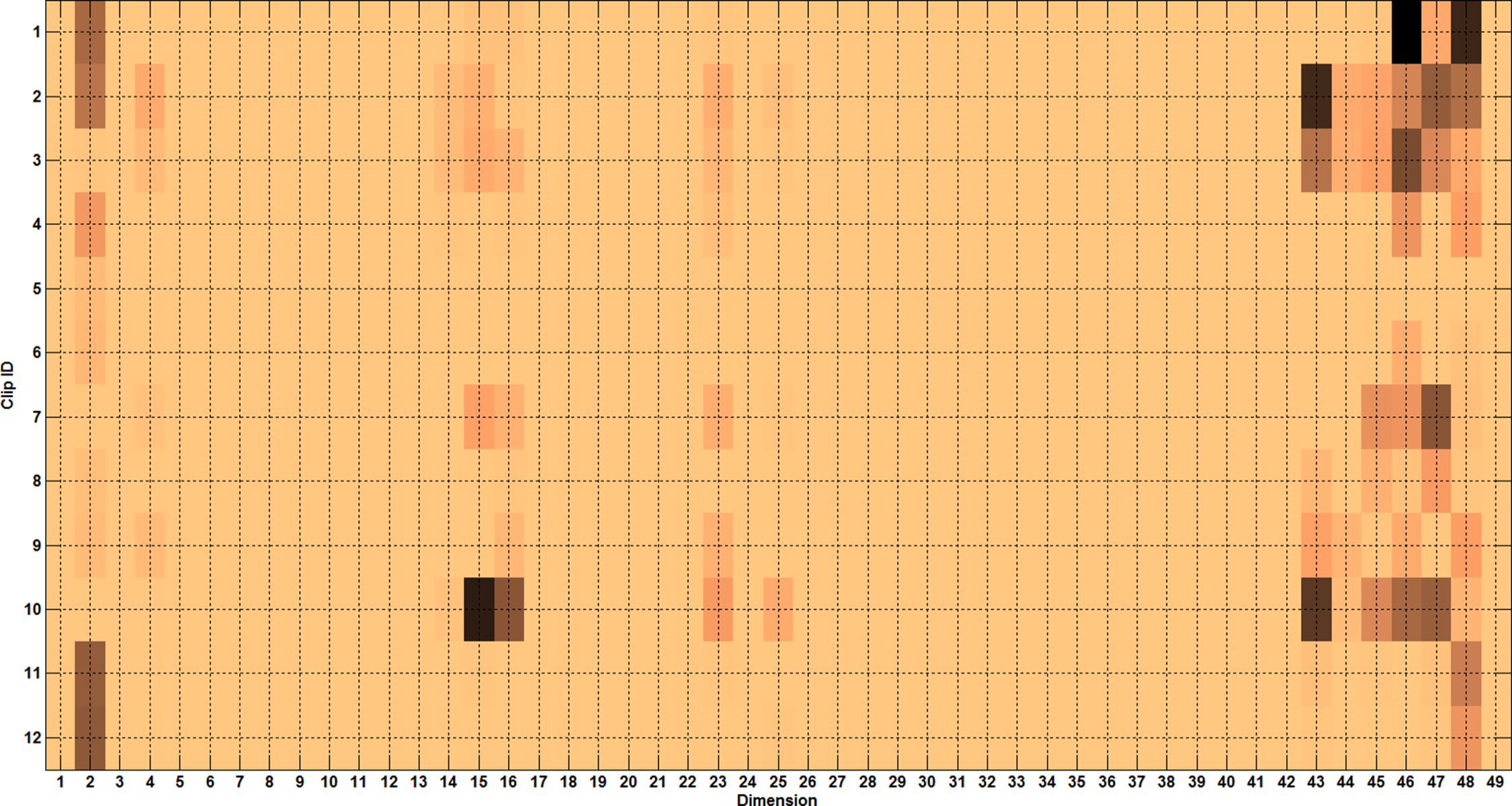}\hspace{.02cm}\includegraphics[width=0.23\linewidth]{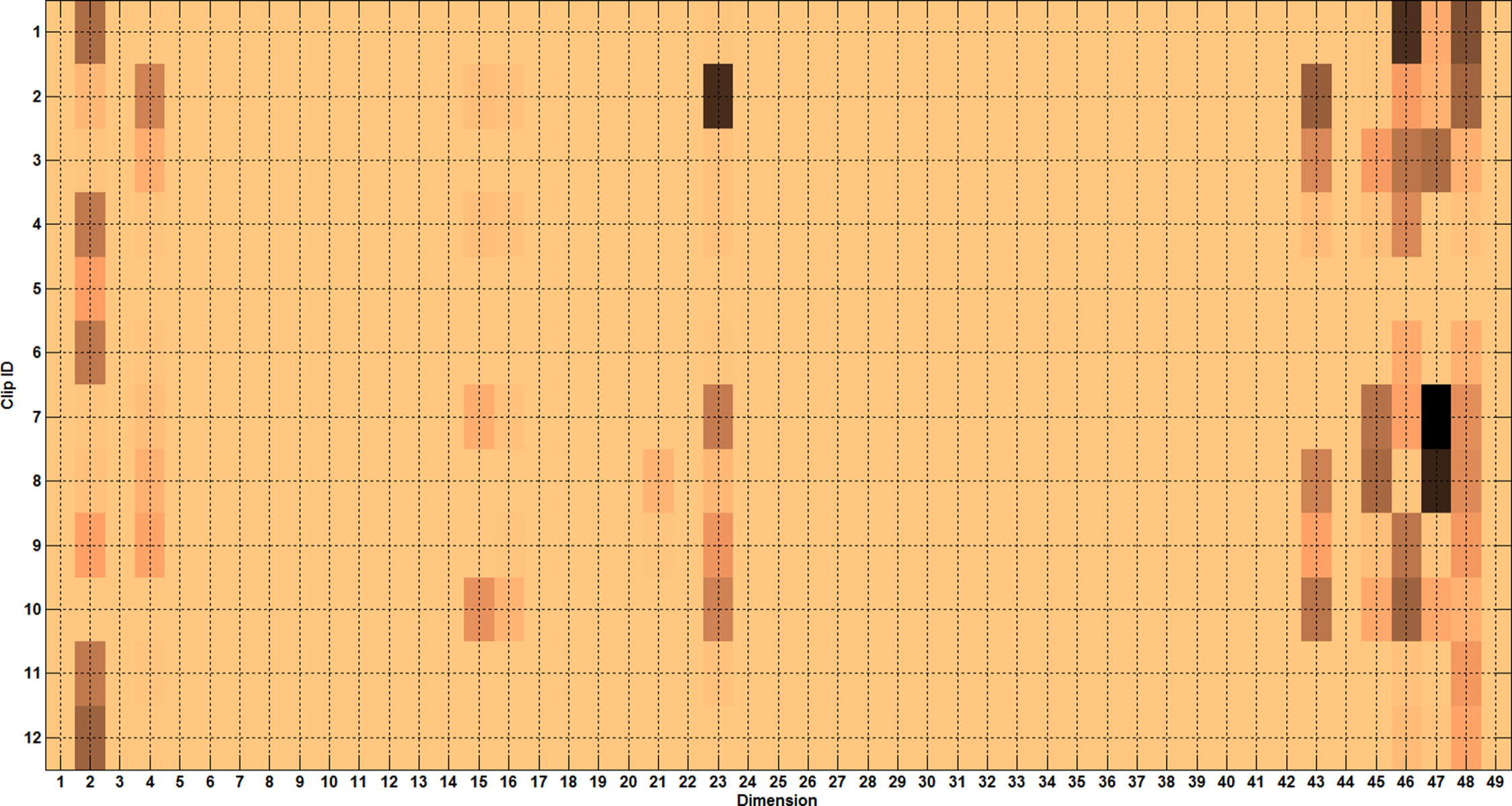}}\vspace{0.1cm}
\centerline{\includegraphics[width=0.23\linewidth]{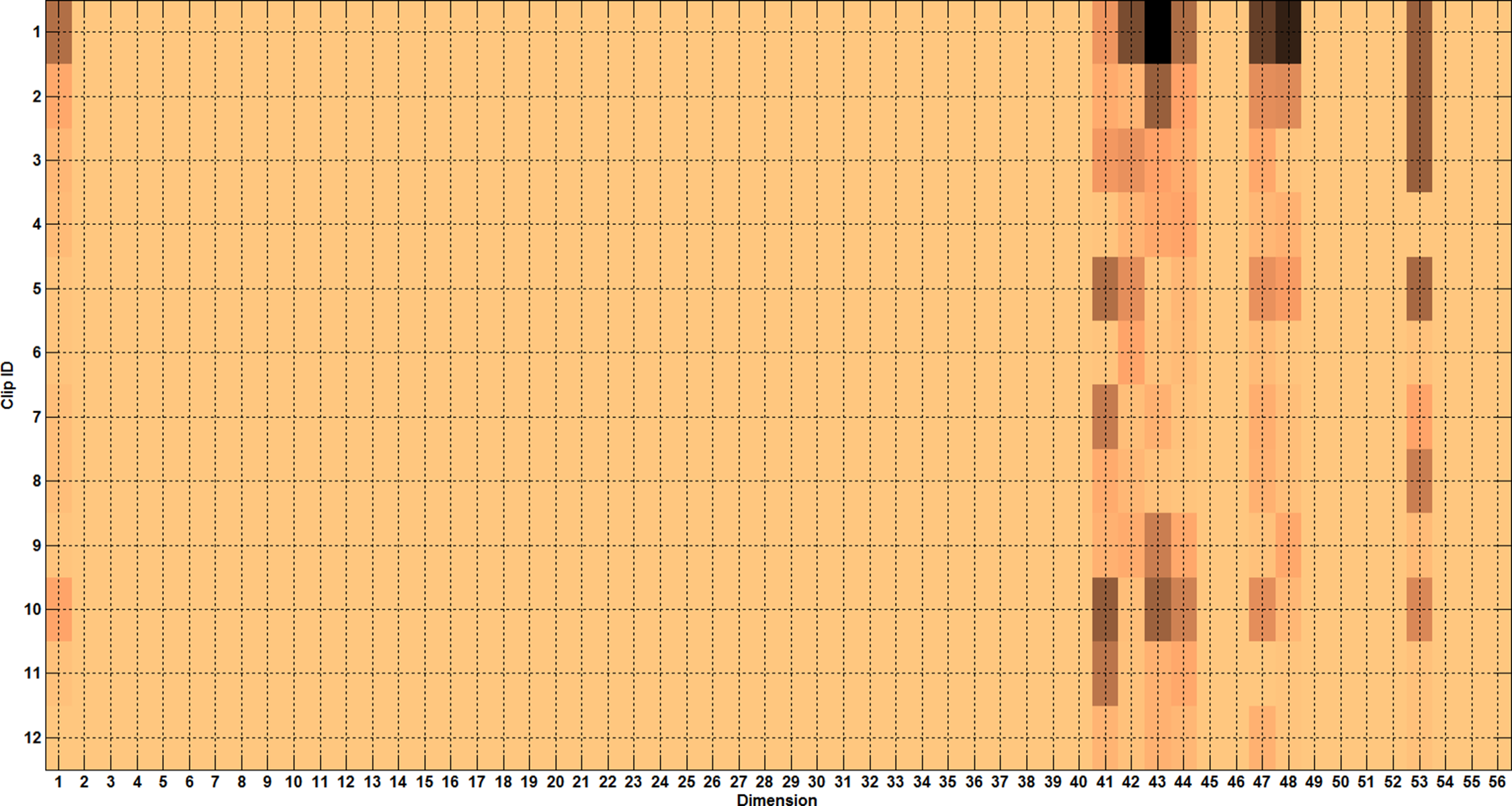}\hspace{.02cm}\includegraphics[width=0.23\linewidth]{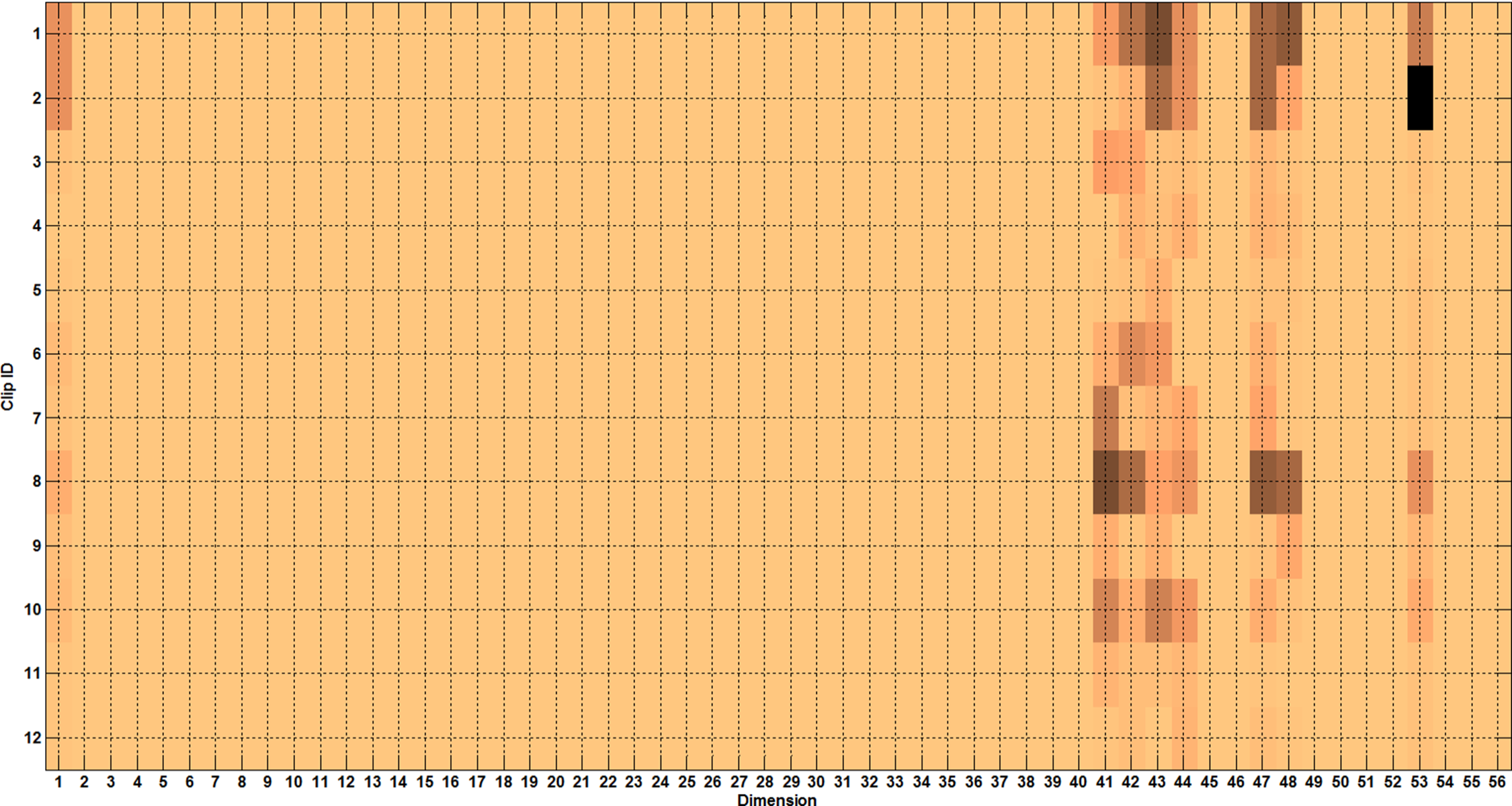}\hspace{.02cm}\includegraphics[width=0.23\linewidth]{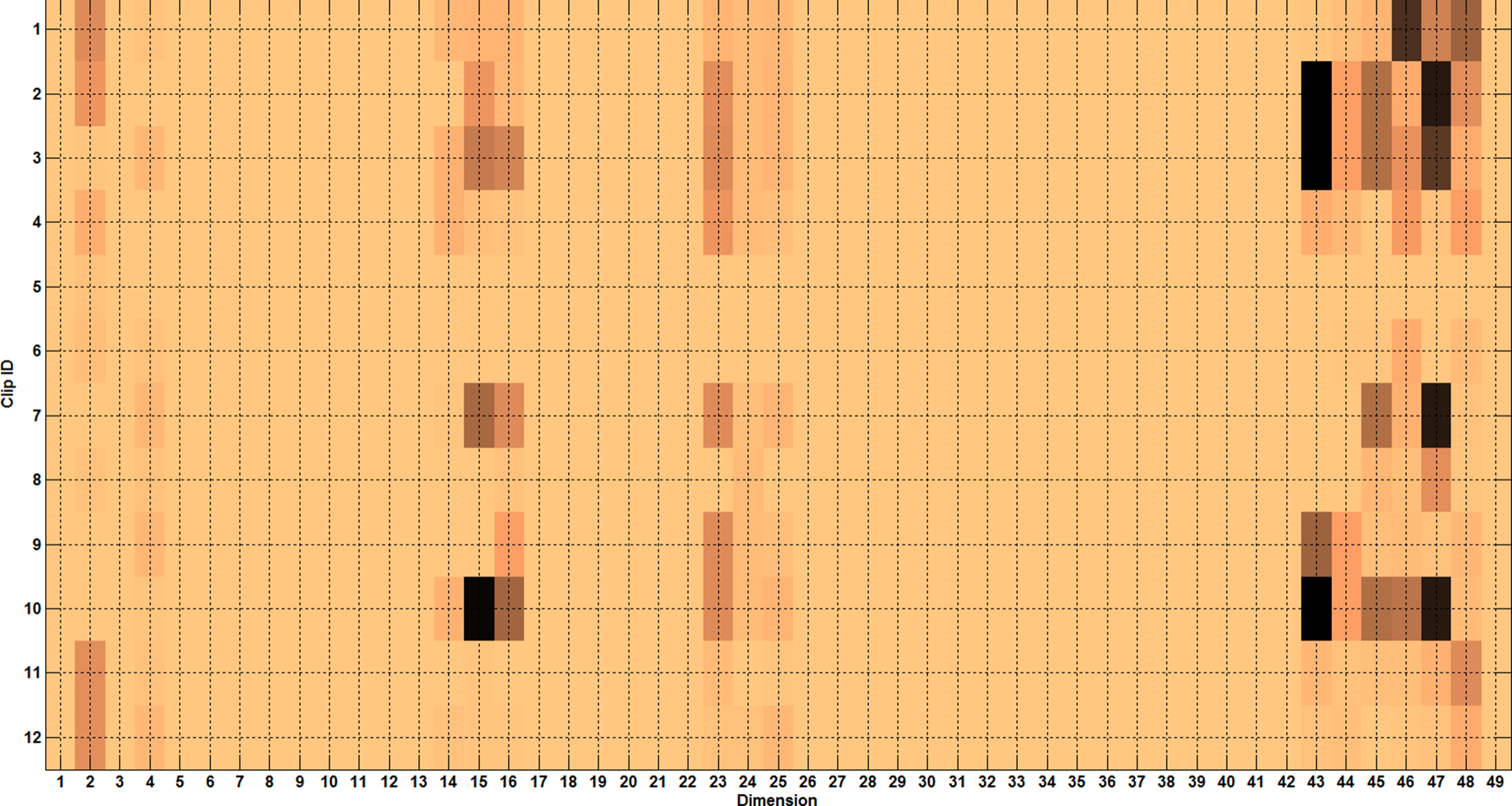}\hspace{.02cm}\includegraphics[width=0.23\linewidth]{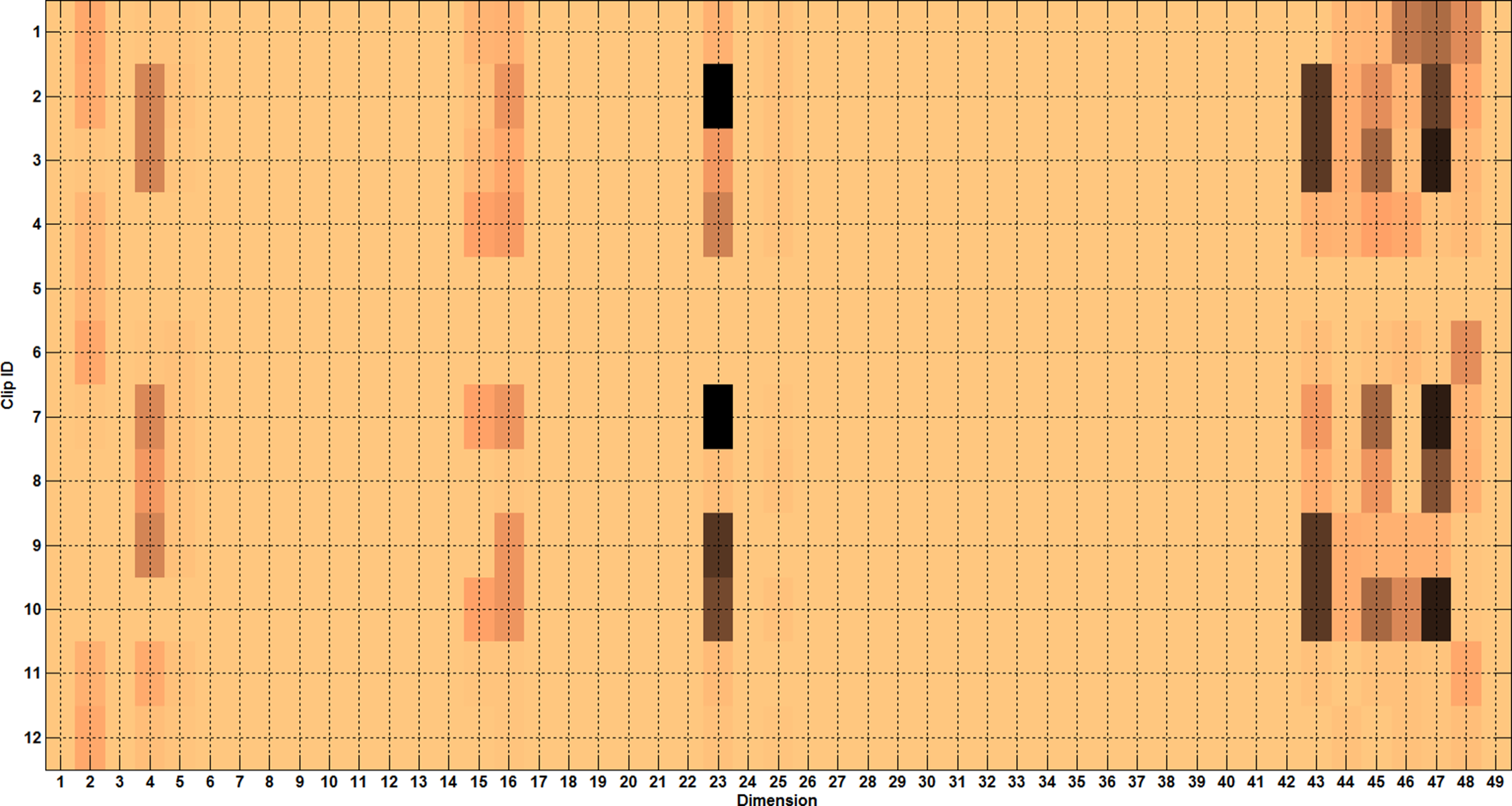}}\vspace{0.1cm}
\centerline{\includegraphics[width=0.23\linewidth]{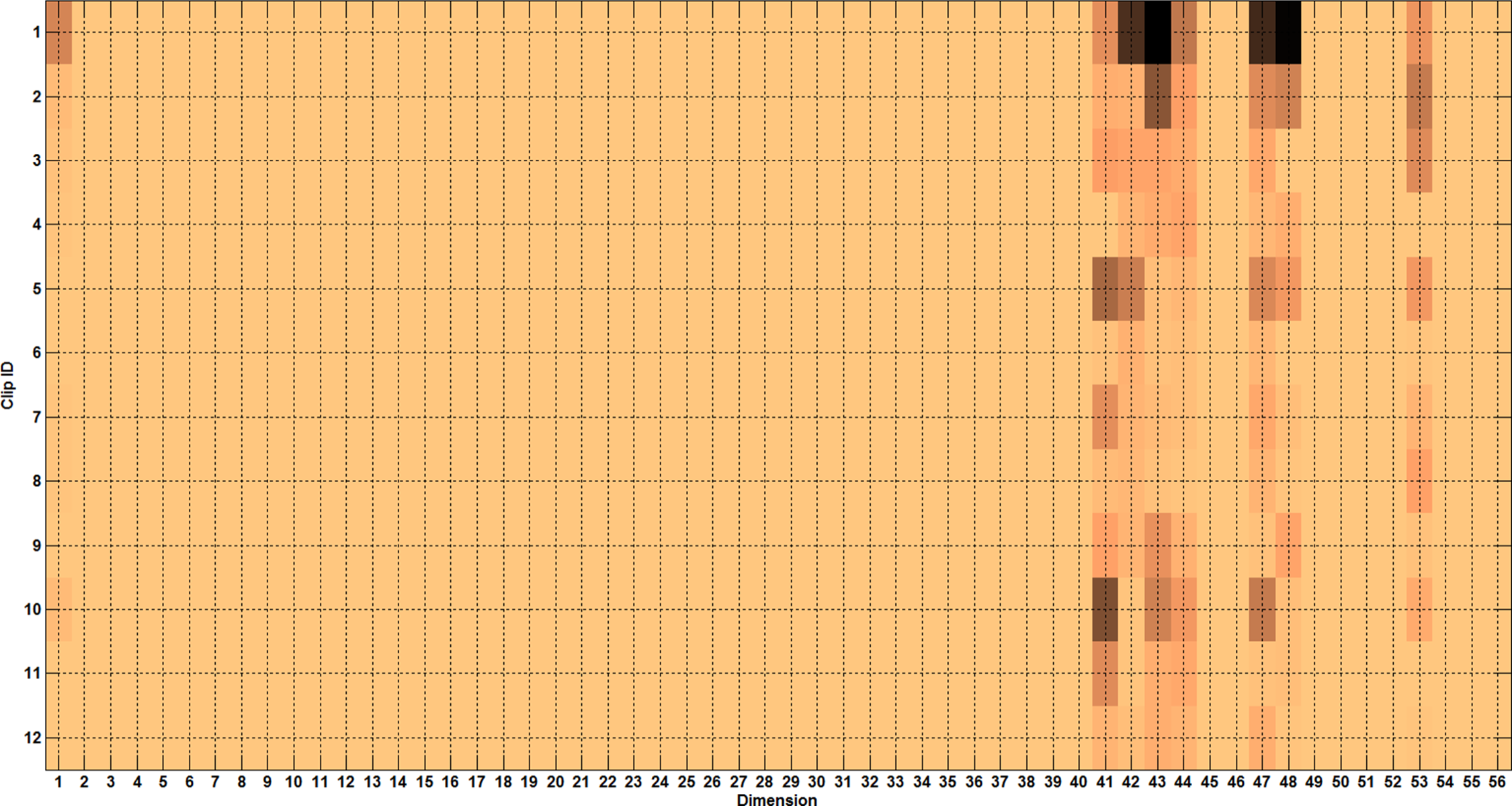}\hspace{.02cm}\includegraphics[width=0.23\linewidth]{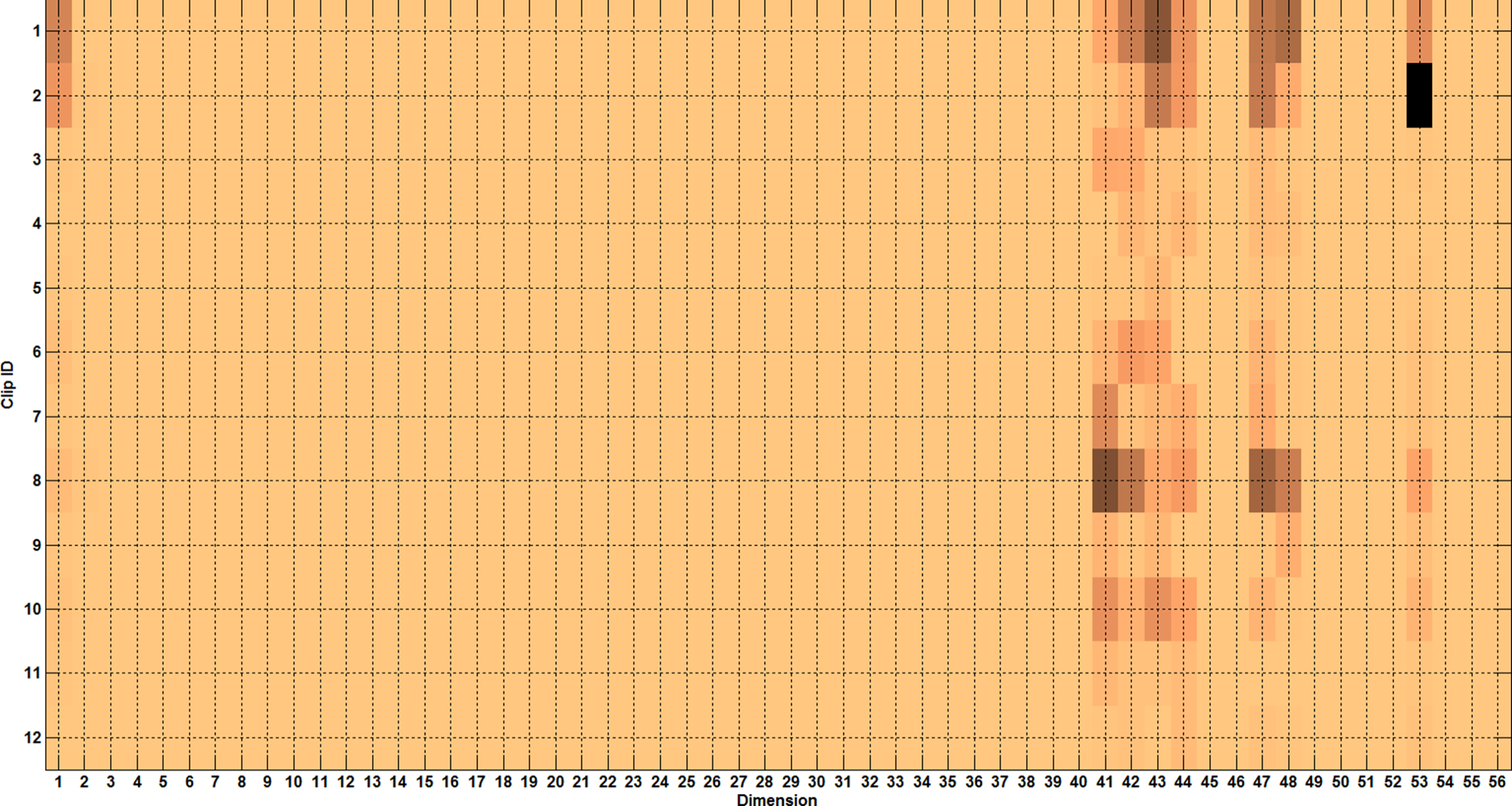}\hspace{.02cm}\includegraphics[width=0.23\linewidth]{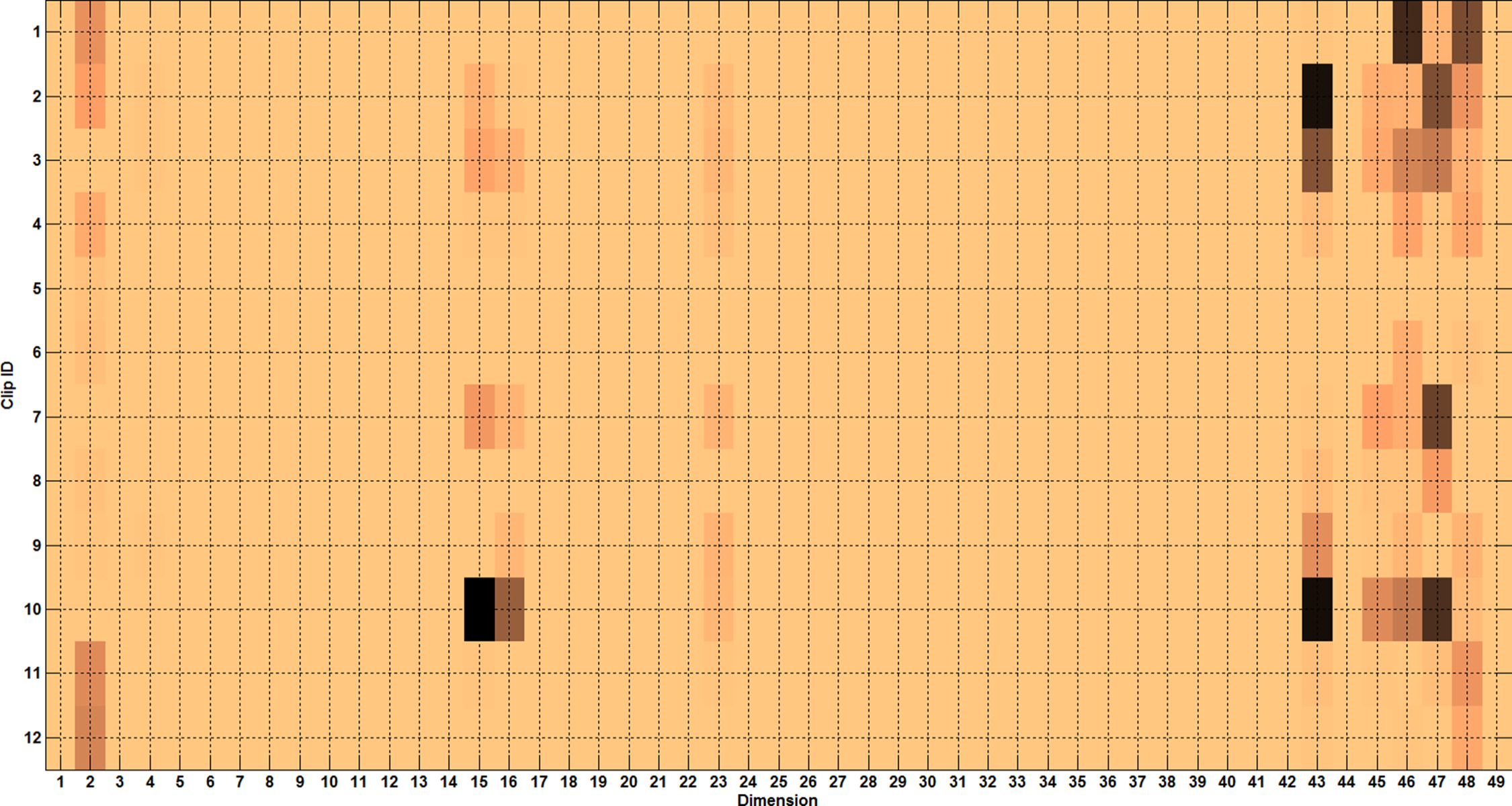}\hspace{.02cm}\includegraphics[width=0.23\linewidth]{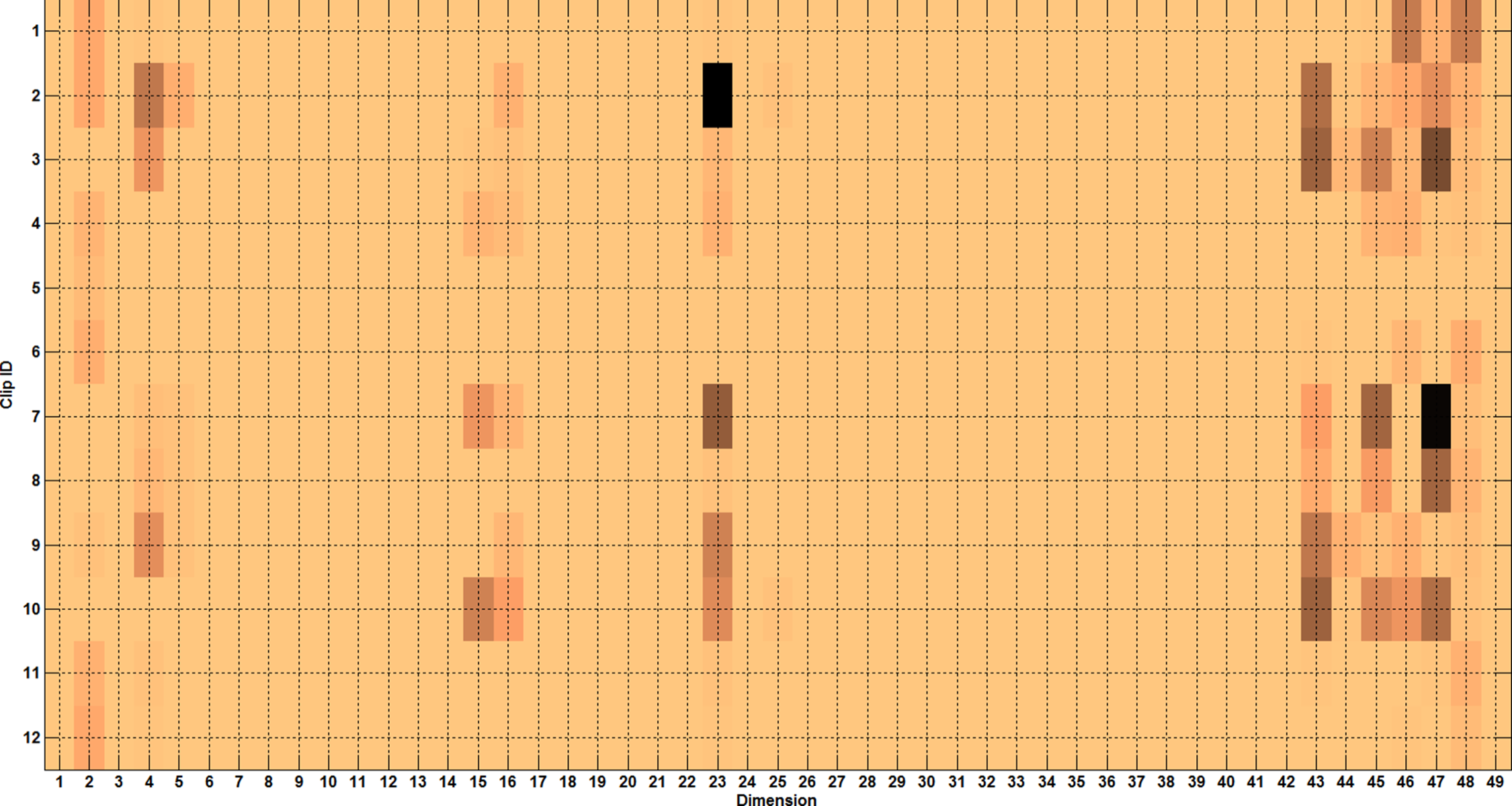}}\vspace{0.1cm}
\centerline{\includegraphics[width=0.23\linewidth]{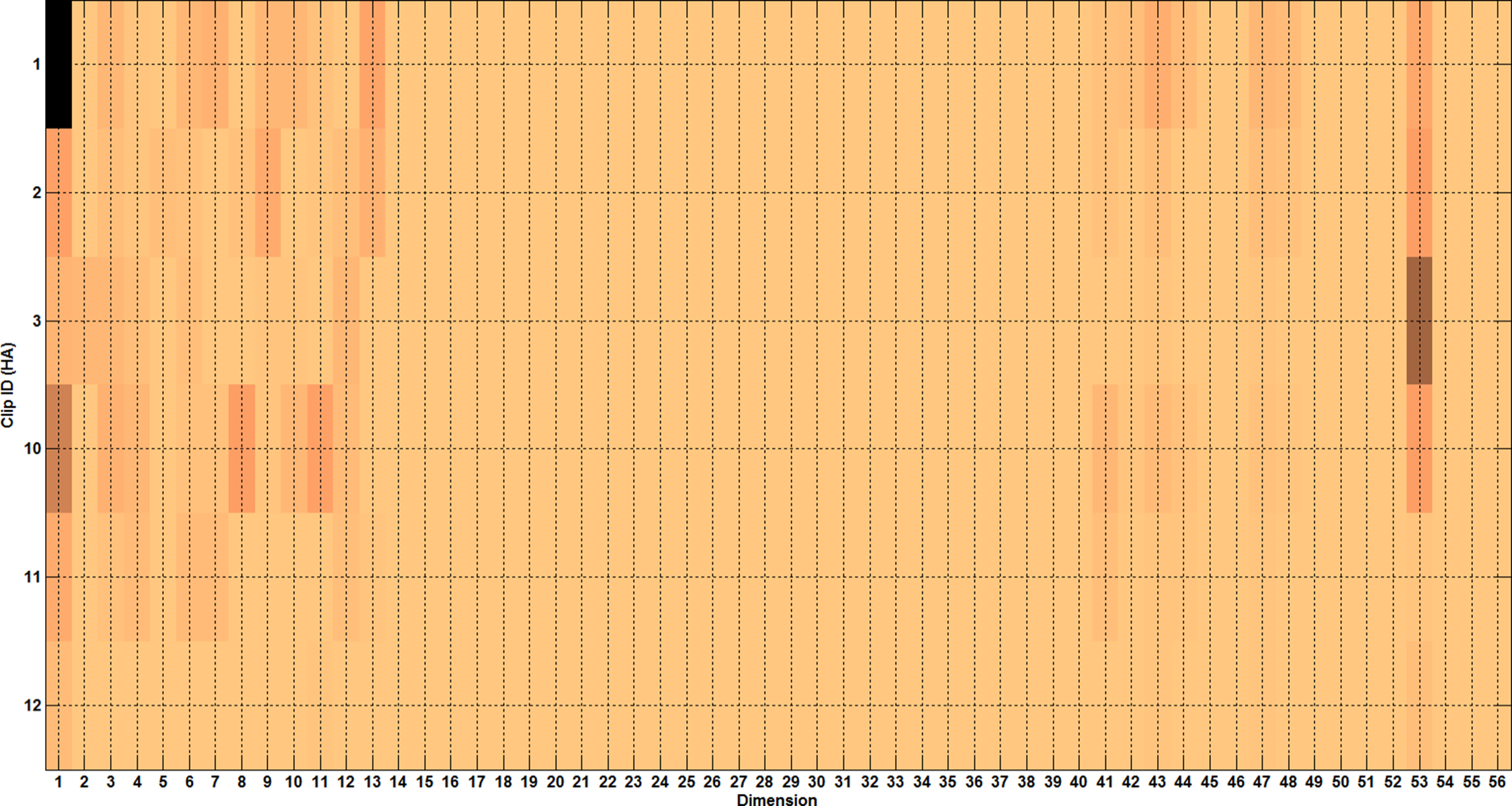}\hspace{.02cm}\includegraphics[width=0.23\linewidth]{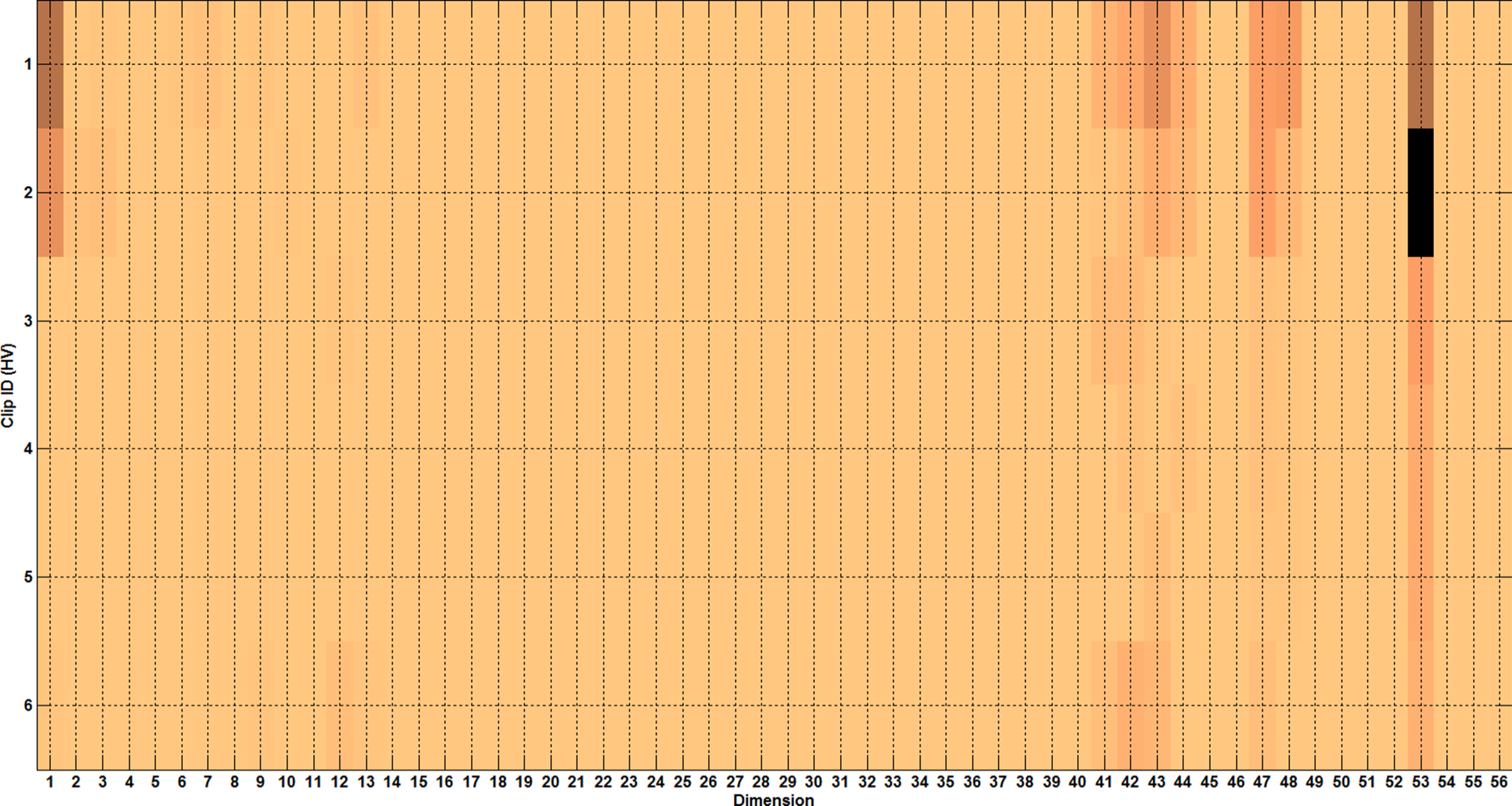}\hspace{.02cm}\includegraphics[width=0.23\linewidth]{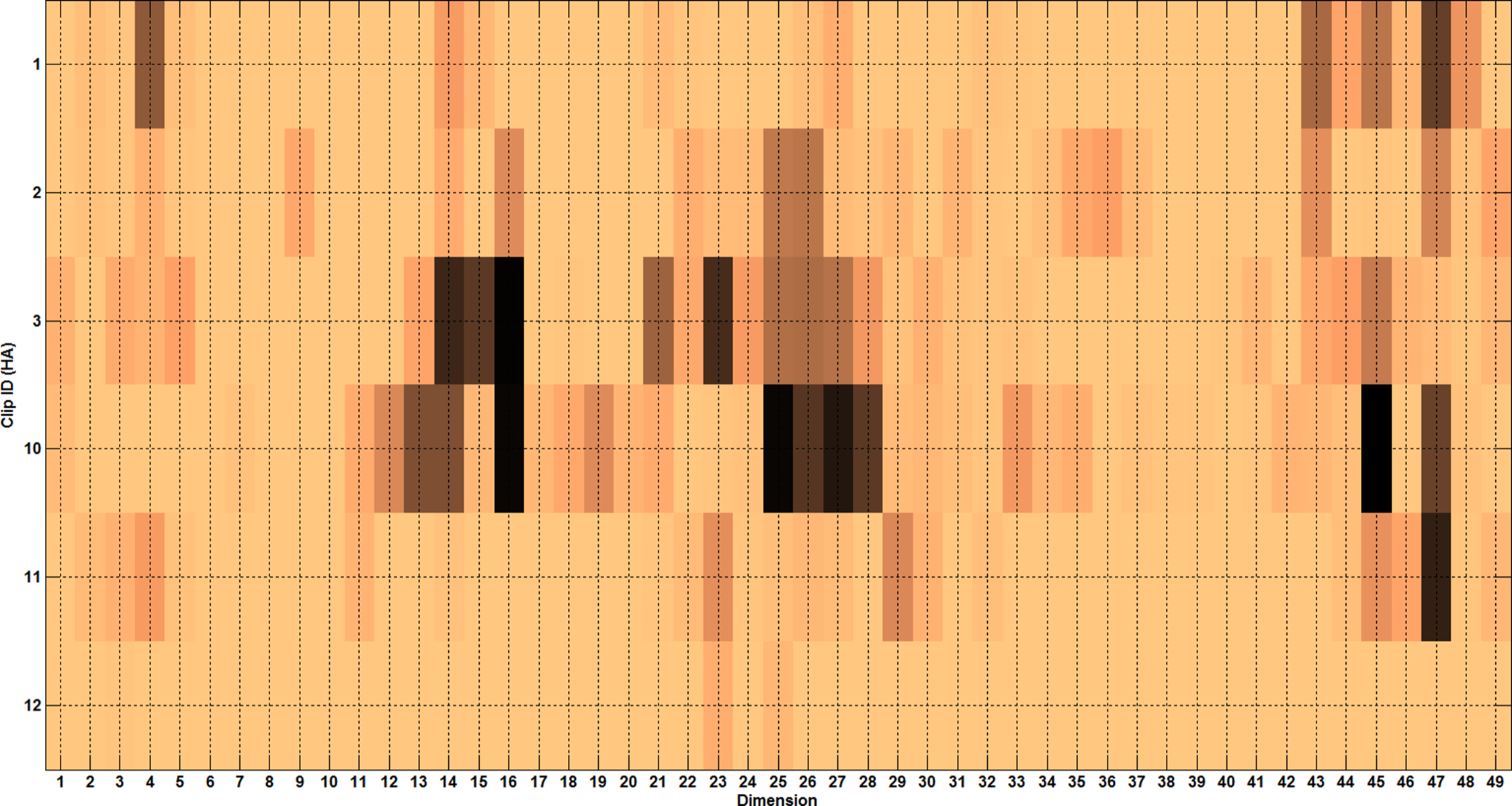}\hspace{.02cm}\includegraphics[width=0.23\linewidth]{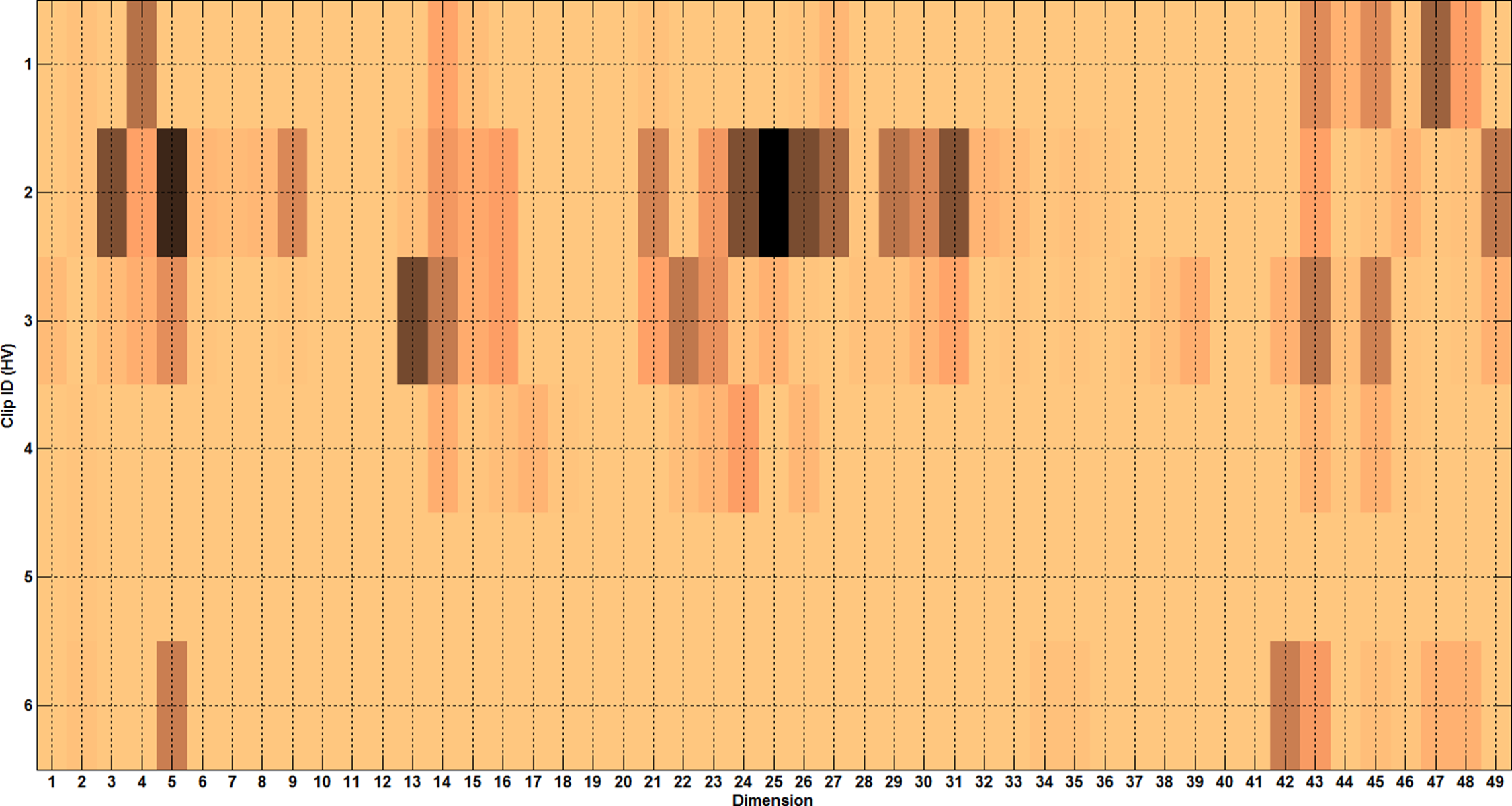}}\vspace{0.1cm}
\centerline{\includegraphics[width=0.23\linewidth]{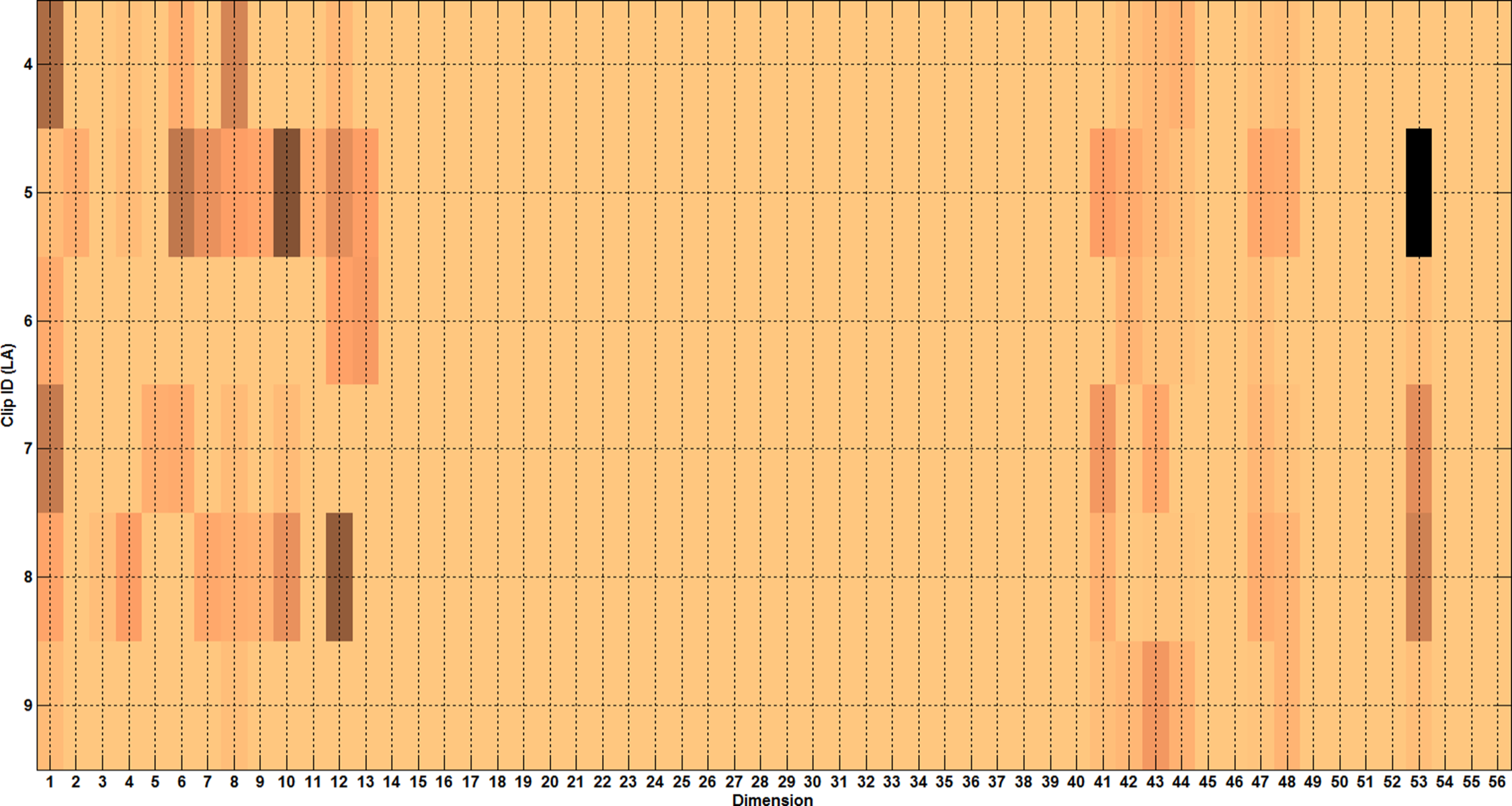}\hspace{.02cm}\includegraphics[width=0.23\linewidth]{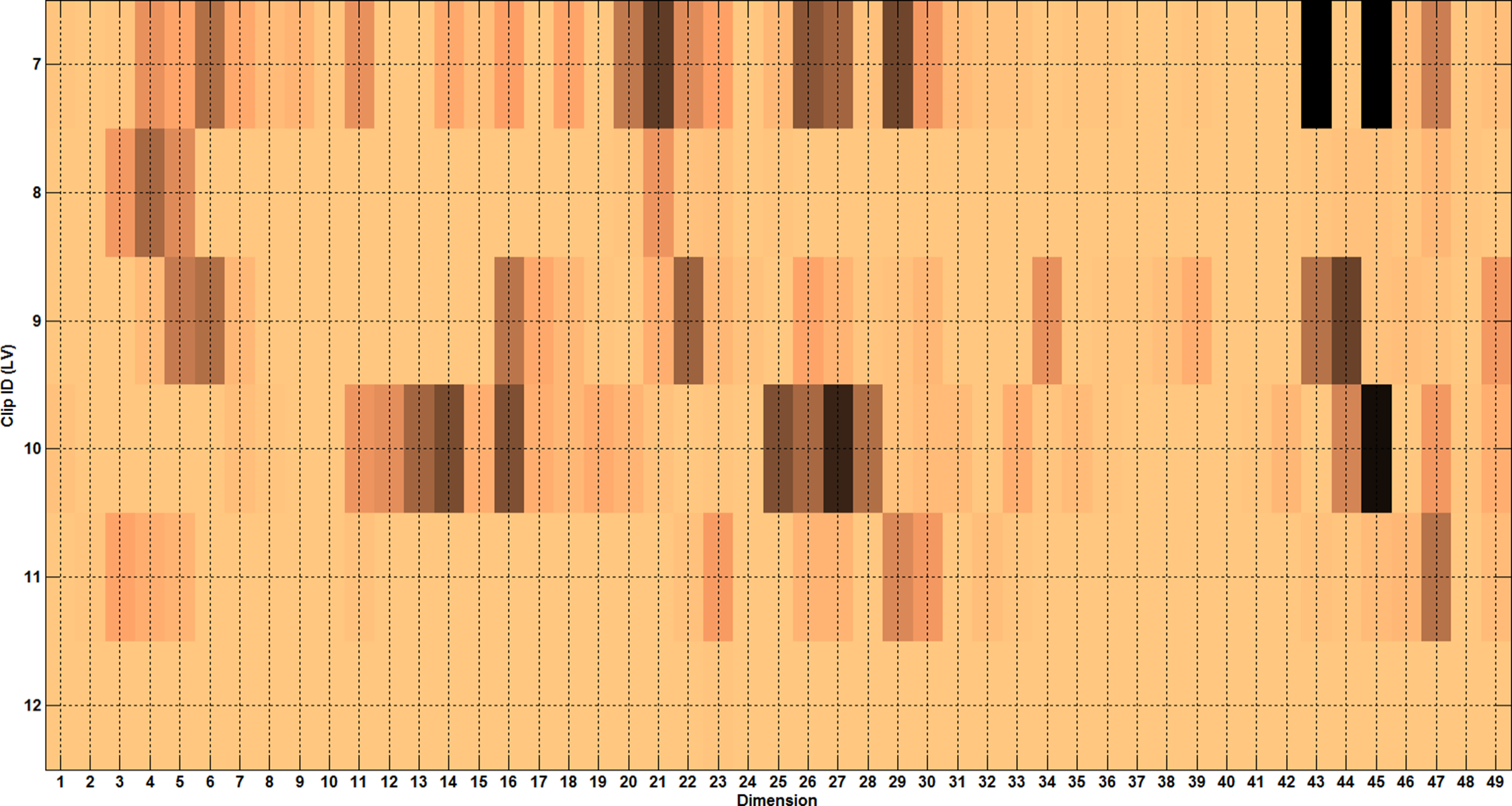}\hspace{.02cm}\includegraphics[width=0.23\linewidth]{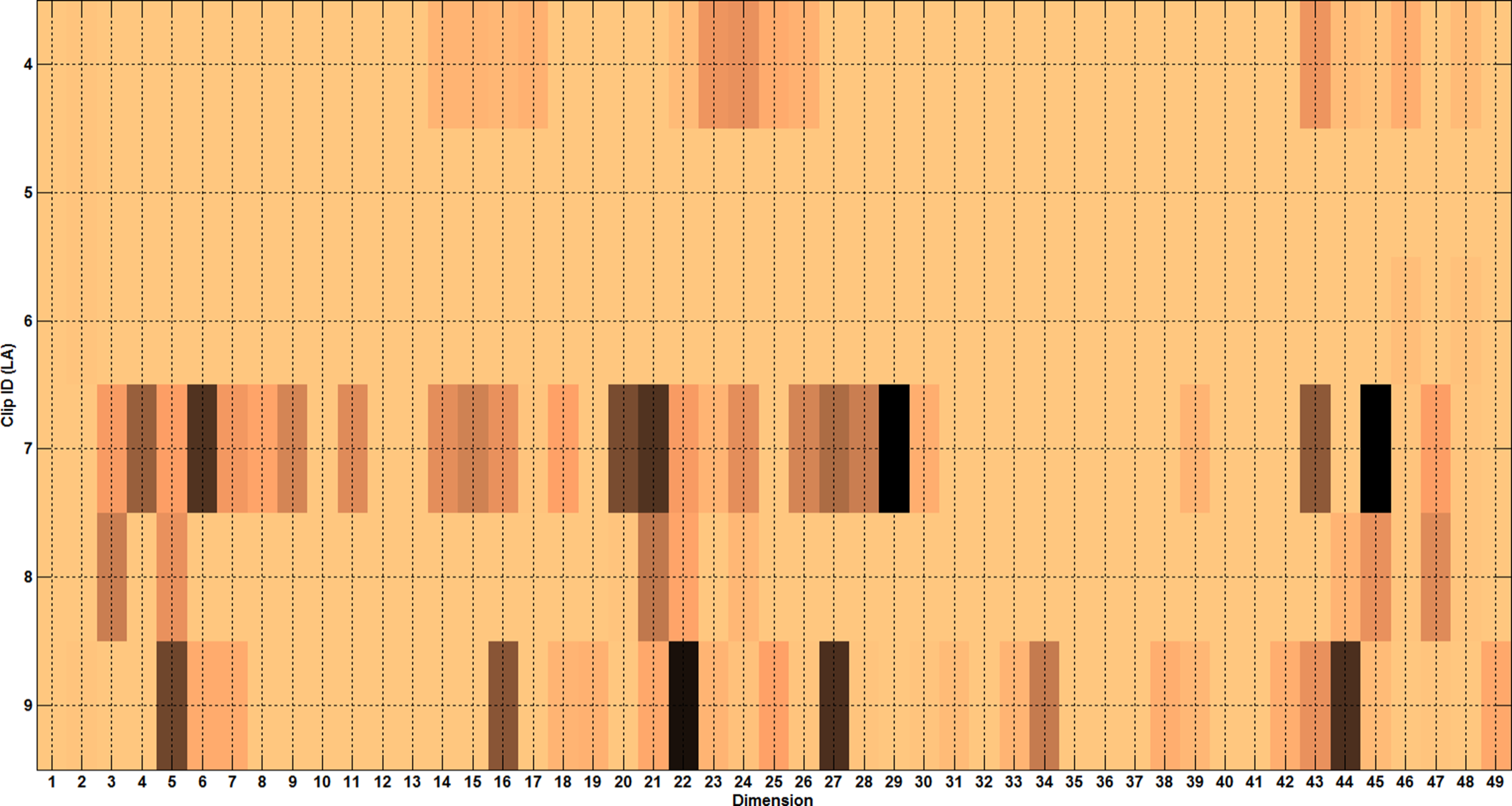}\hspace{.02cm}\includegraphics[width=0.23\linewidth]{SR_LV_vi.png}}\vspace{0.1cm}
\centerline{Arousal-Audio \hspace{0.12\linewidth} Valence-Audio \hspace{0.12\linewidth} Arousal-Video \hspace{0.12\linewidth} Valence-Video} 
\vspace{-.05cm}
\caption{MTL with \textbf{Val} movie clips as tasks and audio-visual descriptors as features. (From top to bottom) $W$ matrices obtained with  $\ell_{2,1}$, \textit{Dirty}, \textit{Robust} and \textit{Graph regularized} MTL (last two rows). Larger weights are denoted using darker shades. Best viewed under zoom.}
\label{Clips_AVfeat}
\end{figure*}

We also show how the movie clips are related in terms of the audio-visual features in Figure~\ref{Clips_AVfeat} as a final illustration. To this end, the various MTL baselines are trained with the extracted audio or visual features and the ground-truth A/V labels. The x-axes of the $W$ matrices shown in Figure~\ref{Clips_AVfeat} depict audio or visual feature dimensions in the same order as listed in Table~\ref{Features}. Formant features appear to be the most important audio descriptors for both static A,V prediction as seen from the $W$ matrices. Looking at the SRMTL $W$'s for high and low V in the second column of the last two rows, it is interesting to note that zero crossing rate and silence ratio make a salient contribution to LV clips but not to HV clips. Concerning video features, brightness and color descriptors 
are indicative of both A and V as expected. From the SRMTL $W$'s in the third and fourth columns of the last two rows, visual excitement and motion descriptors are seen to have a greater role in characterizing HA and HV clips rather than in LA and LV clips. 

Table~\ref{table_avrec} supplements Figure~\ref{Clips_AVfeat} by tabulating the root mean square error (RMSE) in predicted A,V levels for MTL baseline models trained with dynamic expert rating labels and audio/visual descriptors as features\footnote{Here again, RMSE is shown on the training set.}. Superior prediction of dynamic A levels is achieved with both audio and visual features. Also, with $W$ matrices of comparable sparsity, audio features are seen to be better predictors of both A,V as compared to the considered visual descriptors.  

\begin{table}
\scriptsize 
\renewcommand{\arraystretch}{1.2}
\caption{\label{table_ca} Arousal (A) and Valence (V) recognition accuracies achieved on the \textbf{Val} set with different MTL approaches using time-continuous crowd (Cwd) and expert (Exp) annotations as features. Values in parentheses denote sparsity level of the weight matrix $W$.}
\vspace{-.2cm}
\centering
\begin{tabular}{|p{0.25cm}|p{0.35cm}|p{1.1cm}|p{1.1cm}|p{1.1cm}|p{1.1cm}|p{1.1cm}|}
\hline 
& & \textbf{Lasso} & \textbf{$\ell_{21}$} &\textbf{Dirty}&\textbf{Robust}&\textbf{SR} \\ \hline
\multirow{2}{*}{\textbf{A}}&Cwd&0.83 (0.22) & 0.80 (0.16) & 0.79 (0.24) & 0.80 (0.32) & 0.84 (0.69) \\
                       &Exp&0.89 (0.16) & 0.85 (0.14) & 0.85 (0.24) & 0.87 (0.30) & 0.87 (0.66) \\
\multirow{2}{*}{\textbf{V}}&Cwd&0.83 (0.18) & 0.77 (0.16) & 0.82 (0.30) & 0.79 (0.28) & 0.86 (0.68) \\
                       &Exp&1.00 (0.12) & 0.99 (0.16) & 0.98 (0.20) & 0.99 (0.24) & 0.99 (0.61) \\ \hline
\end{tabular}
\end{table}

\begin{table}
\scriptsize 
\renewcommand{\arraystretch}{1.2}
\caption{\label{table_avrec} Time-continuous Arousal (A) and Valence (V) estimation performance achieved on the \textbf{Val} set with different MTL approaches using audio and video features. Model trained/tested with median of the experts' annotations in the range $[-1,1]$. Values in parentheses denote sparsity level of the weight matrix $W$. Lower RMSE indicates better performance.}
\vspace{-.2cm}
\centering
\begin{tabular}{|p{0.25cm}|p{0.35cm}|p{1.1cm}|p{1.1cm}|p{1.1cm}|p{1.1cm}|p{1.1cm}|}
\hline 
& & \textbf{Lasso} & \textbf{$\ell_{21}$} &\textbf{Dirty}&\textbf{Robust}&\textbf{SR} \\ \hline
\multirow{2}{*}{\textbf{A}}&Aud&0.65 (0.40) & 0.66 (0.21) & 0.62 (0.33) & 0.61 (0.37) & 0.57 (0.35) \\
                       &Vid&0.69 (0.32) & 0.71 (0.34) & 0.67 (0.34) & 0.68 (0.26) & 0.54 (0.60) \\
\multirow{2}{*}{\textbf{V}}&Aud&0.88 (0.38) & 0.88 (0.32) & 0.81 (0.21) & 0.81 (0.33) & 0.79 (0.39) \\
                       &Vid&0.93 (0.37) & 0.98 (0.30) & 0.87 (0.34) & 0.88 (0.32) & 0.73 (0.48) \\ \hline
\end{tabular}
\end{table}

\subsection{Experimental Results}

We now evaluate the proposed EG-MTL algorithm against other MTL baselines for problems \textbf{\texttt{P1}} and \textbf{\texttt{P2}}, and begin with \textbf{\texttt{P1}}. 

\subsubsection{Refining a regression model via EG-MTL}

\textbf{\texttt{P1}} \textbf{problem description:} To reiterate, \textbf{\texttt{P1}} involves refining a regression model trained from dynamic \textbf{Val} affective annotations provided by crowdworkers utilizing a small amount of expert knowledge, and employing this refined model to estimate A,V on 5--15 second snippets extracted from the \textbf{Val} clips. Therefore, in the optimization function (Eq.(1)), $\mathbf{X}$ and $\mathbf{P}$ are the audio/visual descriptors. Since a representative sequence of dynamic expert/crowdworker labels is required to denote $\mathbf{Y}$, $\mathbf{V}$, we considered the median of all expert/crowd annotations available for each movie clip to this end, since the median filter is less sensitive to outliers. 

\textbf{Experimental settings:} In our experiments, we randomly extracted 5, 10 and 15 second snippets corresponding to the same time points for all of the movie clips to be used as test data. Given that the latter portions of the movie clips are emotionally salient, we extracted these snippets from either the first or second halves of all movie clips. The remainder of the clips was used as training data. All MTL models were trained using audio/visual features corresponding to the training set time points, and the median crowd ratings corresponding to these time points were considered as the dependent variables. For the EG-MTL algorithm, median expert ratings for the training time points were fed as expert data to the training module. The best regularization parameter $\lambda_1$ controlling group sparsity was chosen from the set $[0.1,1,10,100]$ via five fold cross-validation on the training set for all methods. For the proposed EG-MTL algorithm, $\lambda_2$, $\lambda_3$ were set to 1. \\ \\ 

\textbf{Results and Discussion:} RMSE values (mean and standard deviation of five runs involving randomly selected training/test sets) for dynamic A,V prediction corresponding to the various approaches are tabulated in Table~\ref{table_AVreg}. EG-MTL (7) corresponds to the condition where the rating data of only seven experts, which by itself is insufficient to train a model, is used to refine crowd annotations via the proposed approach. Clearly, all multi-task methods outperform single-task Lasso regressor confirming the utility of MTL. Also, incorporation of a small amount of expert knowledge to refine the crowd model is highly beneficial as exemplified by the superior performance of EG-MTL and EG-MTL (7) with respect to the other methods in a vast majority of conditions. The fact that there is a considerable difference in prediction performance of EG-MTL and other approaches also highlights the amount of noise in the dynamic affective crowd annotations.  

RMSE values increase for longer test snippets (implying lesser training data). Furthermore, RMSE values are higher for snippets extracted from the second half of each movie clip as compared to first half snippets, as the latter part of all clips is known to be emotionally salient. Finally, considering the performance of the best audio and video-based models, audio features produce better prediction performance in general as compared to video descriptors, consistent with the observations from Table~\ref{table_avrec}. Finally, the considered video descriptors predict arousal better than valence, and this result also holds good for Table~\ref{table_avrec}.     

\subsubsection{Refining a classification model via EG-MTL}

We now focus on problem \textbf{\texttt{P2}} and discuss experimental results thereof. \\ \\

\textbf{\texttt{P2}} \textbf{problem description:} \textbf{\texttt{P2}} involves learning a binary classification model with the dynamic \textbf{Val} expert and crowd annotations as features, and corresponding static ground-truth A,V labels as data labels. This model is to be subsequently utilized for improving static A,V recognition performance on the \textbf{Eval} set for which only dynamic crowd annotations are available. In the optimization function (Eq.(1)), we therefore have $\mathbf{X}$ and $\mathbf{P}$ denoting the set of dynamic crowd and expert affective annotations repsectively, while $\mathbf{Y}$ and $\mathbf{V}$ are identically the set of static A,V labels for the \textbf{Val} movie clips. \\ \\

\textbf{Experimental settings:} All MTL models were trained using dynamic \textbf{Val} crowd A,V ratings over the final 50 second duration of the clip and the ground-truth labels available for the \textbf{Val} clips as part of~\cite{abadi2015decaf}. For the EG-MTL algorithm, dynamic expert A,V ratings were fed as expert data. As with \textbf{\texttt{P1}}, the best regularization parameter $\lambda_1$ controlling group sparsity was chosen from the set $[0.1,1,10,100]$ via five fold cross-validation on the training set for all methods. For the proposed EG-MTL algorithm, $\lambda_2$, $\lambda_3$ were set to 1. The learned $W$s from each method were then applied on the dynamic \textbf{Eval} crowd to predict static A,V labels for the \textbf{Eval} clips.  \\ \\ 

\textbf{Results and Discussion:}  Table~\ref{table_cav} presents A,V recognition accuracies obtained with the different MTL approaches on the \textbf{Eval} set. Evidently, all MTL methods outperform single-task Lasso for both A and V recognition, which reconfirms the benefit of employing multi-task learning for affect prediction. Among the various MTL baselines, SRMTL produces best recognition for both arousal and valence (along with $\ell_{2,1}$ MTL) due to the fact that the weights are adapted based on the graph specifying related tasks (clips). Nevertheless, the proposed EG-MTL approach produces the best recognition performance with EG-MTL (7), which incorporates data from only seven experts in the learning process producing very competitive performance. For each of the considered approaches, higher recognition performance is obtained for arousal as compared to valence on the \textbf{Eval} set, which can possibly be attributed to the low agreement among crowdworkers for HV clips (see Table~\ref{table_KV}).

\vspace{-.1in}
\begin{table*}[!htbp]
  \fontsize{8}{8}\selectfont
	\centering 
	\renewcommand{\arraystretch}{1.3}
  \caption{\label{table_AVreg} RMSE-based V,A prediction performance of task-specific vs multi-task methods. RMSE mean, standard deviation over five runs are reported. Best model RMSE is shown in bold.}
    \begin{tabular}{|c|cc|ccc|ccc|}
      \hline
			& & &\multicolumn{3}{c|}{\textbf{Front}}&\multicolumn{3}{c|}{\textbf{Back}} \\ \hline
			& & &5 s&10 s&15 s&5 s&10 s&15 s \\ \hline
			
			\multirow{14}{*}{\textbf{Valence}}&\multirow{7}{*}{\textbf{Video}}&Lasso&0.880$\pm$0.170 & 1.724$\pm$0.531 & 2.111$\pm$0.522 & 1.066$\pm$0.150 & 1.677$\pm$0.219 & 2.004$\pm$0.152\\
			& & MT-Lasso&{0.542$\pm$0.043} &  0.988$\pm$0.129&{1.202$\pm$0.034} & 0.685$\pm$0.078&1.050$\pm$0.202 & 1.331$\pm$0.094\\
			& & $\ell_{21}$ MTL& 0.526$\pm$0.032 & 0.959$\pm$0.148 & 1.183$\pm$0.089 & 0.685$\pm$0.078 &1.053$\pm$0.219 & 1.288$\pm$0.104\\			
			& & Dirty MTL& 0.564$\pm$0.019 & 0.889$\pm$0.076& 1.126$\pm$0.067 & 0.671$\pm$0.071 & 0.986$\pm$0.198 & 1.221$\pm$0.130 \\
			& & SR MTL&  0.494$\pm$0.040 & 0.735$\pm$0.026 & 0.904$\pm$0.043 & 0.666$\pm$0.103 & 1.078$\pm$0.189 & 1.290$\pm$0.124 \\ 
			& & EG-MTL&  \textbf{0.480$\pm$0.052} & \textbf{0.720$\pm$0.015} & \textbf{0.877$\pm$0.026} & \textbf{0.585$\pm$0.093} & \textbf{0.893$\pm$0.166} & \textbf{1.047$\pm$0.097} \\
			& & EG-MTL (7)&0.484$\pm$0.043 & 0.726$\pm$0.062  & 0.893$\pm$0.069 & 0.592$\pm$0.071 & 0.919$\pm$0.062 & 1.086$\pm$0.085 \\ \cline{2-9}
			
			&\multirow{7}{*}{\textbf{Audio}}&Lasso&0.819$\pm$0.083 & 1.227$\pm$0.055 & 1.437$\pm$0.077 & 1.098$\pm$0.143 & 1.541$\pm$0.133 & 2.064$\pm$0.072 \\
			& & MT-Lasso& 0.542$\pm$0.052 & 0.800$\pm$0.053 & 0.961$\pm$0.061 & 0.731$\pm$0.115 & 1.079$\pm$0.064 & 1.489$\pm$0.099 \\
			& & $\ell_{21}$ MTL&0.580$\pm$0.058 & 0.775$\pm$0.083 & 0.894$\pm$0.115 & 0.751$\pm$0.111 & 1.097$\pm$0.097 & 1.409$\pm$0.104 \\
			& & Dirty MTL& 0.528$\pm$0.032 & 0.758$\pm$0.024 &0.942$\pm$0.036 & 0.689$\pm$0.102 & 0.969$\pm$0.089 & 1.313$\pm$0.031 \\
			& & SR MTL&0.482$\pm$0.044 & 0.715$\pm$0.021 &0.829$\pm$0.027& 0.684$\pm$0.111 & 0.976$\pm$0.106 & 1.357$\pm$0.045 \\ 
			& & EG-MTL&\textbf{0.477$\pm$0.048} & \textbf{0.710$\pm$0.023} & \textbf{0.848$\pm$0.036} & \textbf{0.680$\pm$0.117} & \textbf{0.923$\pm$0.060} & \textbf{1.234$\pm$0.038} \\
			& & EG-MTL (7)&0.478$\pm$0.052 &0.711$\pm$0.023 & 0.859$\pm$0.046 & 0.683$\pm$0.110 & 0.972$\pm$0.038 & 1.309$\pm$0.123 \\ \hline
			
				\multirow{14}{*}{\textbf{Arousal}}&\multirow{7}{*}{\textbf{Video}}&Lasso&0.879+-0.205 & 1.661+-0.610 & 2.526+-0.632 & 0.841+-0.047 & 1.250+-0.075 & 1.501+-0.015 \\
				& & MT-Lasso& 0.482+-0.088 & 0.745+-0.066 & 1.055+-0.088 & 0.367+-0.060 & 0.665+-0.088 & 0.947+-0.039 \\
				& & $\ell_{21}$ MTL& 0.465+-0.090 & 0.729+-0.073 & 1.068+-0.112 & 0.355+-0.051 & 0.644+-0.075 & 0.914+-0.066 \\
				& & Dirty MTL&0.391+-0.085 & 0.683+-0.067 & 0.913+-0.059 & 0.334+-0.024 & 0.578+-0.098 & 0.685+-0.045 \\
				& & SR MTL& 0.365+-0.043 & 0.575+-0.028 & 0.761+-0.054 & 0.384+-0.036 & \textbf{0.640+-0.070 }& 0.811+-0.042 \\
				& & EG-MTL& \textbf{0.298+-0.056 }&  \textbf{0.479+-0.015} & \textbf{0.608+-0.010} & \textbf{0.313+-0.025} & 0.811+-0.042 & \textbf{0.549+-0.024} \\
			& & EG-MTL (7)&0.302$\pm$0.052 & 0.523$\pm$0.023 & 0.665$\pm$0.062 & 0.336$\pm$0.027& 0.792$\pm$0.038 & 0.625$\pm$0.052\\ \cline{2-9}
				
			&\multirow{7}{*}{\textbf{Audio}}&Lasso&2.146$\pm$0.122 & 3.705$\pm$0.242 & 4.474$\pm$0.699 & 1.673$\pm$0.241 & 2.240$\pm$0.285 & 2.412$\pm$0.548 \\
			& & MT-Lasso&0.183$\pm$0.070 & 0.307+-0.078 & 0.508+-0.017 & \textbf{0.218+-0.013} & 0.358+-0.040 & 0.460+-0.014 \\		
			& & $\ell_{21}$ MTL&0.183$\pm$0.070 & 0.307$\pm$0.078 & 0.508$\pm$0.017 & \textbf{0.218$\pm$0.013} & 0.358$\pm$0.040 & 0.460$\pm$0.014 \\
			& & Dirty MTL&  0.205$\pm$0.057 & 0.350$\pm$0.078 & 0.538$\pm$0.035 & 0.235$\pm$0.017 & 0.391$\pm$0.038 & 0.493$\pm$0.017 \\
			& & SR MTL&0.174$\pm$0.075 & 0.295$\pm$0.077& 0.505$\pm$0.018 & 0.227$\pm$0.014 & 0.366$\pm$0.038 & 0.486$\pm$0.019 \\
			& & EG-MTL& \textbf{0.161$\pm$0.060} & \textbf{0.252+-0.050} & \textbf{0.359$\pm$0.017} & 0.223$\pm$0.014 & \textbf{0.349$\pm$0.028} & \textbf{0.427$\pm$0.021} \\
			& & EG-MTL (7)&0.172$\pm$0.082 & 0.290$\pm$0.067 & 0.403$\pm$0.010 & 0.225$\pm$0.013 & 0.355$\pm$0.035 & 0.442$\pm$0.018 \\ \hline
         
    \end{tabular}%
  \label{regress}%
\end{table*}%

\begin{table*}[!htbp]
\fontsize{8}{8}\selectfont
\caption{\label{table_cav} Comparison of Arousal (A) and Valence (V) recognition accuracies obtained with different methods on the \textbf{Eval} set.} \vspace{-.3cm}
\begin{center}
\resizebox{0.99\linewidth}{!} {
\begin{tabular}{|l|c|c|c|c|c|c|c|c|c|c|c|}
\hline
  & {\textbf{ST-Lasso}}~\cite{tibshirani1996regression} & \textbf{MT-Lasso} & $\mathbf{\ell_{21}}$ \textbf{MTL}~\cite{Argyrious07} & \textbf{Dirty MTL}~\cite{Jalali} & \textbf{Robust MTL}~\cite{Gong12} & \textbf{SRMTL} & \textbf{EG-MTL} & \textbf{EG-MTL (7)}\\ \hline
\textbf{A} & 0.592 & 0.764 & 0.753 &  0.717 & 0.761 & 0.829 & \textbf{0.893} & {0.862}\\
\textbf{V} & 0.567 & 0.718 & 0.728 &  0.677 & 0.667 & 0.728 & \textbf{0.811} & {0.795}\\ 
\hline
\end{tabular}
}
\end{center}
\vspace{-0.5cm}
\end{table*}

\section{Conclusion}\label{Con}
Given that emotion perception is a highly subjective phenomenon, the difficulty in finding adequate number of reliable annotators has forced most affect recognition approaches to involve only small or mid-sized user populations. Crowdsourcing, which has been successfully utilized in natural language processing and computer vision applications, presents an attractive proposition for the affective computing community in this regard, and can be employed to generate large amounts of training data in an inexpensive manner. Nevertheless, crowd generated data can be extremely noisy and requires robust and efficient machine learning techniques so as to be useful for learning generalizable models. This work, which proposes to use a small amount of expert data in order to improve/refine crowd models via the expert-guided Multi-task learning (EG-MTL) algorithm, is one of the first steps towards making crowdsourced data utilizable for affective computing applications.  

In this paper, we have shown how EG-MTL can be effectively used to (i) refine a crowd-based regression model, and enhance prediction of arousal and valence levels for 5--15 second snippets, and (ii) effectively learn classification weights using crowd-plus-expert data on one (\textbf{Val}) dataset, and utilize this knowledge to improve binary emotion recognition performance on a second (\textbf{Eval}) set for which only crowd data is available. Apart from presenting benchmarking results, we also illustrate how MTL can be utilized to determine discriminative features for arousal and valence prediction, and also identify time points from the dynamic emotion profile that have the greatest contributions towards determining the static arousal and valence for a movie clip. We believe that multi-task learning which attempts to learn the relationships between multiple tasks is naturally suited to (especially dimensional) affect recognition, but has surprisingly not been utilized for the same.  

Even though EG-MTL achieves superior performance by incorporating some expert knowledge in the learning framework, its full potential is yet to fully realized. For example, we did not explicitly define the related tasks (clips) via the graph regularization term in Eq.(1), and an automated framework for determining the optimal $\gamma_{ij}$'s based on the input data needs to be developed. Future work will involve investigation of these aspects, and development of a MTL-based framework to fuse multiple dynamic annotations so as to compute a representative emotional profile for affective media from crowd-generated data.     

\section{Acknowledgements}\label{Ack}

Many thanks to Mojtaba Khomami Abadi and Azad Abad for the crowdsourcing protocol implementation. The authors acknowledge that this research was funded by the research grant for UIUC ADSC's the Human-Centered Cyber-physical Systems Programme from Singapore's Agency for Science, Technology and Research (A*STAR), and the FIRB 2008 project S-PATTERNS.


%

%
%
%
%
%

\ifCLASSOPTIONcaptionsoff
  \newpage
\fi

\bibliographystyle{abbrv}
\bibliography{IEEEabrv,COAF-ACMM14}

\begin{thebibliography}{10}

\bibitem{abadi2015decaf}
M.~Abadi, R.~Subramanian, S.~Kia, P.~Avesani, I.~Patras, and N.~Sebe.
\newblock {DECAF}: {MEG}-based multimodal database for decoding affective
  physiological responses.
\newblock {\em IEEE Transactions on Affective Computing}, 2015.

\bibitem{Abadi2014}
M.~K. Abadi, A.~Abad, R.~Subramanian, N.~Rostamzadeh, E.~Ricci, J.~Varadarajan,
  and N.~Sebe.
\newblock A multi-task learning framework for time-continuous emotion
  estimation from crowd annotations.
\newblock In {\em CrowdMM Workshop}, pages 17--23, 2014.

\bibitem{DBLP:conf/acii/AbadiKRAS13}
M.~K. Abadi, S.~M. Kia, S.~Ramanathan, P.~Avesani, and N.~Sebe.
\newblock User-centric affective video tagging from {MEG} and peripheral
  physiological responses.
\newblock In {\em 2013 Humaine Association Conference on Affective Computing
  and Intelligent Interaction, {ACII} 2013, Geneva, Switzerland, September 2-5,
  2013}, pages 582--587. {IEEE} Computer Society, 2013.

\bibitem{Argyrious07}
A.~Argyriou, T.~Evgeniou, and M.~Pontil.
\newblock Multi-task feature learning.
\newblock In {\em Neural Information Processing Systems}, 2007.

\bibitem{Beck09}
A.~Beck and M.~Teboulle.
\newblock A fast iterative shrinkage- thresholding algorithm for linear inverse
  problems.
\newblock {\em SIAM Journal on Imaging Sciences}, 2(1):183--202, 2009.

\bibitem{Boyd}
S.~Boyd, N.~Parikh, E.~Chu, B.~Peleato, and J.~Eckstein.
\newblock Distributed optimization and statistical learning via the alternating
  direction method of multipliers.
\newblock {\em Foundations and Trends in Machine Learning}, 3(1):1--122, 2010.

\bibitem{caruana1998multitask}
R.~Caruana.
\newblock {\em Multitask learning}.
\newblock Springer, 1998.

\bibitem{chen2006mixed}
L.~Chen, S.~Gunduz, and M.~T. Ozsu.
\newblock Mixed type audio classification with support vector machine.
\newblock In {\em IEEE Int'l Conference on Multimedia and Expo}, pages
  781--784, 2006.

\bibitem{Gtrace}
R.~Cowie, G.~McKeown, and E.~Douglas-Cowie.
\newblock Tracing emotion: an overview.
\newblock {\em Int'l Journal of Synthetic Emotions (IJSE)}, 3(1):1--17, 2012.

\bibitem{Gong12}
P.~Gong, J.~Ye, and C.~Zhang.
\newblock Robust multi-task feature learning.
\newblock In {\em ACM Int'l conference on Knowledge discovery and data mining},
  2012.

\bibitem{gross1995emotion}
J.~J. Gross and R.~W. Levenson.
\newblock Emotion elicitation using films.
\newblock {\em Cognition \& Emotion}, 9(1):87--108, 1995.

\bibitem{HanjalicITM2005}
A.~Hanjalic and L.-Q. Xu.
\newblock Affective video content representation and modeling.
\newblock {\em IEEE Transactions on Multimedia}, 7(1):143--154, 2005.

\bibitem{Jalali}
A.~Jalali, P.~Ravikumar, S.~Sanghavi, and C.~Ruan.
\newblock A dirty model for multi-task learning.
\newblock In {\em Neural Information Processing Systems}, 2010.

\bibitem{AAAIHiro}
H.~Kajino, Y.~Tsuboi, and H.~Kashima.
\newblock A convex formulation for learning from crowds.
\newblock In {\em AAAI Conference on Artificial Intelligence}, 2012.

\bibitem{10.1145/1873951.1874047}
H.~Katti, R.~Subramanian, M.~Kankanhalli, N.~Sebe, T.-S. Chua, and K.~R.
  Ramakrishnan.
\newblock Making computers look the way we look: Exploiting visual attention
  for image understanding.
\newblock In {\em Proceedings of the 18th ACM International Conference on
  Multimedia}, MM '10, pages 667--670, New York, NY, USA, 2010. Association for
  Computing Machinery.

\bibitem{Kehoe2012a}
E.~G. Kehoe, J.~M. Toomey, J.~H. Balsters, and A.~L.~W. Bokde.
\newblock {Personality modulates the effects of emotional arousal and valence
  on brain activation.}
\newblock {\em Social Cognitive \& Affective Neuroscience}, 7:858--70, 2012.

\bibitem{DEAP}
S.~Koelstra, C.~Muhl, M.~Soleymani, J.-S. Lee, A.~Yazdani, T.~Ebrahimi, T.~Pun,
  A.~Nijholt, and I.~Patras.
\newblock Deap: A database for emotion analysis; using physiological signals.
\newblock {\em IEEE Transactions on Affective Computing}, 3(1):18--31, 2012.

\bibitem{PatrasIVC}
S.~Koelstra and I.~Patras.
\newblock Fusion of facial expressions and eeg for implicit affective tagging.
\newblock {\em Image and Vision Computing}, 31(2):164--174, 2013.

\bibitem{lerch2012introduction}
A.~Lerch.
\newblock {\em An introduction to audio content analysis: Applications in
  signal processing and music informatics}.
\newblock John Wiley \& Sons, 2012.

\bibitem{li2001classification}
D.~Li, I.~K. Sethi, N.~Dimitrova, and T.~McGee.
\newblock Classification of general audio data for content-based retrieval.
\newblock {\em Pattern Recognition Letters}, 22(5):533--544, 2001.

\bibitem{lucas1981iterative}
B.~D. Lucas, T.~Kanade, et~al.
\newblock An iterative image registration technique with an application to
  stereo vision.
\newblock In {\em Int'l Joint Conference on Artificial Intelligence},
  volume~81, pages 674--679, 1981.

\bibitem{mcduff2012crowdsourcing}
D.~McDuff, R.~Kaliouby, and R.~W. Picard.
\newblock Crowdsourcing facial responses to online videos.
\newblock {\em IEEE Transactions on Affective Computing}, 3(4):456--468, 2012.

\bibitem{AngelikiFG13}
A.~Metallinou and S.~Narayanan.
\newblock Annotation and processing of continuous emotional attributes:
  Challenges and opportunities.
\newblock In {\em Emo{SPACE} Workshop}, pages 1--8, 2013.

\bibitem{mustafa2006robust}
K.~Mustafa and I.~C. Bruce.
\newblock Robust formant tracking for continuous speech with speaker
  variability.
\newblock {\em IEEE Transactions on Audio, Speech, and Language Processing},
  14(2):435--444, 2006.

\bibitem{Nicolaou2012}
M.~A. Nicolaou, H.~Gunes, and M.~Pantic.
\newblock Output-associative {RVM} regression for dimensional and continuous
  emotion prediction.
\newblock {\em Image and Vision Computing}, 30(3):186--196, 2012.

\bibitem{MihalisTPAMI}
M.~A. {Nicolaou}, V.~{Pavlovic}, and M.~{Pantic}.
\newblock Dynamic probabilistic cca for analysis of affective behaviour and
  fusion of continuous annotations.
\newblock {\em IEEE {T}ransactions on {P}attern {A}nalysis and {M}achine
  {I}ntelligence}, 36(7):1299--1311, 2014.

\bibitem{picard2000affective}
R.~W. Picard.
\newblock {\em Affective computing}.
\newblock MIT press, 2000.

\bibitem{Raykar2010}
V.~C. Raykar, S.~Yu, L.~H. Zhao, G.~H. Valadez, C.~Florin, L.~Bogoni, and
  L.~Moy.
\newblock Learning from crowds.
\newblock {\em Journal of Machine Learning Research}, 11:1297--1322, 2010.

\bibitem{ross2010crowdworkers}
J.~Ross, L.~Irani, M.~Silberman, A.~Zaldivar, and B.~Tomlinson.
\newblock Who are the crowdworkers?: shifting demographics in {M}echanical
  {T}urk.
\newblock In {\em Human Factors in Computing Systems}, pages 2863--2872, 2010.

\bibitem{russell1980circumplex}
J.~Russell.
\newblock {A circumplex model of affect}.
\newblock {\em Journal of personality and social psychology}, 39:1161--1178,
  1980.

\bibitem{10.1145/3242969.3242988}
A.~Shukla, H.~Katti, M.~Kankanhalli, and R.~Subramanian.
\newblock Looking beyond a clever narrative: Visual context and attention are
  primary drivers of affect in video advertisements.
\newblock In {\em Proceedings of the 20th ACM International Conference on
  Multimodal Interaction}, ICMI '18, pages 210--219, New York, NY, USA, 2018.
  Association for Computing Machinery.

\bibitem{soleymaniCMM13}
M.~Soleymani, M.~N. Caro, E.~M. Schmidt, C.-Y. Sha, and Y.-H. Yang.
\newblock 1000 songs for emotional analysis of music.
\newblock In {\em CrowdMM Workshop}, pages 1--6, 2013.

\bibitem{soleymani2008affective}
M.~Soleymani, G.~Chanel, J.~J. Kierkels, and T.~Pun.
\newblock Affective characterization of movie scenes based on multimedia
  content analysis and user's physiological emotional responses.
\newblock In {\em IEEE Int'l Symposium on Multimedia}, pages 228--235, 2008.

\bibitem{soleymani2009collaborative}
M.~Soleymani, J.~Davis, and T.~Pun.
\newblock A collaborative personalized affective video retrieval system.
\newblock In {\em Affective Computing and Intelligent Interaction}, pages 1--2,
  2009.

\bibitem{pun011}
M.~Soleymani and M.~Larson.
\newblock Crowdsourcing for affective annotation of video: development of a
  viewer-reported boredom corpus.
\newblock In {\em Workshop on Crowdsourcing for Search Evaluation}, 2010.

\bibitem{MAHNOB-HCI}
M.~Soleymani, J.~Lichtenauer, T.~Pun, and M.~Pantic.
\newblock A multimodal database for affect recognition and implicit tagging.
\newblock {\em IEEE Transactions on Affective Computing}, 3:42--55, 2012.

\bibitem{Steiner2011}
T.~Steiner, R.~Verborgh, R.~Van~de Walle, M.~Hausenblas, and J.~G. Vall{\'e}s.
\newblock Crowdsourcing event detection in youtube video.
\newblock In M.~Van~Erp, W.~R. Van~Hage, L.~Hollink, A.~Jameson, and R.~Troncy,
  editors, {\em Workshop on detection, representation, and exploitation of
  events in the semantic web}, pages 58--67, 2011.

\bibitem{tibshirani1996regression}
R.~Tibshirani.
\newblock Regression shrinkage and selection via the lasso.
\newblock {\em Journal of the Royal Statistical Society. Series B
  (Methodological)}, pages 267--288, 1996.

\bibitem{valdez1994effects}
P.~Valdez and A.~Mehrabian.
\newblock Effects of color on emotions.
\newblock {\em Journal of Experimental Psychology: General}, 123(4):394, 1994.

\bibitem{VondrickRP10}
C.~Vondrick, D.~Patterson, and D.~Ramanan.
\newblock Efficiently scaling up crowdsourced video annotation.
\newblock {\em Int'l Journal of Computer Vision}, 101(1):184--204, 2013.

\bibitem{wang2006affective}
H.~L. Wang and L.-F. Cheong.
\newblock Affective understanding in film.
\newblock {\em IEEE Transactions on Circuits and Systems for Video Technology},
  16(6):689--704, 2006.

\bibitem{williams2011crowd}
J.~D. Williams, I.~D. Melamed, T.~Alonso, B.~Hollister, and J.~Wilpon.
\newblock Crowd-sourcing for difficult transcription of speech.
\newblock In {\em IEEE Workshop on Automatic Speech Recognition and
  Understanding (ASRU)}, pages 535--540, 2011.

\bibitem{yan2013no}
Y.~Yan, E.~Ricci, R.~Subramanian, O.~Lanz, and N.~Sebe.
\newblock No matter where you are: Flexible graph-guided multi-task learning
  for multi-view head pose classification under target motion.
\newblock In {\em International Conference in Computer Vision}, 2013.

\bibitem{yan2014multi}
Y.~Yan, E.~Ricci, R.~Subramanian, G.~Liu, and N.~Sebe.
\newblock Multi-task linear discriminant analysis for view invariant action
  recognition.
\newblock {\em {IEEE {T}ransactions on Image Processing}}, 2014.

\bibitem{yuan}
X.~Yuan and S.~Yan.
\newblock Visual classification with multi-task joint sparse representation.
\newblock In {\em Computer Vision and Pattern Recognition}, 2010.

\bibitem{YuenRLT09}
J.~Yuen, B.~C. Russell, C.~Liu, and A.~Torralba.
\newblock Labelme video: Building a video database with human annotations.
\newblock In {\em Int'l Conference on Computer Vision}, pages 1451--1458, 2009.

\bibitem{zhou2012mutal}
J.~Zhou, J.~Chen, and J.~Ye.
\newblock {\em MALSAR: Multi-tAsk Learning via StructurAl Regularization}.
\newblock Arizona State University, 2011.

\end{thebibliography}



\vspace*{-2\baselineskip}
\begin{IEEEbiography}[{\includegraphics[width=1in,height=1.25in,clip,keepaspectratio]{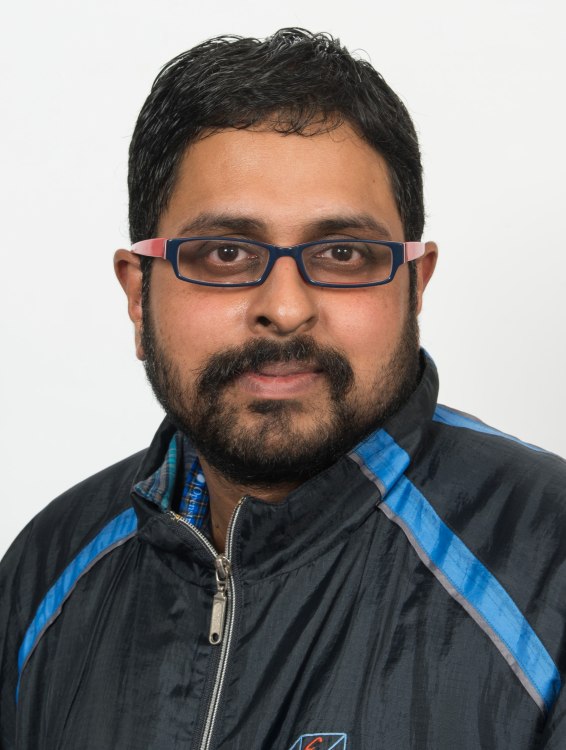}}]{Ramanathan Subramanian} received his Ph.D. in Electrical and Computer Engg. from NUS in 2008. He is an Associate Professor in University of Canberra, Australia. His past affiliations include IHPC (Singapore), U Glasgow (Singapore), IIIT Hyderabad (India), IIT Ropar (India) and UIUC-ADSC (Singapore). His research focuses on Human-centered computing, Interactive analytics and Explainable machine learning. He is an IEEE Senior Member and a member of the ACM and AAAC.
\end{IEEEbiography}
\vspace{-.4in}
\begin{IEEEbiography}[{\includegraphics[width=1in,height=1.25in,clip,keepaspectratio]{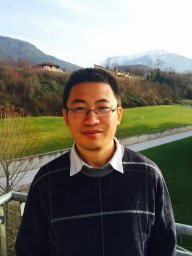}}]{Yan Yan} rreceived his Ph.D. in computer science from the University of Trento and is an Assistant Professor of the Computer Science Department at the Illinois Institute of Technology. His research interests include computer vision, machine learning, and multimedia. He has been PC member for several major conferences and reviewer for referred journals in computer vision and multimedia.
\end{IEEEbiography}
\vspace{-.4in}
\begin{IEEEbiography}[{\includegraphics[width=25mm,height=32mm,clip,keepaspectratio]{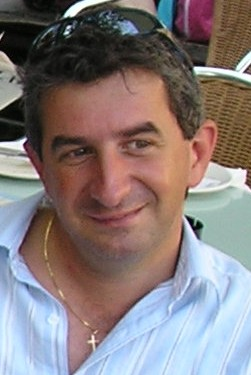}}]{Nicu Sebe}
is with the Faculty of Cognitive Sciences, University of Trento, Italy, where he is leading research in multimedia information retrieval and human-computer interaction in computer vision applications. He has been involved in the organization of major conferences and workshops on computer vision and human-centered aspects of multimedia information retrieval, among which he served as a general co-chair of the IEEE FG08, CIVR2007,2010 and WIAMIS09 and was one of the initiators and a program co-chair for the Human-Centered Multimedia track in ACMMM07. He was general chair of ACMMM2013 and program chair of ACMMM2011. He is a senior member of the IEEE.
\end{IEEEbiography}

\end{document}